\documentclass[a4paper]{article}

\usepackage{mlmodern}
\usepackage{amsfonts,amssymb,amsthm,ellipsis}
\usepackage{upgreek,mathrsfs}

\usepackage{twemojis} 
\usepackage{bold-extra}

\usepackage[normalem]{ulem} 

\usepackage{silence}
\usepackage{microtype}

\usepackage[dvipsnames]{xcolor}
\definecolor{maroon}{RGB}{167,74,74}
\definecolor{magreen}{RGB}{59,125,37}
\definecolor{mablue}{RGB}{59,136,195}
\definecolor{mayellow}{RGB}{242,147,24}
\colorlet{mabrown}{maroon!50!magreen}
\colorlet{maorange}{maroon!50!mayellow}
\colorlet{macyan}{magreen!50!mablue}

\colorlet{red}{maroon}
\colorlet{green}{magreen}
\colorlet{blue}{mablue}
\colorlet{yellow}{mayellow}
\colorlet{brown}{mabrown}
\colorlet{orange}{maorange}
\colorlet{cyan}{macyan}

\definecolor{rred}{RGB}{167,33,74}
\definecolor{bblue}{RGB}{29,136,255}
\definecolor{ppurple}{RGB}{113,84,165}
\definecolor{ppink}{RGB}{255,55,219}

\RequirePackage{hyperref}
\hypersetup{colorlinks,
    linkcolor=mablue,
    citecolor=magreen,
    urlcolor=maroon,
    anchorcolor=mablue,
    filecolor=mablue,
    menucolor=mablue}

\RequirePackage[top=10mm,bottom=12mm,left=30mm,right=30mm,head=12mm,includeheadfoot]{geometry}
\makeatletter
\g@addto@macro\bfseries{\boldmath}
\makeatother

\makeatletter
\renewcommand\section{\@startsection{section}{1}{\z@}%
  {-3.5ex \@plus -1.3ex \@minus -.7ex}%
  {2.3ex \@plus.4ex \@minus .4ex}%
  {\large\bfseries}}
\renewcommand\subsection{\@startsection{subsection}{2}{\z@}%
    {-2.3ex\@plus -1ex \@minus -.5ex}%
    {1.2ex \@plus .3ex \@minus .3ex}%
    {\normalsize\bfseries}}
\renewcommand\subsubsection{\@startsection{subsubsection}{3}{\z@}%
    {-2.3ex\@plus -1ex \@minus -.5ex}%
    {1ex \@plus .2ex \@minus .2ex}%
    {\normalsize\bfseries}}
\renewcommand\paragraph{\@startsection{paragraph}{4}{\z@}%
    {1.75ex \@plus1ex \@minus.2ex}%
    {-1em}%
    {\normalsize\bfseries}}
\renewcommand\subparagraph{\@startsection{subparagraph}{5}{\parindent}%
    {1.75ex \@plus1ex \@minus .2ex}%
    {-1em}%
    {\normalsize\bfseries}}
\makeatother

\RequirePackage{tocloft}

\RequirePackage[nottoc,notlot,notlof]{tocbibind}

\usepackage{comment}

\usepackage{import}
\usepackage{enumitem}
\usepackage{csquotes}

\usepackage[new]{old-arrows}

\renewcommand{\geq}{\geqslant}
\renewcommand{\leq}{\leqslant}

\usepackage{physics}
\usepackage{stackengine}

\newcommand{\pd}{\partial} 
\let\f\relax
\newcommand{\f}[2]{{#1}_{\qty[#2]}} 
\newcommand{\gf}[2]{{#1}^{\qty[#2]}} 

\newcommand{\bm}[1]{\boldsymbol{#1}} 

\newcommand{\tps}[2]{\texorpdfstring{\ensuremath{#1}}{#2}} 

\newcommand{\suchthat}{\;\middle|\;} 

\renewcommand{\vec}{\boldsymbol} 

\newcommand{\w}{\mathbin{\scalebox{0.8}{$\wedge$}}} 
\newcommand{\ex}[1]{\mathrm{e}^{#1}} 
\newcommand{\ii}{i} 

\let\C\undefined
\let\U\undefined

\newcommand{\R}{\mathbb{R}} 
\newcommand{\C}{\mathbb{C}} 
\newcommand{\Z}{\mathbb{Z}} 
\renewcommand{\S}{S} 

\newcommand{\B}{\mathrm{B}} 
\newcommand{\set}[1]{\qty{#1}} 

\newcommand{\U}{\mathrm{U}} 
\renewcommand{\u}{\alg{u}} 

\newcommand{\1}{\mathbf{1}} 

\renewcommand{\t}[1]{{\text{#1}}} 
\newcommand{\xmapsto}[1]{\overset{#1}{\mapsto}} 
\renewcommand{\mod}{\ \text{mod}\ } 

\makeatletter
\renewcommand{\xmapsto}[2][]{\ext@arrow 0599{\mapstofill@}{#1}{#2}}
\def\mapstofill@{\arrowfill@{\mapstochar\relbar}\relbar\rightarrow}
\makeatother

\let\eval\relax
\newcommand*{\eval}[1]{\left.#1\vphantom{\Big|}\right\rvert}

\def \cA {\mathcal{A}}
\def \cB {\mathcal{B}}

\def \cD {\mathcal{D}}

\def \cF {\mathcal{F}}

\def \cH {\mathcal{H}}
\def \cI {\mathcal{I}}
\def \cJ {\mathcal{J}}
\def \cK {\mathcal{K}}

\def \cN {\mathcal{N}}
\def \cO {\mathcal{O}}

\def \cQ {\mathcal{Q}}

\def \cS {\mathcal{S}}

\def \cW {\mathcal{W}}

\def \cY {\mathcal{Y}}

\DeclareSymbolFont{bbold}{U}{bbold}{m}{n}
\DeclareSymbolFontAlphabet{\mathbbold}{bbold}

\def \bbA {\mathbb{A}}

\def \bbW {\mathbb{W}}

\def \bbY {\mathbb{Y}}

\def \bb1 {{\mathbb{1}}}

\def \sfD {\mathsf{D}}

\def \sfH {\mathsf{H}}

\def \sfM {\mathsf{M}}

\def \sfT {\mathsf{T}}

\def \sfm {\mathsf{m}}
\def \sfn {\mathsf{n}}

\def \bR {\mathbb{R}}
\def \bZ {\mathbb{Z}}

\newcommand{\g}{\mathrm{g}} 

\newcommand{\sd}{\dd_\Sigma}

\newcommand{\cdd}{\dd^{\dagger}}
\newcommand{\csd}{\sd^{\dagger}}
\newcommand{\lapl}{\Delta}

\usepackage{tikz} 

\usetikzlibrary{decorations.markings,arrows.meta,svg.path}

\tikzset{line/.style={line width=0.25mm},
curve/.style={line,smooth,tension=1},
->-/.style={decoration={
  markings,
  mark=at position #1 with {\arrow[>=stealth]{>}}},postaction={decorate}},
-<-/.style={decoration={
  markings,
  mark=at position #1 with {\arrow[>=stealth]{<}}},postaction={decorate}},
}

\tikzset{bg/.style={opacity=.5}}

\usepackage{tikz-cd}
\usepackage{multirow}
\RequirePackage{mdframed}
\usepackage{empheq}

\definecolor{pansypurple}{rgb}{0.47, 0.09, 0.29}
\definecolor{patriarch}{rgb}{0.5, 0.0, 0.5}
\definecolor{carmine}{rgb}{0.59, 0.0, 0.09}
\definecolor{blueflag}{rgb}{0.2, 0.2, 0.6}
\usepackage{tikz-3dplot}

\def\fdiffd{\mathcal{D}}
\DeclareDocumentCommand\fdifferential{ o g d() }{ 
    \IfNoValueTF{#2}{
        \IfNoValueTF{#3}
        {\fdiffd\IfNoValueTF{#1}{}{^{#1}}}
        {\mathinner{\fdiffd\IfNoValueTF{#1}{}{^{#1}}\argopen(#3\argclose)}}
    }
    {\mathinner{\fdiffd\IfNoValueTF{#1}{}{^{#1}}#2} \IfNoValueTF{#3}{}{(#3)}}
}
\DeclareDocumentCommand\DD{}{\fdifferential}

\DeclareDocumentCommand\variation{ o g d() }{ 
    \IfNoValueTF{#2}{
        \IfNoValueTF{#3}
        {\updelta \IfNoValueTF{#1}{}{^{#1}}}
        {\mathinner{\updelta \IfNoValueTF{#1}{}{^{#1}}\argopen(#3\argclose)}}
    }
    {\mathinner{\updelta \IfNoValueTF{#1}{}{^{#1}}#2} \IfNoValueTF{#3}{}{(#3)}}
}

\DeclareDocumentCommand\wedgecommutator{ l m m }{\braces#1{\lbrack}{\rbrack}{#2\w #3}} 

\DeclareMathOperator{\volume}{vol}
\DeclareDocumentCommand\vol{}{\opbraces{\volume}}

\renewcommand{\ip}[2]{\left\langle#1,#2\right\rangle}

\RequirePackage{interval}
\intervalconfig{soft open fences}

\newcommand{\closed}[2]{\interval{#1}{#2}}

\usepackage[noabbrev]{cleveref}

\crefname{subsection}{subsection}{subsections}
\crefname{equation}{}{}

\numberwithin{equation}{section}

\usepackage[style=numeric-comp, eprint=true, natbib, backend=biber, maxbibnames=10, giveninits=true, isbn=false, url=false, doi=false, date=year, sorting=none, useprefix=true]{biblatex}

\usepackage{hyperref}

\DeclareFieldFormat*{title}{\emph{#1}}

\DeclareFieldFormat*{journaltitle}{#1}

\DeclareFieldFormat{pages}{#1}

\DeclareFieldFormat{labelalpha}{\textsc{#1}}

\renewbibmacro{in:}{}

\renewbibmacro*{date}{%
  \iffieldundef{year}
    {}
    {\printtext[parens]{\printdate}}}

\renewbibmacro*{issue+date}{%
    \printfield{issue}%
    \setunit*{\addspace}%
    \usebibmacro{date}%
    \newunit}

\renewbibmacro*{publisher+location+date}{%
    \printlist{location}%
    \iflistundef{publisher}
        {\setunit*{\addcomma\space}}
        {\setunit*{\addcolon\space}}%
    \printlist{publisher}%
    \setunit*{\addspace}%
    \usebibmacro{date}%
    \newunit}

\renewbibmacro*{institution+location+date}{%
    \printlist{location}%
    \iflistundef{institution}
        {\setunit*{\addcomma\space}}
        {\setunit*{\addcolon\space}}%
    \printlist{institution}%
    \setunit*{\addspace}%
    \usebibmacro{date}%
    \newunit}

\renewbibmacro*{organization+location+date}{%
    \printlist{location}%
    \iflistundef{organization}
        {\setunit*{\addcomma\space}}
        {\setunit*{\addcolon\space}}%
    \printlist{organization}%
    \setunit*{\addspace}%
    \usebibmacro{date}%
    \newunit}

\renewbibmacro*{location+date}{%
    \printlist{location}%
    \setunit*{\addspace}%
    \usebibmacro{date}%
    \newunit}

\AtEveryBibitem{\clearfield{urldate}}
\AtEveryBibitem{\clearfield{month}}
\AtEveryBibitem{\clearfield{day}}
\AtEveryBibitem{\clearfield{issn}}
\AtEveryBibitem{\clearfield{version}}

\DeclareFieldFormat{doilink}{%
	\iffieldundef{doi}{%
		\iffieldundef{url}{%
			\iffieldundef{isbn}{%
				\iffieldundef{issn}{%
					#1%
				}{%
					\href{http://books.google.com/books?vid=ISSN\thefield{issn}}{#1}%
				}%
			}{%
				\href{http://books.google.com/books?vid=ISBN\thefield{isbn}}{#1}%
			}%
		}{%
			\href{\thefield{url}}{#1}%
		}%
	}{%
		\href{http://dx.doi.org/\thefield{doi}}{#1}%
	}%
}

\DeclareBibliographyDriver{article}{%
	\usebibmacro{bibindex}%
	\usebibmacro{begentry}%
	\usebibmacro{author/translator+others}%
	\setunit{\labelnamepunct}\newblock
	\usebibmacro{title}\addcomma\space%
	\newunit
	\printlist{language}%
	\newunit\newblock
	\usebibmacro{byauthor}%
	\newunit\newblock
	\usebibmacro{bytranslator+others}%
	\newunit\newblock
	\printfield{version}%
	\newunit\newblock
	\printtext[doilink]{%
		\usebibmacro{journal+issuetitle}%
		\newunit
		\usebibmacro{byeditor+others}%
		\newunit
		\usebibmacro{note+pages}%
	}%
	\newunit\newblock
	\iftoggle{bbx:isbn}
	{\printfield{issn}}
	{}%
	\newunit\newblock
	\usebibmacro{doi+eprint+url}%
	\newunit\newblock
	\usebibmacro{addendum+pubstate}%
	\setunit{\bibpagerefpunct}\newblock
	\usebibmacro{pageref}%
	\usebibmacro{finentry}}

\usepackage{slashed}

\newcommand{\J}{\mathtt{J}}
\renewcommand{\B}{\cD}
\newcommand{\parity}{\vec{\sigma}_z}
\let\thintilde\tilde
\renewcommand{\tilde}{\widetilde}
\newcommand{\p}{\mathrm{P}}

\newcommand{\fact}[1]{{#1}!}
\newcommand{\phiu}{\phi}
\newcommand{\phid}{\tilde{\phi}}
\newcommand{\pp}{d-p-1}
\newcommand{\wdeg}{h} 
\newcommand{\inda}{n}

\let\u\relax
\newcommand{\u}{u} 

\definecolor{tubeBlue}{rgb}{0.231369,0.533325,0.764694}
\definecolor{tubeFill}{rgb}{0.72,0.86,0.92}
\definecolor{tubeFillLight}{rgb}{0.88,0.95,0.97}
\definecolor{tubeShade}{rgb}{0.38,0.62,0.73}
\definecolor{warmShadow}{rgb}{0.24,0.20,0.03}

\renewenvironment{quote}{%
  \list{}{%
    \leftmargin0.75cm   
    \rightmargin\leftmargin
  }
  \item\relax
}
{\endlist}
\begin{document}

\begin{center}
	{\Large\textsc{The state/defect correspondence}} \\ 
    \bigskip
	Adrien Arbalestrier,
	Elise Paznokas, and
	Stathis Vitouladitis \\
	\bigskip
	\footnotesize{
		\textrm{Physique Théorique et Mathématique, Université Libre de Bruxelles \& \\ International Solvay Institutes, CP 231, 1050 Brussels, Belgium} \\ \bigskip
		\href{mailto:adrien.arbalestrier@ulb.be}{\small \sf adrien.arbalestrier@ulb.be}, \quad
		\href{mailto:elise.paznokas@ulb.be}{\small \sf elise.paznokas@ulb.be},
		\quad
		\href{mailto:stathis.vitouladitis@ulb.be}{\small \sf stathis.vitouladitis@ulb.be}
	}
\end{center}

\begin{abstract}
    \noindent We formulate a one-to-one correspondence between states and defects for higher-form gauge theories in arbitrary dimensions. The correspondence is not predicated on conformal invariance, as these theories are in general not conformal. Instead, it relies on the existence of infinitely many conserved charges associated with the mixed anomaly of electric and magnetic higher-form symmetries. In $p$-form Maxwell theory, these charges generate an extended Kac--Moody algebra that acts simultaneously on states and on extended operators. We show that this algebra organizes the Hilbert space on $S^p\times S^{d-p-1}$ into highest-weight representations allowing for a direct identification between states and $p$-dimensional defects. In particular, Wilson--'t Hooft defects dressed with local gauge-invariant operators are mapped to squeezed energy eigenstates. We further relate these novel  symmetries to higher-spin currents and demonstrate that their construction persists in a class of interacting non-linear electrodynamics theories.
\end{abstract}

\vspace{10pt}
\tableofcontents
\vspace{10pt}

\section{Introduction}\label{sec:introduction}

To the dismay of theoretical physicists, real materials are not ideal and often contain impurities.%
\footnote{Though sometimes impurities are intentionally added to \emph{dope} materials.}
These impurities can control observable physics. They scatter excitations \cite{Anderson:1958vr,Kondo:1964nea}, modify transport \cite{Kondo:1964nea,Hewson1993TheKP}, seed new phases \cite{Sachdev:1999qin,Vojta:2000tld,Barkeshli:2012dm}, and determine the universal behavior of systems near criticality \cite{Cardy:1984bb,Affleck:1991tk,Cuomo:2021rkm}. In the effective description within quantum field theory (QFT), they are modeled by defects. For example, a point impurity in a material becomes a line defect in QFT \cite{Sachdev:1999qin,Vojta:2000tld}.

Defects and extended excitations now play a central role across theoretical physics, including topological order \cite{Wen:1989iv,Wen:1990zza,Kitaev:2005dm,Nayak:2008zza,Barkeshli:2012dm}, high-energy theory \cite{Wilson:1974sk,tHooft:1977nqb,Polyakov:1978vu,Makeenko:2009dw,Kapustin:2005py,Gukov:2006jk,Billo:2016cpy,Cuomo:2021rkm,Hofman:2018lfz,Hofman:2024oze,Antinucci:2024izg}, and string theory and holography \cite{Polchinski:1995mt,Bachas:1995ik,Giveon:1998sr,Gibbons:1997xz,Maldacena:1998im,Rey:1998ik,Maldacena:1997re,Witten:1998qj,DeWolfe:2001pq,Takayanagi:2011zk,Belov:2006jd}. The advent of generalized symmetries \cite{Gaiotto:2014kfa}%
\footnote{An incomplete list of applications includes: higher-form \cite{McGreevy:2022oyu,Cordova:2022ruw,Brennan:2023mmt,Gomes2023}, higher-group \cite{Benini:2018reh,Cordova:2018cvg,Apruzzi:2021mlh,Arbalestrier:2025poq}, and non-invertible symmetries \cite{Bhardwaj:2017xup,Chang:2018iay,Heidenreich_2021,Kaidi:2021xfk,Choi:2021kmx,Roumpedakis:2022aik, Choi:2022zal,Niro:2022ctq,Antinucci:2022eat,Antinucci:2022vyk,Argurio:2024ewp,Paznokas:2025auw}.}
sharpens this point. Symmetry operators appear as topological defects of various codimensions, with intricate algebraic structures, while charged objects can also be extended. We need tools that identify the role of extended operators in QFT: which states they prepare, and how they act on the Hilbert space. Equivalently, we ask how the spectrum of states in a QFT relates to its spectrum of defect operators.

For local operators, this question is well understood, particularly within conformal field theory (CFT). There, there exists an isomorphism between (a basis of) states on a spatial sphere and local operators inserted at a single point. This one-to-one correspondence was found to hold thanks to the constraints provided by conformal symmetry. In essence, it functions as follows. Using the conformal map of the cylinder $\mathbb{R}\times S^{d-1}$ to $\mathbb{R}^d$, the distant past on the cylinder is mapped to the origin of a ball whose boundary is that $\S^{d-1}$. Thus, defining a state on the spatial slice $S^{d-1}$ in the infinite past is tantamount to inserting a local operator at the origin. Moreover, this conformal mapping identifies the Hamiltonian on the cylinder with the dilatation operator, allowing for states to be time evolved and local operators to be radially evolved, respectively, thus proving a general one-to-one identification between states and local operators. For a more detailed account, we refer the reader to \cite{DiFrancesco:1997nk, Tong:2009np}.

This result has drastically aided progress along various directions. To name a few, it is instrumental for defining 2d CFTs on arbitrary Riemann surfaces \cite{Friedan:1986ua,Moore:1988qv}, and remains a key tool in the conformal bootstrap program \cite{Belov:2005ze,Rattazzi:2008pe,Poland:2018epd,Hartman:2022zik,Poland:2022qrs}. In string theory, this map is used for the calculation of string amplitudes \cite{Polchinski:1998rq}, while in AdS/CFT it provides a matching between states in AdS and operator insertions on the boundary CFT \cite{Maldacena:1997re,Witten:1998qj}, aiding in computations of black hole microstates and holographic entanglement entropy \cite{Sen:2011cn,Benini:2015eyy, Cabo-Bizet:2018ehj}. 

As the above construction relies heavily on the existence of conformal symmetry, it is not expected to hold for general quantum field theories (QFTs).%
\footnote{Topological quantum field theories (TQFTs), such as Chern--Simons theory, do have a state-operator correspondence which follows straightforwardly from their topological nature. These theories have no propagating degrees of freedom and are therefore quite different than your average QFT.}
Indeed, while it is true that acting with an operator on the vacuum $\ket{0}$, or performing a path integral with a local insertion on a manifold with a boundary defines a new state, the converse breaks down. There is no guarantee that given a state, one can identify it with a local operator of the theory.

Nonetheless, recent work has been pushing the bounds to try to define a matching between operators and states in more general cases, motivated by the benefits such a correspondence is known to bestow. For example, \cite{Belin2018} attempts to formulate a correspondence for 3d CFTs between states on a spatial torus and line operators. In \cite{Chen:2025ujx} a correspondence between local operators and states is obtained in a specific class of integrable 3d QFTs owing to a generalization of the Kac--Moody algebra in 2d Wess--Zumino--Witten models. 
More relevant to this work is the story of \cite{Hofman:2024oze} wherein the authors constructed a correspondence between states and line operators for 4d CFTs with continuous symmetries tied by ‘t Hooft anomalies. Building on these methods, \cite{Vitouladitis:2025zoy} found that, in fact, conformal symmetry was not what underpinned this construction, but rather it was the existence of the continuous, anomalous symmetries. With this realization, \cite{Vitouladitis:2025zoy} starts with a $d$-dimensional scalar theory and formulates a \emph{non-conformal state-operator correspondence} between the states on $\S^{d-1}$ and local operators. 

Returning to the question posed in the beginning, the purpose of this paper is to unify these works and push towards a \emph{non-conformal state-defect correspondence} between states and extended operators in general dimensions. More precisely, our main result is the following:
\begin{quote}
    In \(d\)-dimensional%
    \footnote{For technical reasons, the case $d=2p+1$ evades our construction. See \cref{sec: the states} for details.}
    QFTs with $\gf{\U(1)}{p}\times \gf{\U(1)}{d-p-2}$ symmetry with a mixed anomaly, states on $\S^p\times\S^{d-p-1}$ are in one-to-one correspondence with $p$-dimensional defects on $S^p$.
\end{quote}
Interestingly, we find that energy eigenstates are in one-to-one correspondence with \emph{squeezed} defects. In particular, the empty path integral does not prepare the vacuum state but rather a \emph{squeezed vacuum}, in accordance with \cite{Belin2018,Hofman:2024oze,Vitouladitis:2025zoy}.

But how, one could ask, is such a correspondence possible while forsaking conformal invariance? The answer was already provided in \cite{Vitouladitis:2025zoy}, and the key is the symmetry structure of these theories. Indeed, we find that the $\gf{\U(1)}{p}\times \gf{\U(1)}{d-p-2}$ symmetry with its mixed anomaly---described by the respective conserved currents $J$ and $\tilde{J}$---implies the existence of an \emph{infinite family of conserved charges} termed \emph{dressed charges},
\begin{equation}
    \cQ_\eta
    =\int_{\Sigma}\left(\eta \w *J+\tilde{\eta}\w*\tilde{J}\right)~,
\end{equation}
by virtue of introducing a pair of auxiliary $p$- and $(d-p-2)$-forms, $\eta$ and $\tilde{\eta}$, respectively, which satisfy a certain relation. The above are codimension-1 charges, meaning that they define 0-form symmetries. Moreover, they are found to have non-trivial commutation relations,
\begin{equation}
    \comm{\cQ_\eta}{\cQ_{\zeta}}=-\frac{\ii}{2\pi}\int_{\Sigma}\qty(\eta\w \dd\tilde{\zeta}- \zeta \w \dd\tilde{\eta})~.
\end{equation}
This algebra is precisely an \emph{extended Abelian Kac--Moody algebra}. As we will show, the Hilbert space of states is organized under representations of this algebra, while extended operators also transform under this algebra. This is not entirely surprising, as it functions similarly to that for 2d unitary CFTs with continuous symmetry, where the Kac--Moody algebra naturally organizes the spectrum of operators and states and allows for easy identification on either side. 

These infinite families of conserved charges, essential for enabling the state-defect correspondence, are also of interest in and of themselves. For instance, we show that they are related to higher-spin currents, which are of particular interest in integrability. Moreover, they relate to the entanglement spectrum of topological orders \cite{Fliss:2023uiv}, and to soft theorems and asymptotic symmetries \cite{He:2014cra,Hofman:2018lfz,Tizzano:2026rgr}.

Lastly, we wish to emphasize that while the majority of our computations and explicit examples are worked out for $p$-form generalized Maxwell theory, which is a free theory, the construction employed here also holds for more general, interacting theories. We demonstrate that in non-linear higher-form electrodynamics, our infinite symmetries and their algebra persist. Presumably then, while perhaps technically intractable, there is a state-defect correspondence for this class of theories as well.

This paper is organized as follows. In \cref{sec: infinite symmetries}, we introduce $p$-form Maxwell theory in general dimensions, defining the conserved currents and their mixed anomaly. This allows us then to construct the dressed charges $\cQ_\eta$. We also comment on how to construct higher-spin currents starting from our dressed currents, as well as the possibility of including interactions. Subsequently, in \cref{sec: algebra}, we compute the commutators of the dressed charges and show that this behaves as a generalization of a Kac--Moody algebra. We then canonically quantize the theory in \cref{sec: the states}, determining the spectrum of states. In \cref{sec: the ops}, we turn to the other side of the coin, detailing the available extended operators of the theory and how these are acted on by combinations of $\cQ_\eta$. Lastly, in \cref{sec: the correspondence}, we tie everything together, making explicit the identification between states and defects.  \Cref{app: transversal laplacian,app: generalized curl} contain details on the transversal Laplacian and the so-called generalized curl operator that acts as a ``square-root'' of the Laplacian. We collect further details on the radial evolution and construction of higher-spin currents in \cref{app: exact solutions,app:spin-four}, respectively.

\paragraph{A note on notation.} For the reader's convenience we collect here some of our frequently used notational conventions.

\begin{enumerate}
    \itemsep0em
	\item The degree of forms is inside square brackets, e.g. $\f{f}{k}$.
	\item Higher-form symmetry groups are denoted with their degree in square brackets, as a superscript, for example: \(\gf{G}{k}\).
	\item \(\Sigma\) will denote a \((d-1)\)-dimensional Cauchy slice. 
	\item The Hodge star in $d$-dimensions is written as $*$, and that in $(d-1)$-dimensions is $*_{\Sigma}$. Those defined on $n$-spheres $S^n$ are written as $*_n$.
	\item
    The eigenforms of the Laplacian on $\S^p\times \S^{\pp}$ are denoted as $\Psi$ with eigenvalues $\lambda$, while those of the generalized curl operator (cf. \cref{app: generalized curl}) are denoted as $\Phi$ with eigenvalues $\mu=\pm\sqrt{\lambda}$.
    \item The Hodge inner product on $\Sigma$ is $\ip{\,\cdot\,}{\,\cdot\,}_\Sigma = \int_\Sigma \cdot\w *_\Sigma\; \cdot$.
   	\item The exterior derivative on the Cauchy slice \(\Sigma\) is denoted as $\sd$. 
    \item When decomposing into components along $\Sigma$ and $r$, we use the convention $\f{f}{k}=\dd{r} \w f_r+f_{\Sigma}$, always moving the $\dd{r}$ to the left-hand side.     \item When decomposing the manifold into $\mathbb{R}\times \Sigma_r$, we employ the formula:
   \begin{equation}
    *\qty(\f{\alpha_r}{i}\w \f{\beta_{\Sigma}}{j})=(-1)^{(1-i)j}*_{\bR}  \f{\alpha_r}{i}\w *_\Sigma\f{\beta_{\Sigma}}{j}
    \end{equation}
where here $\f{\alpha_r}{i}\in \Omega^i(\R)$, while $\f{\beta_{\Sigma}}{j}\in \Omega^j(\Sigma)$. 
\item We denote vectors (often a doublet of $p$-forms and $(d-p-2)$-forms) by a boldface letter, like $\vec{X}$.
\item Poincaré duals of cycles, are denoted with a hat.
\end{enumerate}

\section{Infinite symmetries of \texorpdfstring{\(p\)}{p}-form gauge theories}\label{sec: infinite symmetries}
We are interested in defining a state-defect correspondence for $p$-form gauge theories in $d$ dimensions. Concretely, we consider generalized Maxwell theories,
\begin{align}
	S = \frac{1}{2\g^2}\int_{X} \f{f}{p+1}\w*\f{f}{p+1}~,
\end{align}
where \(\f{f}{p+1}\) is the curvature of a \(p\)-form \(\U(1)\) connection \(\f{a}{p}\). In this section, we work in Euclidean signature. We will state it explicitly when we switch to Lorentzian.

The coupling constant \(\g\) has mass dimension \([\g]=\frac{2p+2-d}{2}\). The theory is conformal only in \(d=2p+2\) dimensions. Otherwise, the theory runs classically.%
\footnote{Defining the dimensionless coupling at scale \(\mu\) as \(\lambda(\mu) = \g^2 \mu^{d-2p-2}\), the renormalization group (RG) equation is trivially \(\mu \dv*{\lambda}{\mu} = (d-2p-2)\lambda\).}
When \(d<2p+2\), the ultraviolet (UV) fixed point corresponds to \(\g\to 0\), where the gauge group decompactifies to \(\R\). Unless \(p=0\), the fixed point describes a scale-invariant but not conformally invariant theory \cite{El-Showk:2011xbs,Jackiw:2011vz,Nakayama:2013is}. The infrared (IR) fixed point at \(\g\to\infty\) is, dually, a non-compact \((d-p-2)\)-form gauge theory. In this case it becomes conformal only when \(p=\frac{d-2}{2}\). When \(d>2p+2\), the two fixed points are flipped. We stay in general dimensions $d\neq 2p+2$, and at finite coupling. Thus, our advertised state-defect correspondence will not be predicated on conformal invariance.

For completeness, we point out that when $d=2p+2$, $p$ is odd, and \(X\) has non-vanishing middle cohomology, a theta term becomes available:
\begin{align}\label{eq:theta-term}
	S_\theta = \frac{\ii \theta}{8\pi^2}\int_{X} \f{f}{p+1}\w\f{f}{p+1}~.
\end{align}
However, the presence of such a theta term does not change anything qualitative, so we will neglect it. When applicable, we will comment in passing on its quantitative effect.

As we hinted at, what is important for us are the symmetries of the theory. As is well appreciated, this theory has an ``electric'' \(p\)-form \(\gf{\U(1)}{p}\) symmetry with the associated conserved current,
\begin{equation}\label{electric current}
\f{J}{p+1} =\frac{\ii}{\g^2}\f{\dd{a}}{p}~.
\end{equation}
The conservation of the current, \(\dd*\f{J}{p+1}=0\), follows from the equations of motion.%
\footnote{The prefactor $\ii/\g^2$ (and \(1/(2\pi)\) for the magnetic current) is chosen such that the charges are integers: $\int *J\in \Z$.} 
Moreover, there is a dual ``magnetic'' \((d-p-2)\)-form \(\gf{\U(1)}{d-p-2}\) symmetry with the current
\begin{equation}\label{magnetic current}
\f{\tilde{J}}{d-p-1}=\frac{1}{2\pi}*\f{\dd{a}}{p}~.
\end{equation}
This current is conserved \emph{off-shell}, $\dd*\tilde{J}=0$, simply due to the Bianchi identities of \(\f{f}{p+1}\). For this reason, the dual symmetry is sometimes called topological. With these currents, we can define the usual electric and magnetic charges respectively,
\begin{equation}\label{electric charge}
    Q_e\qty(C_{\pp})= \int_{C_{\pp}}*J~, \qquad Q_m\qty(C_{p+1}) = \int_{C_{p+1}} *\tilde{J}~.
\end{equation}
The electric and magnetic symmetries are tied by a \emph{mixed anomaly}, meaning one cannot simultaneously gauge the two symmetries. The anomaly can be characterized by the anomaly inflow action in $d+1$ dimensions,
\begin{equation}\label{anom inflow}
    S_{\t{anomaly}}=\frac{i}{2\pi}\int_{M_{d+1}}\f{A}{\pp}\w \dd\f{A}{p+1}~, 
\end{equation}
where $\f{A}{p+1}$ is the background field for the electric symmetry and $\f{A}{\pp}$ that for the magnetic symmetry, respectively.

Besides these two well-known symmetries, the theory actually enjoys infinitely many more. Such symmetries can be constructed in terms of a current which is a combination of $J$ and $\tilde{J}$,
\begin{equation} \label{infinite current}
    *\mathcal{J}_{\eta}=\eta \w *J+\tilde{\eta}\w*\tilde{J}~.
\end{equation}
Here we have introduced a pair of \(p\)- and \((d-p-2)\)-forms, \(\f{\eta}{p}\) and \(\f{\widetilde{\eta}}{d-p-2}\), respectively. We see that $\mathcal{J}_\eta$ is conserved if they satisfy the condition,
\begin{equation}\label{condition eta}
    \frac{\g}{\sqrt{2\pi}}\dd \tilde{\eta}+\ii \frac{ \sqrt{2\pi} }{\g}*\dd\eta =0~.
\end{equation}
Following the terminology of \cite{Vitouladitis:2025zoy}, we call these \(\cJ_\eta\) \emph{dressed currents}. 

The auxiliary forms \(\eta\) and \(\tilde{\eta}\) behave exactly as a pair of \(S\)-dual \(\U(1)\) gauge fields; they are electric and magnetic (higher-form) photons. Indeed, the above equation is known in the literature as \emph{twisted self-duality} \cite{Cremmer:1997ct,Cremmer:1998px}, satisfied in ``democratic'' formulations of electrodynamics and its generalizations. Accordingly, there is a gauge redundancy baked in \cref{infinite current,condition eta}. We see that shifting either of the auxiliary forms by an exact form, say  \(\dd{\lambda}\) and \(\dd{\tilde{\lambda}}\), generates a gauge equivalent \(\cJ_\eta\). In other words, the charge associated with \(\eta=\dd{\lambda}\) and \(\tilde{\eta}=\dd{\tilde{\lambda}}\) is zero. Furthermore, due to the fact that $\int * J$ and $\int * \tilde{J}$ are quantized, $\eta$ and $\tilde{\eta}$ admit large gauge transformations. This, however, does not necessarily imply that all the associated charges $\cQ_\eta$---defined in the following---are themselves quantized, a point that we expand upon later, in \cref{sec: the ops}. With the condition in \cref{condition eta} satisfied, and the gauge equivalent modes projected out, this family of conserved currents contains \(\binom{d-2}{p}\) degrees of freedom.%
\footnote{Note that a \(p\)-form gauge field and a \((d-p-2)\)-form gauge field have the same number of degrees of freedom since \(\binom{d-2}{p}=\binom{d-2}{d-p-2}\).}  

Equation \cref{condition eta} admits infinitely many solutions on non-compact manifolds or manifolds with boundary. In contrast, on closed manifolds it becomes a problem of cohomology. Its globally well-defined solutions are therefore harmonic forms, which reduce the infinite symmetries back to the ordinary electric and magnetic symmetries. It is still possible, however, to solve this differential equation locally, but these solutions will necessarily be singular at some points. These singularities will play an important role in the following.

Integrating the dressed currents along a codimension-1 surface, one gets a set of conserved, \emph{dressed charges}:
\begin{equation} \label{conserved charges}
\begin{split}
	\cQ_\eta &\coloneqq \int_{\Sigma} * \cJ_\eta
    =\int_{\Sigma}\left(\eta \w *J+\tilde{\eta}\w*\tilde{J}\right)~.
    \end{split}
\end{equation}
Note that due to the nontrivial profile of the forms \(\eta\) and \(\tilde{\eta}\) these charges are spatially \emph{modulated}, resembling multipole or subsystem symmetries \cite{Gorantla:2021bda,Gorantla:2022eema,Gromov:2020yoc,Pace:2022wgl,Pace:2024tgka,Sala:2021}. 
From the definition of $\cJ_\eta$, it is clear that when either \(\widetilde{\eta}\) or \(\eta\) vanishes, the extended charges reduce to the usual electric and magnetic ones, respectively. More explicitly,
\begin{equation}	
\cQ_{\eta}=\begin{cases}
Q_e~, & \text{if}\quad \eta=\widehat{C}_{\pp}~,\ \t{and}\ \tilde{\eta}=0~, \\[2ex] 
Q_m ~, & \text{if}\quad \eta=0, \ \t{and}\ \tilde{\eta}=\widehat{C}_{p+1}~,
\end{cases} 
\end{equation}
where $\widehat{C}_x$ refers to (a representative of) the Poincaré dual of $C_x\subset \Sigma$.

Let us discuss a few equivalent rewritings of our dressed charges, each of which clarifies different structural or physical properties they subsume. All of these rewritings rely on the fact that the dressed symmetries are zero-form symmetries (and hence supported in codimension-1) and reveal a bit of the philosophy we will adopt in the rest of the paper. We would like to think of \(\Sigma\)---the surface on which the dressed charges are supported---as a Cauchy slice on which the phase space of the theory resides. To that end, it is useful to introduce a local neighborhood $I\times \Sigma$, where $I$ is an interval of Euclidean time. In anticipation of \cref{sec: the ops}, where the role of time will be played by a radial coordinate (and in an effort to use consistent notation throughout), we parametrize the interval by $r$. This allows us to decompose differential forms into their $r$ and $\Sigma$ components. For example, a \(k\)-form \(\f{\omega}{k}\) can be written as
\begin{equation}
    \f{\omega}{k} = \dd{r}\w\omega_r + \omega_\Sigma~,
\end{equation}
where \(\omega_r\) and \(\omega_\Sigma\) are (\(r\)-dependent) \((k-1)\)- and \(k\)-forms on \(\Sigma\), respectively. In this decomposition the dressed charges read:
\begin{equation} \label{conserved charges decomp}
	\cQ_\eta =\int_{\Sigma}\left(\eta_{\Sigma} \w *_{\Sigma}J_r+\tilde{\eta}_{\Sigma}\w *_{\Sigma}\tilde{J}_r\right)~,
\end{equation}
while the condition for conservation and topologicalness of the charges becomes,
\begin{equation}\label{condition eta decomp}
\begin{split}
 \frac{\sqrt{2\pi}}{\g}\left( \pd_r \eta_{\Sigma}- \dd_{\Sigma}\eta_r \right) &=\ii(-1)^{dp}\frac{\g}{\sqrt{2\pi}}*_\Sigma\dd_{\Sigma}\tilde{\eta}_{\Sigma} \\
\frac{g}{\sqrt{2\pi}}(\pd_r\tilde{\eta}_{\Sigma}- \dd_{\Sigma}\tilde{\eta}_r)&=\ii \frac{ \sqrt{2\pi} }{g}(-1)^{p} *_{\Sigma}\dd_{\Sigma}\eta_{\Sigma} ~.
\end{split}
\end{equation}

Let us now try to build some intuition for these dressed charges by relating them to the usual electric and magnetic fields.   Using the decomposition of the field strength into electric and magnetic components,
\begin{align} \label{field strength E and B}
	\f{f}{p+1} =  -i\dd{r}\w \f{E}{p}+(-1)^{d(p+1)}*_{\Sigma}\f{B}{d-p-2}~,
\end{align}
this implies
\begin{equation}
    J_r = \frac{1}{g^2} E \qq{and} \tilde{J}_r = \frac{1}{2\pi} B ~.
\end{equation}
From \cref{conserved charges decomp} it then follows that the dressed charges can be expressed simply as:
\begin{align} \label{Q new expression}
	\cQ_\eta = \frac{1}{\g^2}\langle \eta_\Sigma ,E\rangle_\Sigma +\frac{1}{2\pi}\langle \tilde{\eta}_\Sigma ,B\rangle_\Sigma~,
\end{align}
where \(\ip{\cdot}{\cdot}_\Sigma=\int_\Sigma \cdot\w *_\Sigma\, \cdot\) is the Hodge inner product on \(\Sigma\).%
\footnote{To define this product, we consider $\Sigma$ to be closed and compact.}
The charge \(\cQ_\eta\) therefore measures the overlap of the electromagnetic field along the smearing profiles \(\eta_\Sigma\) and \(\tilde{\eta}_\Sigma\). In ordinary \(d=4\), \(p=1\) Maxwell theory, the same kind of projection is used in optics to decompose radiation into polarization and helicity modes, and is related to optical helicity and optical chirality charges \cite{Cameron:2012em,Cameron:2012optical,Bliokh:2011zz,Philbin:2013zz}.

An alternative way to write the dressed charges is as follows. Starting from \cref{conserved charges decomp} we write,
\begin{equation}
    \begin{split}
        \cQ_\eta &=\int_{\Sigma}\left(\frac{\sqrt{2\pi}}{\g}\eta_{\Sigma} \w *_{\Sigma}\frac{\g}{\sqrt{2\pi}}J_r+\frac{\g}{\sqrt{2\pi}}\tilde{\eta}_{\Sigma}\w *_{\Sigma}\frac{\sqrt{2\pi}}{\g}\tilde{J}_r\right)
        =
        \ip{
            \mqty(
                \frac{\sqrt{2\pi}}{\g}\eta_\Sigma \\[0.7em]
                \frac{\g}{\sqrt{2\pi}}\tilde{\eta}_\Sigma 
            )
        }
        {
            \mqty(
                \frac{\g}{\sqrt{2\pi}}J_r \\[0.7em] \frac{\sqrt{2\pi}}{\g}\tilde{J}_r
           )
        }_\Sigma~. 
    \end{split}
\end{equation}
On the right-hand-side, we have repacked the $p$- and $(d-p-2)$-forms into one object, living in
\begin{equation}
    \Omega^{p,d-p-2}(\Sigma) \coloneqq \Omega^p(\Sigma)\oplus \Omega^{d-p-2}(\Sigma)~.
\end{equation}
A priori, this may seem arbitrary, but it will prove to be useful to facilitate a mode expansion of the currents. The electric and magnetic Gauss laws require that the two currents be co-closed on \(\Sigma\). 
\begin{equation}\label{coclosed currents}
    \begin{cases}
        \dd{*J} =0~, \\
        \dd{*\tilde{J}}=0~, \\
    \end{cases}\implies\begin{cases}
        \pd_r\qty(*_{\Sigma}J_r)+(-1)^{p} \sd *_{\Sigma}J_{\Sigma}=0~, \\[0.2em]
        \pd_r\qty(*_{\Sigma}\tilde{J}_r)+(-1)^{d-p} \sd *_{\Sigma}\tilde{J}_{\Sigma}=0~, \\[0.2em]
        \sd *_{\Sigma} J_r = 0~, \\[0.2em]
        \sd *_{\Sigma} \tilde{J}_r=0~.
    \end{cases}
\end{equation}
Likewise, we can fix $\eta$ and $\tilde{\eta}$ in Coulomb gauge, in which they are also co-closed on \(\Sigma\). Acting on co-closed forms, the Laplacian on $\Sigma$ degenerates to the transversal Laplacian, $\dd^\dagger_\Sigma\dd^{\phantom{\dagger}}_\Sigma$. We collect some useful details on the transversal Laplacian and its eigenforms in \cref{app: transversal laplacian}.

A useful tool in our analysis will be to decompose our vector of forms into eigenforms of the \emph{generalized curl operator}:
\begin{align}\label{generalized curl}
	\B_p \coloneqq \mqty(\admat[0]{(-1)^{dp+p+1}\,*_{\Sigma}{\dd}_\Sigma ,*_{\Sigma}{\dd}_\Sigma}):\quad \Omega^{p,d-p-2}(\Sigma)\to\Omega^{p,d-p-2}(\Sigma).
\end{align}
When \(d=4\) and \(p=1\), this operator acts as two copies of the conventional curl of vector calculus, which motivates the name. More generally \(\cD_p\) is a particular restriction of the Hodge--de Rham--Dirac  operator \(\sd +\csd\) and it squares to two copies of the transversal Laplacian, acting independently on \(\Omega^p\) and \(\Omega^{d-p-2}\).  For more details on the generalized curl operator and its spectrum, we refer the reader to \cref{app: generalized curl}. 

What matters, for the time being, is that the generalized curl operator is a self-adjoint operator on co-closed forms on \(\Omega^{p,d-p-2}(\Sigma)\). Furthermore, \(\B_p\) anticommutes with a helicity operator, given here by the Pauli \(z\)-matrix, \(\parity\). Self-adjointness of \(\B_p\) implies that we can expand any element of \(\Omega^{p,d-p-2}(\Sigma)\) in terms of its eigenmodes $\Phi_{\inda\sigma}$ (that we will occasionally refer to as \emph{curly eigenforms}). As we will go into more detail later, these are labeled by a discrete (so long as \(\Sigma\) is compact) collection of indices, which we will denote as \(\inda\), and a sign, \(\sigma=\pm\) that labels the eigenvalue under $\parity$. The eigenvalue equation is therefore,
\begin{equation}
    \B_p \Phi_{\inda\sigma} = \mu_{\inda\sigma} \Phi_{\inda\sigma}~,
\end{equation}
where the eigenvalue is \(\mu_{\inda\sigma}=\sigma\, \mu_{\inda}\) with \(\mu_\inda\geq 0\). In essence, \(\mu_\inda\) is the positive square root of the eigenvalue of the Laplacian on \(\Sigma\). The eigenforms, \(\Phi_{\inda\sigma}\), are taken to be orthonormal. We again stress that $\Phi_{\inda \sigma}=\qty\big(\phi_{\inda\sigma},\tilde{\phi}_{\inda\sigma})^\sfT$ is a vector, whose components are the $p$-form $\phi_{\inda\sigma}$ and the $(d-p-2)$-form $\tilde{\phi}_{\inda\sigma}$.

The decompositions of the currents and auxiliary forms into eigenmodes of the generalized curl operator are,
\begin{equation} \label{decomp to eigenmodes}
    \begin{split}
        \begin{pmatrix}
\frac{\g}{\sqrt{2\pi}}J_r\\[0.5em]
\frac{\sqrt{2\pi}}{\g}\tilde{J}_r
\end{pmatrix}=\sum_{\inda, \sigma}\J_{\inda,\sigma}\Phi_{\inda \sigma} 
\qq{and}
\begin{pmatrix}
\frac{\sqrt{2\pi}}{\g}\eta_{\Sigma}\\[0.3em]
\frac{\g}{\sqrt{2\pi}}\tilde{\eta}_{\Sigma}
\end{pmatrix}=\vec{\eta}=\sum_{\inda,\sigma}\mathtt{\eta}_{\inda \sigma}\Phi_{\inda \sigma}
    \end{split}~,
\end{equation}
where the factors of \(\g\) in the above are dictated by dimensional analysis, while the factors of \(2\pi\) are chosen for convenience. Now, $\cQ_\eta$ becomes
\begin{equation}\label{Q eta and modes}
    \cQ_\eta =\sum_{\inda, \sigma}\mathtt{\eta}_{\inda \sigma}\J_{\inda \sigma}~,
\end{equation}
where we have used the orthonormality of the eigenforms $\Phi_{\inda\sigma}$.

\subsection{Towards higher-spin currents}

Before moving on to the main question we set out to answer, let us take a small detour to discuss higher-spin currents. In quantum field theory, such currents are highly constraining. In a massive theory, the Coleman--Mandula theorem \cite{Coleman:1967ad} implies that higher-spin symmetries make scattering trivial. CFTs do not have an \(S\)-matrix, but the same obstruction reappears in correlation functions. The Maldacena--Zhiboedov theorem \cite{Maldacena:2011jn} and its generalizations \cite{Alba:2013yda,Alba:2015upa} imply that higher-spin symmetry forces the theory to be free. Here we explicitly construct higher-spin currents, starting from the dressed family of charges. In this passage, we restrict to the conformal case \(d=2p+2\), and for simplicity to flat space, with metric \(g_{\mu \nu} = \delta_{\mu \nu}\). 

At the level of the classical theory, we can let the parameters depend on the fields.
This turns out to be a very good idea in order to make contact with higher-spin currents. The dressed currents are linear in \(J,\ \widetilde{J}\) when \(\eta,\ \widetilde{\eta}\) are fixed. Taking the parameters instead to be linear in the fields results in bilinear conserved currents, as is typically the case with higher-spin symmetries \cite{Klebanov:2002ja,Gelfond:2006be,Bekaert:2010hw}.

The first example to reproduce is the stress tensor. Take \(\xi^\alpha\) to be a constant vector and consider
\begin{align}
	\eta_{\mu_1\cdots\mu_p}
	&=
	-\frac{\g^2}{2}\,\xi^\alpha J_{\alpha\mu_1\cdots\mu_p}~,
    \\
	\widetilde{\eta}_{\mu_1\cdots\mu_p} 
    &=
	-\frac{2\pi^2}{\g^2}\,\xi^\alpha\widetilde{J}_{\alpha\mu_1\cdots\mu_p}~.
\end{align}
One can readily check that these obey the necessary condition \cref{condition eta}. The corresponding dressed current is (here and in the following, we drop the subscript \(\eta\) from the dressed current to ease notation)
\begin{align}
	\cJ_\nu = \xi^\mu T_{\mu\nu}~,
\end{align}
where \(T_{\mu\nu}\) is the stress tensor of the theory,
\begin{align}\label{eq:stress-tensor}
	T_{\mu\nu}
	=
	\frac{1}{\g^2\, p!}
	\qty(
	f_{\mu\alpha_1\cdots \alpha_p}f_{\nu}^{~~\alpha_1\cdots \alpha_p}
	-
	\frac{1}{2(p+1)}\,\delta_{\mu\nu}f_{\beta_1\cdots\beta_{p+1}}f^{\beta_1\cdots\beta_{p+1}}
    )~.
\end{align}
Note that it is symmetric, and in \(d=2p+2\) also traceless. The conservation of \(\cJ_\nu\) then yields a conserved spin-2 current.

To construct a spin-3 current, we start with a constant symmetric tensor, \(\xi^{\alpha\beta}\) and define
\begin{align}\label{eq:eta-zilch}
\begin{split}
	\eta_{\vec{\kappa}}                 
    & = 
    \xi^{\alpha\beta}\pd_\alpha \widetilde{J}_{\beta\vec{\kappa}}~, \\
	\widetilde{\eta}_{\vec{\kappa}} 
    & = 
    (-1)^p\,\xi^{\alpha\beta}\pd_\alpha J_{\beta\vec{\kappa}}~,
\end{split}
\end{align}
where we have defined for notational simplicity the fully antisymmetric multi-index \(\vec{\kappa}=[\kappa_1\cdots \kappa_p]\). One can again check that \(\eta\) and \(\widetilde{\eta}\) satisfy \cref{condition eta}. With this choice, the dressed conserved currents have components
\begin{align}
	\cJ_\mu = \xi^{\alpha\beta} Z_{\mu\alpha\beta},
\end{align}
where
\begin{align}
	Z^{\mu}_{~\alpha\beta}
	=
	J^{\mu \vec{\kappa}}\pd_{(\alpha}\widetilde{J}_{\beta)\vec{\kappa}}
	+
	(-1)^p\,\widetilde{J}^{\mu \vec{\kappa}}\pd_{(\alpha}J_{\beta)\vec{\kappa}}~.
\end{align}
The tensor \(Z^{\mu}_{~\alpha\beta}\) is symmetric in \(\alpha,\beta\), and has no symmetry relating these indices to \(\mu\). This is a higher-dimensional generalization of the zilch tensor introduced by Lipkin \cite{Lipkin:1964zz}.%
\footnote{
This conserved quantity contains the optical chirality density as one of its components \cite{Tang:2010zz}, which controls the leading chiral asymmetry in the excitation rate of a small chiral molecule, measured experimentally in \cite{Tang:2011zz}. See also \cite{Bliokh:2011zz,Philbin:2013zz} relating the zilch to other optical quantities, and \cite{Letsios:2023zz} for a direct derivation of full zilch current directly from the free electromagnetic action.
}
This tensor is a conserved current with mixed symmetry:
\begin{align}
	\pd_\mu Z^\mu_{~\alpha\beta}=0~,
\end{align}
so it is not yet a symmetric spin-\(3\) current. The fix is straightforward; one must first symmetrize%
\footnote{Strictly speaking, the symmetric tensor \cref{eq:symZilch} is the symmetrization of the zilch tensor \(\tilde{Z}_{\mu \nu \rho} = Z_{(\mu \nu \rho)}\) when \(d=0\mod 4\) (equivalently when \(p\) is odd) in Euclidean signature or when \(d=2\mod 4\) (equivalently when \(p\) is even) in Lorentzian signature. For a unified treatment, it is more convenient to switch to a ``chiral'' basis as in \cite{Hofman:2024oze}.}
and then subtract the trace to get:
\begin{align}
    \tilde{Z}_{\mu \nu \rho} &\coloneqq (-1)^p\qty(J_{\vec{\kappa}(\mu}\pd_{\nu}\widetilde{J}_{\rho)}^{~~\vec{\kappa}}- \widetilde{J}_{\vec{\kappa}(\mu}\pd_{\nu}J_{\rho)}^{~~\vec{\kappa}})~, \label{eq:symZilch} \\ 
    T_{\mu \nu \rho}^{(3)} &= \tilde{Z}_{\mu \nu \rho} - \frac{3}{d+2}\, \delta^{\vphantom{\alpha}}_{(\mu \nu}\, \tilde{Z}^{\lambda}_{~~\rho)\lambda}~. 
\end{align}
The tensor \(T^{(3)}_{\mu \nu \rho}\) is a genuine spin-3 current. It obeys
\begin{align}
	T^{(3)}_{\mu\nu\rho}
	=
	T^{(3)}_{(\mu\nu\rho)}~,
	\qquad
	\delta^{\mu\nu}T^{(3)}_{\mu\nu\rho}=0~,
	\qquad
	\pd^\mu T^{(3)}_{\mu\nu\rho}=0~.
\end{align}
The last equality follows directly from the source-free Maxwell equations. Thus, the theory contains both the mixed-symmetry zilch current and the symmetric spin-\(3\) current \(T^{(3)}\).

Starting from spin \(s\geq 4\), the original dressed currents no longer determine the symmetric higher-spin primary current by themselves. Let us illustrate this for \(s=4\). We will be very schematic in this passage, but we provide all the details in \cref{app:spin-four}. It is tempting to take, continuing the pattern,
\begin{align}
	\eta \sim \xi_1\, \pd^2 J + \xi_2\, \pd^2 \tilde{J} \qq{and} \tilde{\eta} \sim \xi_3\, \pd^2 J + \xi_4\, \pd^2 \tilde{J}~,
\end{align}
where \(\xi_i\) \((i=1,\ldots,4)\)  are rank-3 symmetric tensors whose tensor structure we suppress. They are fixed by requiring that \(\eta\) and \(\tilde{\eta}\) satisfy \cref{condition eta}. This choice produces a conserved mixed-symmetry tensor, analogous to \cref{eq:symZilch}, with derivative structure \(J \pd^2 J\) (by which we mean any combination of \(J \pd^2 J\), \(\tilde{J}\pd^2 J\), \(J\pd^2 \tilde{J}\), or \(\tilde{J}\pd^2 \tilde{J}\)). However, after symmetrizing and subtracting traces, this does not lead to a conserved spin-\(4\) current.

A spin-\(4\) current built from field strengths contains two derivatives in total. Higher-spin currents built out of \((p+1)\)-form field strengths were worked out explicitly in \cite{Anselmi:1999bb}. The derivative splittings that appear in the spin-4 current are
\begin{align}\label{eq:spin-4-current-schematic}
	T^{(4)}
	\sim
	\pd^2 J\,J
	+
	\pd J\,\pd J
	+
	J\,\pd^2 J~.
\end{align}
The family \cref{infinite current} only accounts for the endpoint splittings where one field strength is undifferentiated, while the field-dependent parameter contains two derivatives of the other field strength. It does not account for the middle splitting \(\pd J\,\pd J\). 

The remedy to this comes in the form of another infinite family of conserved currents enjoyed by the theory. With the goal of reproducing this middle splitting, we make the following ansatz:
\begin{align}\label{eq:family-2-ansatz}
	*\cJ^{(1)}_{\eta} = \eta^{(1)}\w *\, \pounds_v J + \tilde{\eta}^{(1)}\w *\, \pounds_v \tilde{J}~. 
\end{align} 
Here \(\eta^{(1)}\) and \(\tilde{\eta}^{(1)}\) are a new pair of \(p\)-forms (recall we are in \(d=2p+2\), so the two forms have the same degree). Moreover, \(v\) is an arbitrary constant vector and \(\pounds_v\) denotes the Lie derivative along \(v\). Since \(J\) and \(\tilde{J}\) are closed on-shell, \(\pounds_v\) acts as \(\dd\, \iota_v\), where \(\iota_v\) denotes the interior product. Further, on flat space and with constant \(v\),%
\footnote{The generalization of \cref{eq:family-2-ansatz} to a general curved spacetime requires that \(v\) be a Killing vector instead.}
the Lie derivative commutes with the Hodge star and so it follows that if \(\eta^{(1)}\) and \(\tilde{\eta}^{(1)}\) also satisfy \cref{condition eta}, then \(\cJ^{(1)}_\eta\) is conserved. Thus, this identifies a second infinite family of conserved currents, that starts at one order higher in derivatives of \(J\). 

For spin \(4\) then, choose \(\eta^{(1)}\) and \(\tilde{\eta}^{(1)}\) to be linear in one derivative of \(J\) and \(\tilde{J}\), as in \cref{eq:eta-zilch}. Then, \(\pounds_v\) differentiates the current in the second slot, and the resulting bilinear contains the missing \(\pd J\,\pd J\) structure. Overall, this gives all the ingredients for a  spin-\(4\) conserved current. The coefficients with which the two families need to be added and the trace subtractions are then fixed by the usual requirements of complete symmetry, tracelessness, and conservation, as in \cite{Anselmi:1999bb}.

The construction extends to arbitrarily high spins. The trouble is one has to identify a new infinite dimensional family of conserved currents for each additional spin. Fortunately, such a collection of families exists. Indeed, consider the ansatz, 
\begin{align}
	* \cJ_\eta^{(k)}
	=
	\eta^{(k)}\w *\, D^{(k)} J
	+
	\tilde{\eta}^{(k)}\w *\, D^{(k)} \tilde{J}~, 	
\end{align}
for a new pair of \(p\)-forms, and with a differential operator \(D^{(k)}\) that reads in local coordinates  
\begin{align}
	D^{(k)} = \xi^{\alpha_1\cdots\alpha_k} \pd_{\alpha_1}\cdots \pd_{\alpha_k}~.
\end{align}
Here \(\xi\) is an arbitrary constant symmetric tensor. It is immediately clear that \(D^{(k)}\) commutes with the exterior derivative and the Hodge star and hence again
\begin{align}
	\dd *\cJ_\eta^{(k)}=0~,
\end{align}
provided that \(\eta^{(k)}\) and \(\tilde{\eta}^{(k)}\) satisfy \cref{condition eta}. The original dressed currents \cref{infinite current} correspond to $k=0$.

This explains how the full derivative structure of the free higher-spin currents is recovered. A spin-\(s\) bilinear current built from field strengths has total derivative order \(s-2\). The family labeled by \(k\) gives terms in which \(k\) derivatives act on the second field strength. If the parameters \(\eta^{(k)}\) and \(\tilde{\eta}^{(k)}\) are chosen linear in \(s-2-k\) derivatives of \(J\) and \(\tilde{J}\), the resulting current contains the schematic structure
\begin{align}
	\pd^{s-2-k}J\,\pd^kJ~,
	\qquad
	\pd^{s-2-k}\tilde{J}\,\pd^kJ~,
	\qquad
	\pd^{s-2-k}J\,\pd^k\tilde{J}~,
	\qquad
	\pd^{s-2-k}\tilde{J}\,\pd^k\tilde{J}~.
\end{align}
Letting \(k=0,\ldots,s-2\) gives all derivative splittings that appear in the free spin-\(s\) currents of \cite{Anselmi:1999bb}, before the trace subtractions. The coefficients are then fixed by imposing complete symmetry, tracelessness, and conservation. With such a construction, we can build an infinite tower of infinite-dimensional families of conserved currents, that together determine all higher-spin currents.

\subsection{Including interactions}\label{sec: interactions}

At first sight, the infinite family of dressed charges constructed above may look like an artifact of the free theory. This is a reasonable suspicion. We have just seen that the conserved currents \cref{infinite current} are closely related to higher-spin currents, whose existence strongly constrains a theory to be free, at least in conformal theories \cite{Maldacena:2011jn,Alba:2013yda,Alba:2015upa}. We will now show that the dressed charges survive the presence of interactions---at least those of a specific kind. 

The story extends most directly to non-linear (higher-form) electrodynamics \cite{Gibbons:1995cv,Gibbons:1997xz,Dunne:2004nc,Buratti:2019cbm,Bandos:2020hgy,Avetisyan:2021heg,Avetisyan:2022zza,Sorokin:2021tge,Mkrtchyan:2022ulc,Babaei-Aghbolagh:2022uij,Russo:2024ptw,Kuzenko:2024zra}, or, relatedly, to higher-form superfluids \cite{Armas:2018zbe,Delacretaz:2019brr,Armas:2023tyx}. The action of such theories can be written in the general form
\begin{align}\label{eq:NLQED action}
    S = \int_X *\p\qty(\u)~, \qquad \u = *(f\w *f) = \frac{1}{(p+1)!} f_{\mu_1\cdots \mu_{p+1}} f^{\mu_1\cdots \mu_{p+1}}~.  
\end{align}
Here \(\p(\cdot)\) is a smooth function of \(\u\) away from the origin. Many applications further assume that it is smooth at the origin and reduces to ordinary electrodynamics in the weak-field limit. We will not need either of these assumptions. This class of theories has two standard origins. In gauge theory, it appears as the leading local effective action after integrating out heavy charged modes, as in the Euler--Heisenberg and Born--Infeld actions \cite{Dunne:2004nc,Gibbons:1997xz}. In the symmetry-breaking description, the same functional \(\p(\u)\) gives the leading-derivative action for the Goldstone mode of a broken higher-form symmetry at finite charge density. In that case, \(\p\) plays the role of the equation of state.

In the allowed dimensions, namely \(d=2p+2\) with \(p\) odd, one may also add non-linear versions of the theta term \cref{eq:theta-term}. The most general interacting action is then
\begin{align}
        S = \int_X *\p\qty(\u,w)~, \qquad w = *(f\w f)~.
\end{align}
This applies, for instance, to a photon (\(p=1\)) in four dimensions. Notable examples of non-linear electrodynamics with a theta term include Born--Infeld theory \cite{Gibbons:1997xz}, Euler--Heisenberg theory \cite{Dunne:2004nc}, and ModMax electrodynamics \cite{Bandos:2020jsw}. Similarly to the ordinary theta term, the \(w\)-dependence does not change the discussion qualitatively. We will therefore proceed without \(w\)-dependence, and return to it briefly at the end of the section.

Let us now repeat the construction of dressed charges. The equations of motion are
\begin{align}
    \dd*\qty(\kappa f) = 0~,
\end{align}
where the response coefficient \(\kappa\qty(\u)\) is
\begin{align}
	\kappa\qty(\u)=2\,\p'(\u)~.
\end{align}
In the Maxwell limit, where \(\p(\u)=\u/(2\g^2)\), this reduces to \(\kappa=1/\g^2\) and is thus field independent. The electric and magnetic currents are therefore
\begin{align}
	J = \ii\,\kappa\,f~, \qquad \widetilde{J}=\frac{1}{2\pi}*f~.
\end{align}
As in the free theory, the electric current is conserved on-shell, while the magnetic one is conserved off-shell.
This gives the same ansatz for the dressed currents as in \cref{infinite current}:
\begin{align}
    *\cJ_\eta = \eta\w *J + \widetilde{\eta}\w *\widetilde{J}~.
\end{align}
With this ansatz, the new dressed current is conserved provided
\begin{align}\label{eq:NLED-dressing-condition}
	\dd\widetilde{\eta}
	+
	2\pi\ii\,\kappa\qty(\u)\,*\dd\eta
	=0~.
\end{align}
For \(\kappa=1/\g^2\), this reduces to \cref{condition eta} as expected.

There is one important difference from the free theory. The condition \cref{eq:NLED-dressing-condition} must be evaluated on a solution of the non-linear equations of motion, meaning the dressing data are necessarily field-dependent. When \(\kappa\) has fixed sign, and in particular does not vanish, it is useful to redefine
\begin{align}
	h=\kappa^{1/2}\;\eta~, \qquad
	\widetilde{h}=\kappa^{-1/2}\;*\widetilde{\eta}~.
\end{align}
Then \cref{eq:NLED-dressing-condition} takes the form
\begin{align}\label{eq:NLED-Witten-condition}
	\dd_\kappa^\dagger\widetilde{h} + 2\pi\ii\,(-1)^{pd}\, \cdd_\kappa h = 0~. 
\end{align}
where \(\dd_\kappa=\kappa^{1/2}\,\dd\,\kappa^{-1/2}=\dd-\dd\log\kappa^{1/2}\w\cdot\) and its adjoint is \(\cdd_\kappa = \kappa^{-1/2}\,\cdd\,\kappa^{1/2}\). Since \(\dd\log\kappa\) is exact, the corresponding Morse--Novikov cohomology is isomorphic to the ordinary de Rham cohomology. Hence, as long as \(\kappa\) does not vanish, the non-linear theory has the same local space of dressing data as the free theory. In particular, modulo the exact shifts of \(\eta\) and \(\widetilde{\eta}\), the dressed currents again carry \(\binom{d-2}{p}\) local degrees of freedom.

Decomposing the dressing data in a local neighborhood of \(\Sigma\), as in \cref{condition eta decomp}, gives the corresponding charges
\begin{align}\label{eq:NLED-dressed-charge}
	\cQ_\eta
	&=
	\int_{\Sigma} *\cJ_\eta 	=
	\ip{\eta_\Sigma}{\kappa\, E}_\Sigma
	+
	\frac{1}{2\pi}\ip{\widetilde{\eta}_\Sigma}{B}_\Sigma~.
\end{align}
In the interacting version of the dressed charges, the electric field is replaced by the non-linear displacement field \(\kappa\, E\), while the magnetic part remains as in the free theory.

Finally, let us reinstate the dependence on \(w=*(f\w f)\), in dimensions that allow it. The theta angle is promoted to a response coefficient
\begin{align}
    \theta\qty(\u,w) = 2\,\pd_w \p\qty(\u,w)~,
\end{align}
together with
\begin{align}
    \kappa\qty(\u,w) = 2\,\pd_{\u}\p\qty(\u,w)~.
\end{align}
The constitutive relation now mixes electric and magnetic field strengths. When \(\theta\) is constant, this is the usual Witten effect. In general, the electric current in this case becomes
\begin{align}
    J = \ii \qty(\kappa\,f+\theta *f)~,
\end{align}
while the magnetic current is unchanged. The dressed current again has the same form, now defined using the new electric current. The condition on \(\eta\) and \(\widetilde{\eta}\) becomes
\begin{align}
    \dd{\tilde{\eta}}+2\pi\ii \qty( \kappa *\dd{\eta}+ \theta \dd{\eta}) = 0~,
\end{align} 
and the Witten effect reappears in the dressed charges:
\begin{align}
	    \cQ_\eta =	\ip{\eta_\Sigma}{\kappa\, E + \ii \theta B}_\Sigma
		+
		\frac{1}{2\pi}\ip{\widetilde{\eta}_\Sigma}{B}_\Sigma~.
\end{align}

More generally, any interaction that changes only the local constitutive relation between currents and field strengths, while preserving current conservation, preserves the infinite family of charges in a deformed form. The examples above do not exhaust such interactions, but they cover a large class. By contrast, dynamical charged matter sources the electric current, while dynamical magnetic defects source the Bianchi identity. Such couplings typically disrupt the construction, and the infinite family of charges ceases to exist.%
\footnote{Of course, it can reappear in the IR, where the charged objects are integrated out.}
Nonetheless, such a coupling may still source the currents in a way that preserves the infinite family of charges. We will not discuss such couplings here; see, however, \cref{sec:conclusions}. For now, we turn all interactions back off and return to the free generalized Maxwell theory.

\section{The dressed-charge algebra}\label{sec: algebra}

Returning to the dressed charges constructed in \cref{conserved charges}, we note that since these are supported in codimension-1, they can have non-trivial commutation relations. We now show that this is indeed the case, and that the algebra of these dressed charges forms a \emph{generalized Kac--Moody algebra}. 

The commutator of $\cQ_{\eta}$ follows straightforwardly from the canonical commutation relations between the gauge field and its conjugate momentum. In Euclidean signature these read:
\begin{equation}
    \comm{a^{\vec{k}}(x)}{\frac{1}{\g^2}\pd_{[r}a_{\vec{\ell}]}(y)} = \delta_{\vec{\ell}}^{\vec{k}}\delta(x-y)~,
\end{equation}
where $\vec{k} = [k_1\cdots k_p]$ and $\vec{\ell}=[\ell_1\cdots \ell_p]$ denote collections of $p$ antisymmetric indices on \(\Sigma\), and we remind the reader that our Euclidean temporal coordinate is \(r\).

From here on, and for the remainder of the paper, we return to taking the dressing functions, \(\eta\), \(\zeta\) and their tilded counterparts to be field-independent. This is sufficient for our goal of reaching the state-defect map.  We find:
\begin{equation}\label{eq:Q-algebra}
    \comm{\cQ_\eta}{\cQ_{\zeta}}=-\ii\,\Gamma(\eta,\zeta)~,
\end{equation}
where the central ``commutator map'' \(\Gamma(\eta,\zeta)\) is:
\begin{equation}\label{eq:comm=map}
     \Gamma(\eta,\zeta)= \frac{1}{2\pi}\int_{\Sigma}\qty(\eta\w \dd\tilde{\zeta}- \zeta \w \dd\tilde{\eta})~.
\end{equation}
This algebra of dressed charges is to be expected. First, note that these charges are, really, modulated refinements of Abelian symmetries. Thus, one does not expect any operator dependence on the right-hand-side. At the same time, we know that the electric and magnetic \(\U(1)\) symmetries of the theory are tied by a mixed 't Hooft anomaly. The central term in \cref{eq:Q-algebra} is precisely an indicator of this anomaly. 

Indeed, the structure that appears on the right-hand-side of the above equation is precisely a centrally extended Abelian Kac--Moody-like algebra. Adapting the methods of \cite{Delacretaz:2019brr,Vitouladitis:2025zoy}, this algebra can be arrived at directly from the anomaly, by a path-integral calculation of the two-point function of electric and magnetic currents. In fact, with the same tools, one can prove an even stronger statement. Any time one has a \(\gf{\U(1)}{p}\times\gf{\U(1)}{d-p-2}\) symmetry with a mixed anomaly like \cref{anom inflow}, one can then construct infinitely many charges like those above with the algebra \cref{eq:Q-algebra}. This includes conformal cases \cite{Hofman:2018lfz,Hofman:2024oze}, relative theories%
\footnote{By relative theories, we are referring to theories that live on the boundary of a non-trivial TQFT, see \cite{Freed:2012bs}. A paradigmatic example of this is of course the chiral boson.}
and edge modes \cite{Fliss:2023uiv}, superfluids \cite{Vitouladitis:2025zoy}, or limits of quantum electrodynamics (QED) \cite{Tizzano:2026rgr}, among others.

For practical purposes, it is convenient to re-express \cref{eq:Q-algebra} as an algebra of modes. This is facilitated by first expressing the algebra in terms of the vector of $\vec{\eta}$ defined in \cref{decomp to eigenmodes}. In terms of these vectors the commutator map \cref{eq:comm=map} reads:
\begin{equation}
    \Gamma(\eta,\zeta)=\frac{(-1)^{p}}{2\pi}\ip{\vec{\eta}}{\B_p\,\parity \vec{\zeta}}_{\Sigma}~.
\end{equation}
Along with \cref{Q eta and modes}, this implies the commutation relation of the eigenmodes of the currents,
\begin{equation} \label{mode commutator}
	\comm{\J_{\inda\sigma}}{\J_{\inda'\sigma'}}=\frac{i(-1)^{p+1}}{2\pi}\sigma\mu_{\inda}\,\delta_{\inda,\inda'}\delta_{\sigma,-\sigma'}~.
\end{equation}
This mode algebra depends on \(\Sigma\) through the spectrum of the generalized curl operator. The label \(\inda\) indexes an eigenmode of \(\B_p\), and \(\mu_\inda\) is its eigenvalue (chosen conventionally positive, with the sign absorbed in \(\sigma\)).%
\footnote{For example, on a rectangular flat torus with periods \(L_i\), the eigenvalues are labeled by a non-zero lattice momentum \(\bm{k}\in\Z^{d-1}\), with
\begin{equation}
    \mu_\inda = 2\pi\qty(\sum_{i=1}^{d-1}\frac{k_i^2}{L_i^2})^{1/2}~.
\end{equation}}
For the product of spheres used below, the mode label is \(\inda=(i,\wdeg,\sfn_1,\sfn_2)\). Here, \(\sfn_1\) and \(\sfn_2\) collect the spherical-harmonic quantum numbers on the two factors, while \(i\) and  \(\wdeg\) are extra labels of the eigenforms. We describe these labels and the corresponding eigenvalues in \cref{app: generalized curl}.

We end this section with a preliminary comment which we will expand upon in \cref{sec: the ops}. Having constructed this infinite family of 0-form symmetries, the natural question to ask is what are the local operators that are charged under this symmetry. The answer can be essentially read off from commutator in \cref{eq:Q-algebra}. The charged operators are built from exponentiating the dressed charges themselves. Indeed, defining $U_\eta = \ex{\ii Q_\eta}$, the algebra \cref{eq:Q-algebra} reads
\begin{equation}
    U_\eta U_\zeta = \ex{\ii \Gamma(\eta,\zeta)} U_\zeta U_\eta~.
\end{equation} 
We thus see that the dressed charges act more naturally on themselves. Even though the dressed charges are topological, solutions to the condition \cref{condition eta} can be singular, and hence $U_\eta$, when shrunk to a point, can define non-trivial local operators. When both $\eta$ and $\zeta$ are regular, one can readily show that the commutator map $\Gamma(\eta,\zeta)$ vanishes. This is sensible, as the operators can be deformed past each other in this case without crossing. In contrast, when $\zeta$ corresponds to a singular solution and $\eta$ a regular one, $\Gamma(\eta,\zeta)$ can be thought of as the linking phase of the symmetry operator $U_\eta$ acting on the charged operator $U_\zeta$.

\section{Quantization and the Hilbert space}\label{sec: the states}

The current algebra of the previous section plays an essential role in organizing the spectrum of the theory. We now put these ingredients to work to build up the first half of the story: the spectrum of states living on $\Sigma$. To this end, we now switch to Lorentzian signature (with mostly-plus convention), considering $X=\mathbb{R}_t \times \Sigma$, where $\Sigma$ is a closed, compact Cauchy slice upon which we quantize the theory.

For $p$-form Maxwell theories, canonical quantization is straightforward \cite{Fliss:2023uiv,Benini:2025hbj}.  From the action one can immediately read off the Hamiltonian as:
\begin{align} \label{hamiltonian}
	H_\Sigma = \frac{1}{2\g^2}\qty(\norm{\f{E}{p}}^2_\Sigma+\norm{\f{B}{d-p-2}}^2_\Sigma),
\end{align}
where the norm here is with respect to the usual inner product on $\Sigma$. From the (now Lorentzian) relation between the currents and electric and magnetic fields, this can be equivalently written in terms of $J_r~,\tilde{J}_r$,
\begin{equation}
    H_\Sigma =\pi\qty(\frac{\g^2}{2\pi}\norm{J_r}_\Sigma^2+\frac{2\pi}{\g^2}\norm*{\tilde{J_r}}_\Sigma^2)~.
\end{equation}
We point out that although here we are working in Lorentzian signature, we still refer to the temporal component of the currents as $J_r$ and $\tilde{J}_r$, respectively, in order to seamlessly import the decomposition into modes from the previous section. Employing said decomposition as in \cref{decomp to eigenmodes}, the Hamiltonian becomes
\begin{equation} \label{ham in terms of modes}
	\begin{split}
		H_\Sigma & = 2\pi \sum_{\inda\sigma} \frac{1}{2}(\J_{\inda\sigma})^2 \norm{\Phi_{\inda\sigma}}_\Sigma^2  \\
	 & = 2\pi\qty[ \sum_{i=1}^{b_{\pp}} j_{0,i}^2 +\sum_{k=1}^{b_{p+1}} \tilde{\jmath}_{0,k}^{\kern2pt 2} +\frac{1}{2}\sum_{\substack{\inda \\ \mu_\inda\neq 0}} \qty\Big(\qty(\J_{\inda+})^2 + \qty(\J_{\inda-})^2)]~.\\
	\end{split}
\end{equation}
We have written $j_0$ and $\tilde{\jmath}_0$ as the zero modes associated to the first and second entries of a zero-mode eigenform $\Phi_{0} =\qty\big(\phiu_0, \phid_0)^{\sfT}$ respectively. The zero modes are related to the electric and magnetic charges, respectively, and their number is determined by the Betti numbers, $b_k(\Sigma)$, of $\Sigma$. From the above, we see that the current algebra enables a higher-dimensional version of the Sugawara construction for the Hamiltonian.

The above expression of the Hamiltonian motivates defining ladder operators as combinations of the current modes:
\begin{equation}\label{creation and annih}
    \begin{split}
		\cA_{\inda}=\frac{1}{\sqrt{2}}\qty\big(\J_{\inda+}-i(-1)^p \J_{\inda-})  \qq{and} \cA_{\inda}^\dagger =\frac{1}{\sqrt{2}} \qty\big(\J_{\inda+}+i(-1)^p \J_{\inda-})~,
	\end{split}
\end{equation}
so that
\begin{equation}\label{ham with A}
    H_\Sigma =2\pi\left( \sum_{i=1}^{b_{\pp}} j_{0,i}^2 +\sum_{k=1}^{b_{p+1}} \tilde{\jmath}_{0,k}^{\kern2pt 2} +\sum_{\inda\neq0} \cA^\dagger_\inda \cA_\inda \right)~.
\end{equation}
These satisfy the expected commutation relations for ladder operators:
\begin{equation}
    \comm{\cA_\inda}{\cA^\dagger_{\inda'}} = \frac{\mu_\inda}{2\pi} \delta_{\inda,\inda'}~,
\end{equation}
which follows directly from the commutator of the modes in \cref{mode commutator}.

From the above decomposition of the Hamiltonian into modes, we see that the states are labeled by their electric and magnetic charges. More specifically, the Hilbert space is graded by the charges under the $\U(1)_e$ and $\U(1)_m$---which we label as $e_i$ and $m_k$, where $i$ and $k$ are again labels given by the aforementioned Betti numbers. The full Hilbert space is then a direct sum over all possible sectors,%
\footnote{In principle, the Hilbert space should be labeled by a maximally commuting set of generators. This theory contains also composite symmetries \cite{Brauner:2020rtz,Heidenreich:2020pkc} with currents $*J_n=*J\w*J\w\cdots\w *J=(*J)^{\w n}$, or $*\tilde{J}_{\thintilde{n}}=(*\tilde{J})^{\w \thintilde{n}}$, whose associated charges commute with everything. However, in the Abelian case, these symmetries are not independent from $\U(1)_e$ and $\U(1)_m$ and the would-be associated sectors are fully determined by $e_i$ and $m_k$. Such sectors may become important in a non-Abelian generalization of the story presented here.}
\begin{equation}
	\mathcal{H}_\Sigma =\bigoplus_{e_i,m_k\in \Z} \mathcal{H}_{e_i,m_k}~.
\end{equation}
The first states to populate the Hilbert space---denoted as \(\ket{e_i,m_k}\)---are those annihilated by all lowering operators
\begin{equation}
	\cA_{\inda} \ket{e_i,m_k}=0 \qquad \forall \inda~.
\end{equation}
As we will see shortly, these states are primary in the sense that they form the highest-weight representations of the generalized Kac--Moody algebra. Acting with creation operators, $\cA^\dagger_{\inda}$, on these states raises the energy by $\mu_\inda$ and takes us to excited sectors.

While the above discussion holds for generic (closed, compact, torsion-free) $\Sigma$, in the following, we will focus on a class of Cauchy slices that can be realized as the boundary of a $d$-dimensional manifold $\Xi$ with the following restriction on its topology:
\begin{equation}\label{Betti condition}
    b_{\pp}(\Xi)=0~, \quad b_{p+1}(\Xi)=0~.
\end{equation}
We require the above condition so that any non-trivial $\U(1)_e$ or $\U(1)_m$ charge measured in $\Xi$ can be obtained by the insertion of a charged operator in $\Xi$. In other words, it guarantees that all possible charge sectors of the Hilbert space can be accessed by a bulk path integral. As a simple example, we will focus on $\Sigma = S^p\times S^{\pp}$, whose Betti numbers are given by
\begin{alignat}{2}\label{betti number}
	b_k(\Sigma) & =\begin{cases}
		       1 & \t{if}\ k=0,\;p,\;\pp,\;d-1 \\
		       0 & \t{else}
	       \end{cases} \; & \qquad d\neq 2p+1~,        \\
	b_k(\Sigma) & = \begin{cases}
		        1 & \t{if}\ k=0,\;d-1 \\
		        2 & \t{if}\ k=p     \\
		        0 & \t{else}
	        \end{cases} \;        & \qquad d=2p+1~.
\end{alignat}
In particular, $b_{\pp}(\Sigma)\neq 0$, which allows non-trivial charge sectors for the $U(1)_e$ symmetry. We can define the corresponding $\Xi$ by filling the bulk of $S^{\pp}$, such that $\Xi=S^p\times B^{d-p}$. With this choice, we have $b_{\pp}(\Xi)=b_{\pp}(\Sigma)-1$ and the condition \cref{Betti condition} is satisfied if $d\neq 2p+1$. As such, we will assume that $d\neq 2p+1$ in the following. In particular, this omits the case of $d=3$, with $p=1$. This case is especially interesting because a Chern--Simons term is allowed, see also \cref{sec:conclusions}.

Interestingly, only in the conformal case $d=2p+2$ do we have both $j_0$ and $\tilde{\jmath}_0$ zero modes; more plainly, it is only then that we have both electric and magnetic charges.%
\footnote{We will see later that this also reflects the fact that in $d=2p+2$ there can exist dyonic defects.}
Otherwise, if $d\neq 2p+2$, $b_{p+1}(\Sigma)=0$ and no $\tilde{\jmath}_0$ can appear. In what follows, we primarily also restrict to \(d\neq 2p+2\). We stress that this does not limit the applicability of our construction. Explicitly, our state-defect correspondence will persist in this case and can be obtained by a combination of the arguments presented here and those in \cite{Hofman:2024oze}. We simply impose this restriction for simplicity of the calculations involved, as well as to emphasize our construction in the non-conformal case. When appropriate, we comment on the $d=2p+2$ case.

The structure of the Betti numbers in \cref{betti number} reveals the motivation for this choice of $\Sigma=S^p \times S^{\pp}$. In this case, the states are labeled by just one electric charge $e$, instead of multiple $e_i$ along different cycles. More generic choices of $\Sigma$ that still satisfy \cref{Betti condition} are, in principle, possible, and our construction of the state-defect correspondence should go through similarly, but with the added responsibility of keeping track of which primaries appear on which cycles of $\Sigma$. 
Thus, we argue that $\Sigma =S^p\times S^{\pp}$ is both sufficient and pedagogical. 

Returning to the primary states, we see from \cref{ham with A} that the energy of $\ket{e}$---being a primary and thus annihilated by $\cA_\inda$---comes solely from the $j_0$ term. Inverting the mode expansion of the currents in \cref{decomp to eigenmodes},
\begin{equation}
    j_{0}=\frac{\g}{\sqrt{2\pi}} \langle \phi_0,J_r \rangle_\Sigma.
\end{equation}
The zero mode $\phi_0$ can be related to the Poincaré dual of $S^{\pp}$, denoted as $\widehat{\S}^{\pp}$,%
\footnote{To be more precise, $\phi_0$ is proportional to the (unique) harmonic representative of the Poincaré dual of $\pp$. 
This is related to other representatives, such as a delta-function localized form, by a gauge transformation of $\eta$. The choice of representative will play no role in our story, so by an abuse of terminology we will refer to it as ``the'' Poincaré dual. \label{foot:poincare-dual=gauge}}
\begin{equation}\label{zero modes}
    \phi_0=\sqrt{\frac{\vol(S^p)}{\vol(S^{\pp})}}\,\widehat{\S}^{\pp}
\end{equation}
where the volume factors of the two spheres ensure the eigenform is well-normalized. Thus,
\begin{equation}
    j_{0}=\frac{\g}{\sqrt{2}\pi} \sqrt{\frac{\vol(S^p)}{\vol(S^{\pp})}} \int_\Sigma  \widehat{\S}^{\pp} \w *_\Sigma J_r~.
\end{equation}
However, the quantity inside the integral is simply the electric charge  \cref{electric charge}. We conclude
\begin{equation}\label{j0 on states}
    j_0 \ket{e} = \frac{\g}{\sqrt{2\pi}} \sqrt{\frac{\vol(S^p)}{\vol(S^{\pp})}} \, e \ket{e}~.
\end{equation}
The energy of the primary states is therefore,
\begin{equation}
	H_\Sigma \ket{e} =\Delta_e \ket{e} = \g^2\frac{\vol(\S^p)}{\vol(\S^{\pp})}\,e^2\,\ket{e}~.
\end{equation}
On top of the primary states lie descendants. They are obtained by successively acting with raising operators,
\begin{equation}
	\ket{N_{\inda}, e} = \prod_{\inda} \qty(\cA^{\dagger}_{\inda})^{N_{\inda}}\ket{e}~.
\end{equation}
These states have energies given by:
\begin{equation}
	H_\Sigma \ket{N_{\inda},e} = \qty(\Delta_e +\sum_{\inda}N_{\inda} \mu_{\inda})\ket{N_{\inda},e}~.
\end{equation}
This concludes the canonical quantization of the theory and the construction of its Hilbert space. Let us briefly mention a few different ways one can view this Hilbert space. From the fact that both the ordinary electric and magnetic charges, and the ladder operators are generators of the Kac--Moody-like algebra, it follows that the Hilbert space furnishes a direct sum of the integrable representations of this algebra. Indeed, the thermal trace, \(\tr_{\cH}\ex{-\beta H}\), is precisely an extended character of the Kac--Moody-like algebra \cite{Fliss:2023uiv}. An alternative but equivalent interpretation is to view the zero-modes as furnishing a quantum mechanical Hilbert space. Concretely, the primary states, \(\ket{e}\), build precisely the Hilbert space of a particle on a ring, \(L^2(\S^1)\). More generally, on a generic $\Sigma$, the primary states build the Hilbert space of a quantum mechanical particle propagating on \(\mathbb{T}^{b}\), with \(b=b_{\pp}(\Sigma)+b_{p+1}(\Sigma)\) and the total Hilbert space is \(\cH_\Sigma = L^2(\mathbb{T}^b)\otimes\cF\), where \(\cF\) is the ordinary Fock space, built out of the oscillators. See also \cite{Kirsten:1993ug,Letsios_Vitouladitis} for an analogous interpretation concerning massless scalars and \cite{Meynet:2025zem} for similar comments in four-dimensional Maxwell theory.

\subsection{Interlude: spontaneous symmetry breaking} 

The by-now-standard understanding of a free $p$-form gauge field is as a Goldstone mode for the spontaneous breaking of a $p$-form symmetry \cite{Gaiotto:2014kfa,Lake:2018dqm,Hofman:2018lfz,McGreevy:2022oyu}. In other words, the higher-form Maxwell theory we have been considering is best seen as the Goldstone effective field theory for the corresponding higher-form symmetry. This effective theory is valid in the low-energy regime, below some UV cutoff $\Lambda$, here encoded in the coupling constant, $\g^2\sim \Lambda^{2p+2-d}$. However, a generalization of the Coleman--Mermin--Wagner (CMW) theorem \cite{Gaiotto:2014kfa,Lake:2018dqm} (see also \cite{Benini:2025hbj} for a recent entropic proof) forbids the symmetry breaking of continuous $p$-form symmetries in dimensions $d\leq p+2$. The quantization on \(\S^p\times\S^{\pp}\) presented above already hints at the higher-CMW theorem. Concretely, let us compare the energy of the primary states to those of the first excited states. Let \(R_1\) and \(R_2\) denote the radii of \(\S^p\) and \(\S^{\pp}\), respectively, and take \(R_2\gg R_1\). We then have: 
\begin{equation}
    \begin{split}
        \t{energy of primary} \, \, \Delta_e  &\sim \frac{1}{R_2}\frac{\qty(\Lambda R_1)^p}{(\Lambda R_2)^{d-p-2}}\, e^2 \\         
		\t{energy of 1\textsuperscript{st} excited state} &\sim \frac{1}{R_2}\frac{\qty(\Lambda R_1)^p}{(\Lambda R_2)^{d-p-2}}\, e^2 + \frac{\sqrt{d-p-1}}{R_2}~,
	\end{split}
\end{equation}
where we also used the fact that the first excited state is obtained by exciting a single oscillator mode with unit angular momentum on \(\S^{\pp}\), for which \(\mu_\inda = \sqrt{\pp/R_2^2}\) (cf. \cref{app: generalized curl}). In the thermodynamic limit, where \(\Lambda R_2 \gg 1\) with \(\Lambda R_1\) kept fixed, we see that the primary states are parametrically lighter than the first excited states if and only if \(d>d_\t{crit} = p+2\), establishing the correct lower critical dimension. See also \cite{Maeda:2025ycr,Benini:2025hbj} for similar arguments.

At finite volume, the symmetry is restored (when \(d>d_\t{crit}\), and trivially preserved otherwise) and there is a single vacuum $\ket{e=0}$. In the strict infinite volume limit, and in \(d>d_\t{crit}\), while the primary states become gapless, they do not constitute vacua. Instead, the primaries can be rearranged into the symmetry-breaking vacua,
\begin{equation}
	\ket{\theta} = \sum_{e\in\Z} \ex{ie\theta}\ket{e}~,
\end{equation}
where
\begin{equation}
	U_\alpha \ket{\theta} = \ket{\theta+\alpha}~.
\end{equation}
In the case \(p=0\), one can explicitly check that the states \(\ket{\theta}\) satisfy cluster decomposition. The analogous statement for \(p>0\) is that large Wilson defects, serving as order parameters for the breaking of the \(p\)-form symmetry, exhibit perimeter-law behavior on these states.

\section{Defects and their descendants}\label{sec: the ops}

Having described the states of our theory and how they are organized, we can move on to look at the other half of the coin, the (extended) operators, with the aim of similarly defining primaries and descendants as well as the action of operators on these defects. We return in this section to Euclidean signature, where the geometry of the full spacetime is a $\Xi^d=S^p\times B^{d-p}$, where \(B^{d-p}\) is a ball whose boundary is the \(S^{\pp}\) of the previous section.

For $p$-form Maxwell theory in $d$ dimensions, the available operators are $p$-dimensional Wilson operators,
\begin{equation}
    W_e(\gamma_p) =\exp(\ii\,e\int_{\gamma_p}a_{[p]})~, \qquad e\in \Z~.
\end{equation}
To ensure gauge invariance under shifts of $a$, $\gamma_p$ must be closed. Moreover, for invariance under large gauge transformations, the charge of the Wilson surface is required to be quantized, $e\in \Z$. These are the charged objects under the $p$-form electric symmetry generated by the topological operators,
\begin{equation}
    U^{\t{E}}_{\alpha}(C_{\pp})=\exp(\ii\,\alpha\int_{C_{\pp}}*J)~.
\end{equation}
The other operators at our disposal are $(d-p-2)$-dimensional 't Hooft operators, $H_m(\gamma_{d-p-2})$. In the electric picture they are obtained as disorder operators, i.e. by performing the path integral on spacetime apart from a $(d-p-2)$-dimensional locus \(\gamma_{d-p-2}\) around which an integer magnetic flux is prescribed
\begin{equation}
    \int_{S^{p+1}(\gamma)} *\f{\tilde{J}}{\pp} = m\in\Z~,
\end{equation}
where \(S^{p+1}(\gamma)\) is a tiny sphere linking with \(\gamma_{d-p-2}\). 

Equivalently, the 't Hooft operators can be viewed as Wilson operators of an $S$-dual gauge field. This can be introduced most cleanly into the action by a ``trivial gauging'' of the electric symmetry $\Z_1\subset \gf{\U(1)}{p}$ \cite{Paznokas:2025auw}. In practice, this is done by introducing a $(p+1)$-form \(\U(1)\) gauge field \(\f{\psi}{p+1}\) together with a $(d-p-2)$-form \(\U(1)\) gauge field $\f{\tilde{a}}{d-p-2}$ so that the action becomes%
\footnote{If there is a theta term present in the action, necessarily $d=2p+2$ with $p$ odd, one gauges $\Z_1\subset \U(1)_e$ by introducing $\f{\psi}{p+1}$ and $\f{\tilde{a}}{d-p-2}$ into the action as,
\begin{equation*}
    S = \frac{1}{2\g^2}\int (f-\psi)\w *(f-\psi) +\frac{i\theta}{8\pi^2}\int (f-\psi)\w (f-\psi)- \frac{\ii}{2\pi}\int \psi\w\dd{\tilde{a}}~,
\end{equation*}
and similarly integrating out $\psi$ and $a$.
}
\begin{equation}
    S = \frac{1}{2\g^2}\int (f-\psi)\w *(f-\psi) - \frac{\ii}{2\pi}\int \psi\w\dd{\tilde{a}}~,
\end{equation}
and integrating out $\psi$ and the original photon $a$. In this picture, \(H_m(\gamma_{d-p-2})\) are simply:
\begin{equation}
    H_m (\gamma_{d-p-2})=\exp(\ii\,m\int_{\gamma_{d-p-2}}\tilde{a}_{[d-p-2]})~, \qquad m \in \Z~.
\end{equation}
Hence, the magnetic charges are also quantized, $m\in\Z$, for the same reason as the Wilson defects. The 't Hooft defects are charged under the generator of the magnetic symmetry,
\begin{equation}
    U^{\t{M}}_\alpha\qty(C_{p+1})=\exp(\ii \alpha\int_{C_{p+1}} *\tilde{J})~.
\end{equation}

In the special case of $d=2p+2$, Wilson and 't Hooft operators are both $p$-dimensional, and as such, we can consider dyonic operators,
\begin{equation}\label{dyonic line}
    V_{e,m}=W_e H_m~.
\end{equation}
Additionally, if $d=2p+2$ and $p$ is odd, we can include a theta term in the action, and in this case the 't Hooft operators will also be charged under the electric symmetry via the Witten effect.

\begin{figure}[thb]
    \centering
    \input{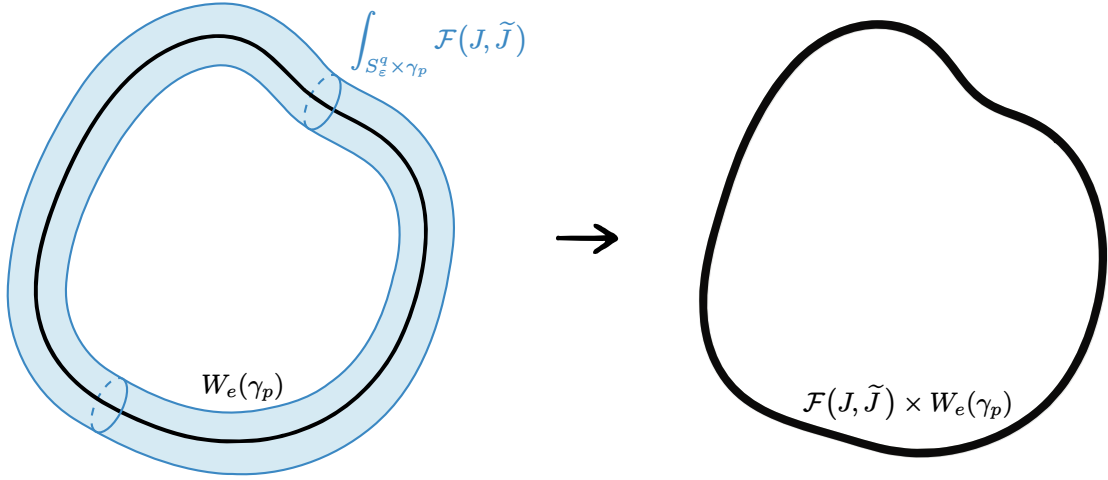}
    \caption{The construction of descendant defects. On the left-hand side, we surround the Wilson surface (drawn here as a line for illustration) by a small cutout tube, $\S_\varepsilon^{\pp}\times \gamma_p$---in blue---upon which we integrate over a function of the currents, $\cF\qty\big(J,\tilde{J}\,)$. This results in a new $p$-dimensional defect on $\gamma_p$, shown on the right-hand side,  after the limit $\varepsilon\to 0$.}
    \label{fig: smearing}
\end{figure}

The Wilson and 't Hooft defects are the   ``elementary'' or \emph{primary} operators of $p$-form Maxwell theory. On top of these, we can define additional \emph{descendant} defects by smearing polynomials of gauge-invariant local operators, $J$, $\tilde{J}$ and their derivatives, along the worldvolume of the primary operators. This is in complete analogy to how one defines descendant operators in 2d CFT. Take, for example, a Wilson operator $W_e(\gamma_p)$. Such smearing over the Wilson operator can be achieved by cutting out a small tube $S^{\pp}_\varepsilon \times \gamma_p$ which surrounds $\gamma_p$, inserting the local operator along this tube, and then taking the limit $\varepsilon\to 0$, as illustrated in \cref{fig: smearing}.

These operators prepare states on the Hilbert space over \(\S^p\times \S^{\pp}\) by performing the Euclidean path integral on \(\Xi^d = \S^p\times B^{d-p}\), with the insertion of such an operator on \(S^p\).%
\footnote{One does not need to wrap a non-trivial cycle, as these operators are not topological. However, the class of states prepared this way will be the relevant ones for the state-defect correspondence.}
For example, a Wilson operator prepares a state as: 

\begin{equation}\label{eq:Wilson-state}
    \ket{W_e}= \int_{\Xi^{d}} \DD{a}\; W_e\qty(\S^p)\, \ex{-S[a]}~,
\end{equation}
and similarly for the 't Hooft operators or the descendant operators. In the above we mean, abusing notation, that we integrate over $p$-form gauge fields on \(\Xi^d\), modulo gauge (and gauge-for-gauge-\dots) transformations. Acting on a state prepared in this way with the operator $\mathcal{O}$ (which can be extended or local) corresponds to inserting this operator in the path integral:
\begin{equation}\label{op in path int}
    \mathcal{O}\ket{W_e}= \int_{\Xi^{d}} \DD{a}\, \mathcal{O}\, W_e(\gamma_p)\, \ex{-S[a]}~.
\end{equation}
In the following we will take the radius of the $(d-p)$-dimensional ball, $B^{d-p}$, to be $r_0$. The radius of the $S^p$ will not play a role in what follows so we suppress it. The states prepared this way live, thus, on the Hilbert space over
\begin{equation}
    \Sigma_{r_0} = \S^p \times S^{\pp}_{r_0}~.
\end{equation}

With this construction at hand, the main idea is the following. Since the Kac--Moody-like algebra of \cref{sec: infinite symmetries} completely characterizes the spectrum of states, as shown in \cref{sec: the states}, we can use it to compare the states prepared as in \cref{eq:Wilson-state} to the states constructed previously. Thus, we shall prepare states with this path integral prescription and use the Kac--Moody charges to determine if this procedure is sufficient to reach the entire Hilbert space. 

\subsection{Zero modes and charged sectors}\label{sec: acting with zero modes}

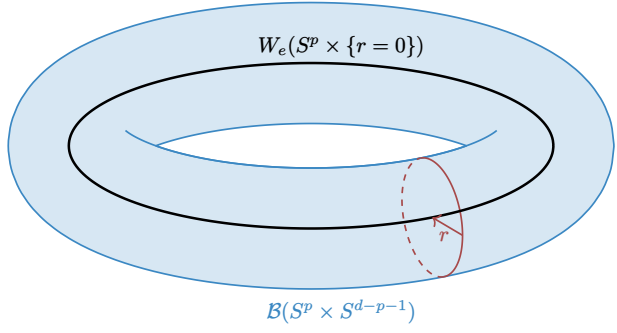
\begin{figure}
    \centering
    \tdplotsetmaincoords{70}{0}
\tikzset{declare function={torusx(\u,\v,\R,\r)=cos(\u)*(\R + \r*cos(\v)); 
torusy(\u,\v,\R,\r)=(\R + \r*cos(\v))*sin(\u);
torusz(\u,\v,\R,\r)=\r*sin(\v);
vcrit1(\u,\th)=atan(tan(\th)*sin(\u));
vcrit2(\u,\th)=180+atan(tan(\th)*sin(\u));
disc(\th,\R,\r)=((pow(\r,2)-pow(\R,2))*pow(cot(\th),2)+%
pow(\r,2)*(2+pow(tan(\th),2)))/pow(\R,2);
umax(\th,\R,\r)=ifthenelse(disc(\th,\R,\r)>0,asin(sqrt(abs(disc(\th,\R,\r)))),0);
}}
\scalebox{.8}{
\begin{tikzpicture}[tdplot_main_coords]
\pgfmathsetmacro{\R}{4}
\pgfmathsetmacro{\r}{1}
 \draw[thick,blue,fill=blue,even odd rule,fill opacity=0.2]
 plot[variable=\x,domain=0:360,smooth,samples=71]
 ({torusx(\x,vcrit1(\x,\tdplotmaintheta),\R,\r)},
 {torusy(\x,vcrit1(\x,\tdplotmaintheta),\R,\r)},
 {torusz(\x,vcrit1(\x,\tdplotmaintheta),\R,\r)}) 
 plot[variable=\x,
 domain={-180+umax(\tdplotmaintheta,\R,\r)}:{-umax(\tdplotmaintheta,\R,\r)},smooth,samples=51]
 ({torusx(\x,vcrit2(\x,\tdplotmaintheta),\R,\r)},
 {torusy(\x,vcrit2(\x,\tdplotmaintheta),\R,\r)},
 {torusz(\x,vcrit2(\x,\tdplotmaintheta),\R,\r)})
 plot[variable=\x,
 domain={umax(\tdplotmaintheta,\R,\r)}:{180-umax(\tdplotmaintheta,\R,\r)},smooth,samples=51]
 ({torusx(\x,vcrit2(\x,\tdplotmaintheta),\R,\r)},
 {torusy(\x,vcrit2(\x,\tdplotmaintheta),\R,\r)},
 {torusz(\x,vcrit2(\x,\tdplotmaintheta),\R,\r)});
 \draw[thick, blue] plot[variable=\x,
 domain={-180+umax(\tdplotmaintheta,\R,\r)/2}:{-umax(\tdplotmaintheta,\R,\r)/2},smooth,samples=51]
 ({torusx(\x,vcrit2(\x,\tdplotmaintheta),\R,\r)},
 {torusy(\x,vcrit2(\x,\tdplotmaintheta),\R,\r)},
 {torusz(\x,vcrit2(\x,\tdplotmaintheta),\R,\r)});
 \foreach \X  in {300}  
 {\draw[thick,dashed,red] 
  plot[smooth,variable=\x,domain={360+vcrit1(\X,\tdplotmaintheta)}:{vcrit2(\X,\tdplotmaintheta)},samples=71]   
 ({torusx(\X,\x,\R,\r)},{torusy(\X,\x,\R,\r)},{torusz(\X,\x,\R,\r)});
 \draw[thick,red] 
  plot[smooth,variable=\x,domain={vcrit2(\X,\tdplotmaintheta)}:{vcrit1(\X,\tdplotmaintheta)},samples=71]   
 ({torusx(\X,\x,\R,\r)},{torusy(\X,\x,\R,\r)},{torusz(\X,\x,\R,\r)});
 }
 \pgfmathsetmacro{\rho}{4} 

\draw[very thick,black]
 plot[smooth,variable=\t,domain=0:360,samples=100]
 ({\rho*cos(\t)},{\rho*sin(\t)},0);
 \node[black] at (.5,4.8) {$W_e\qty(S^p\times \{r=0\})$};
 \node[blue] at (.5,-8) {$\mathcal{B}(S^p\times S^{\pp})$};
 
 \pgfmathsetmacro{\u}{300}
\pgfmathsetmacro{\v}{0}

\draw[->,thick,red]
({torusx(\u,\v,\R,\r)},
 {torusy(\u,\v,\R,\r)},
 {torusz(\u,\v,\R,\r)})
--
({0.8*torusx(\u,\v,\R,\r)},
 {0.8*torusy(\u,\v,\R,\r)},
 {0.8*torusz(\u,\v,\R,\r)});
    \node[red] at (2.2,-4.4) {$r$};
\end{tikzpicture}}
    \caption{The operator \(\cB\), constructed on \(\Sigma =S^p\times S^{\pp}\), acts on the Wilson surface by radially evolving inward toward \(r=0\). The final configuration \(\cB\ket{W_e}\) is supported on \(S^p\) and defines a \(p\)-dimensional operator. This mechanism applies to \(\cB_0\), \(\cB_\inda\), and \(\cB_\inda^\dagger\), so we label the operator simply by \(\cB\). Acting with \(\cB^\dagger_\inda\) will define the descendant defects.}
    \label{fig: Baction}
\end{figure}

In order to determine how (combinations of) the Kac--Moody generators act on operator insertions, the same mechanism will appear several times below. We start with an operator written on \(\Sigma_{r_0}\), identify it with a dressed charge, and use the topological condition of \cref{condition eta decomp} to evolve its dressing data into the bulk, as in \cref{fig: Baction}. If the radial evolution is regular at the origin, the operator can be shrunk through the ball without leaving an additional insertion. However, if the radial evolution is singular, the singularity produces a genuine \(p\)-dimensional defect on \(S^p\). This is the picture replacing the familiar geometric radial quantization, as it applies to CFTs.

The first test is the electric current zero mode. When \(d\neq 2p+2\), only electric zero modes, and not magnetic ones, are present. We therefore act with
\begin{equation}\label{B0 defintion}
\begin{split}
    \cB_0 =v_0\ j_0= v_0 \left\langle \phi_0, \frac{\g}{\sqrt{2\pi}}J_r\right\rangle_{\Sigma_{r_0}}
    ~,
\end{split}
\end{equation}
where \(J_r\) is the radial component of the electric current,
\begin{equation}
    \f{J}{p+1} = \dd{r} \w J_r + J_{\Sigma_r}~,
\end{equation}
while \(\phi_0\) is the electric component of the zero mode of the generalized curl operator (cf. \cref{decomp to eigenmodes,app: generalized curl}). We also allowed for a generally complex coefficient \(v_0\). Comparing with \cref{j0 on states}, we already see that \(\cB_0\) is proportional to the electric charge.

The operator \(\cB_0\) is a dressed charge \(\cQ_\eta\), with
\begin{equation}
    v_0\,\phi_0=\frac{\sqrt{2\pi}}{\g} \eta_\Sigma~.
\end{equation}
To move this charge into the bulk, we allow \(v_0\) and \(\phi_0\) to depend on the radial coordinate \(r\in\closed{0}{r_0}\), with boundary condition \(v_0(r_0)=v_0\). We extend the curly eigenforms so that they remain orthonormal on each \(\Sigma_r\). As reviewed in \cref{app: inner prod computation}, this fixes their radial profile. For the zero mode this gives:
\begin{equation}
    \pd_r\phi_0=-\frac{\pp}{2r}\phi_0~.
\end{equation}
The topological condition \cref{condition eta decomp} then implies
\begin{equation}\label{cond v zero}
    \pd_r v_0(r)  -\frac{\pp}{2r}v_0(r)=0~.
\end{equation}
Together with the boundary condition \(v_0(r_0)=1\),%
\footnote{We drop the arbitrary proportionality factor from \cref{B0 defintion}.}
this gives
\begin{equation}
    v_0(r)=\qty(\frac{r}{r_0})^{\pp/2}~,
\end{equation}
and thus fixes the radial evolution of the zero-mode operator.

We now evaluate this operator on the Wilson state. To do so, it is convenient to rewrite \(\cB_0\) as a bulk integral on \(\Xi^d_{r_0}\), using Stokes' theorem:
\begin{equation}
    \cB_0 = \frac{\g}{\sqrt{2\pi}}\int_{\Xi^d_{r_0}} \dd{r}\w \qty\Big(\qty\big(\pd_r v_0\; \phi_0 +v_0\, \pd_r \phi_0 )\w *_\Sigma J_r +v_0 \phi_0 \w \pd_r *_\Sigma J_r)~.
\end{equation}
Using the zero-mode radial profile, this becomes
\begin{equation}\label{B0 intermediate}
    \cB_0= \frac{\g}{\sqrt{2\pi}} \int_{\Xi^d_{r_0}} \dd{r}\w \qty(\qty(\pd_r v_0  -\frac{\pp}{2r}v_0)\phi_0\w *_\Sigma\, J_r +v_0\,\phi_0 \w \pd_r *_\Sigma J_r)~. 
\end{equation}
Clearly, the first term in brackets vanishes due to \cref{cond v zero}. In the presence of a Wilson operator \(W_e(\gamma)=\ex{\ii e\int_\gamma a}\), the equation of motion for \(J\) is altered to:
\begin{equation}\label{eq:EOM-Wilson}
    \dd{r} \w \qty\big(\pd_r *_\Sigma J_r +(-1)^{p}~ \sd *_\Sigma J_\Sigma) + \sd *_\Sigma J_r = (-1)^{pd} e\,\hat{\gamma},
\end{equation}
where \(\hat{\gamma}\) is the Poincaré dual of \(\gamma\) in \(\Xi^d\) and we consider $\gamma=S^p$ at $r=0$. The term \(\sd *_\Sigma J_\Sigma\) does not contribute after integrating by parts and using \(*_\Sigma \sd \phi_0=0\). The term $\sd *_\Sigma J_r$ also vanishes, due to the Gauss law and the fact that with our choice of $\gamma$, $\hat{\gamma}$ is proportional to $\dd{r}$. Thus,
\begin{equation}
    \begin{split}
        \eval{\cB_0}_{W_e}= (-1)^{p(d-p)}\frac{\g\,e}{\sqrt{2\pi}} \int_{\Xi^d_{r_0}} \qty(\frac{r}{r_0})^{\pp/2} \,\phi_0 \w\hat{\gamma}~.
    \end{split}
\end{equation}
Here, \((\cdot)\big\vert_{W_e}\) denotes the on-shell value in the presence of the Wilson operator.
Finally, the zero mode is related to the Poincaré dual of \(S^{\pp}\) by%
\footnote{Again, to be precise the zero mode is gauge equivalent to \(\widehat{\S}^{\pp}\), but this is sufficient for our purposes. We refer the reader back to the discussion in \cref{foot:poincare-dual=gauge}.}
\begin{equation}\label{phi 0}
    \phi_0(r)=\sqrt{\frac{\vol(S^p)}{\vol(S^{\pp}_r)}}\,\widehat{\S}^{\pp}~.
\end{equation}
The combination \(v_0(r)\phi_0(r)\) has no residual radial dependence. Therefore, \(\cB_0\) is topological, and its action on \(\ket{W_e}\) is
\begin{equation}
    {\cB_0} \ket{W_e} 
    = \frac{\g\,e}{\sqrt{2\pi}}\sqrt{\frac{\vol(S^p)}{\vol(S^{\pp}_{r_0})}}\int_{\Xi^d_{r_0}} \hat{\gamma}\w \widehat{\S}^{\pp}\;\ket{W_e}
    = \frac{\g}{\sqrt{2\pi}}\sqrt{\frac{\vol(S^p)}{\vol(S^{\pp}_{r_0})}}\,e\ket{W_e}~.
\end{equation}
Here we have picked up a sign from exchanging \(\hat{\gamma}\) and \(\widehat{\S}^{\pp}\). 

This long computation is perhaps overkill for this simple case, but it still teaches us a couple of things. First, it proves that the state $\ket{W_e}$ lies in the sector $\cH_e$, of electric charge $e$. This is by no means surprising, and could be inferred without a computation. At the same time, it serves as a warm-up for the subsequent section where the same logic will be applied to higher Kac--Moody modes.

\subsection{Regular modes annihilate primary defects}

We now turn to the higher Kac--Moody modes. Wilson operators are the elementary electrically charged defects of the theory, so the states that they prepare should be primary. We will show this explicitly by constructing operators \(\cB_\inda\) that annihilate them:
\begin{equation}
    \cB_{\inda}\ket{W_e}= 0\qquad  \forall \inda \neq 0 ~.
\end{equation}
We start with a generic linear combination of the current modes,
\begin{equation}\label{eq:Ba-def}
    \begin{split}
    \cB_{\inda}
    &= v_{\inda}\ip{\phi_{\inda+ }}{\frac{\g}{\sqrt{2\pi}}J_r}_{\Sigma_{r_0}}+\tilde{v}_{\inda}\ip{\tilde{\phi}_{\inda+}}{\frac{\sqrt{2\pi}}{\g}\tilde{J}_r}_{\Sigma_{r_0}} \\
    &=\frac{v_{\inda}+\tilde{v}_{\inda}}{2}\,\J_{\inda +}+\frac{v_{\inda}-\tilde{v}_{\inda}}{2}\,\J_{\inda -}~,
\end{split}
\end{equation}
with \(v_\inda,\,\tilde{v}_\inda\in \mathbb{C}\). The components of the curly eigenforms satisfy \(\phi_{\inda-}=\phi_{\inda+}\) and \(\tilde{\phi}_{\inda-}=-\tilde{\phi}_{\inda+}\), as reviewed in \cref{app: generalized curl}. Below we keep the notation \(\phi_{\inda+},\,\tilde{\phi}_{\inda+}\) when useful, but usually drop the \(+\) label, to lighten the notation.

The operator \(\cB_\inda\) is again a dressed charge \(\cQ_\eta\), now with dressing data
\begin{equation}\label{identification v eta}
    \frac{\sqrt{2\pi}}{\g} \eta_{\Sigma}= v_\inda\phi_\inda~, \qq{and} \frac{\g}{\sqrt{2\pi}} \tilde{\eta}_{\Sigma}= \tilde{v}_\inda\tilde{\phi}_\inda~.
\end{equation}
The topological condition \cref{condition eta decomp} becomes
\begin{equation}\label{diffeq vs etar}
    \pd_r \mqty(
        v_\inda\phi_\inda\\
        \tilde{v}_\inda\tilde{\phi}_\inda
    )
    =\dd _{\Sigma}
    \mqty(
        \frac{\sqrt{2\pi}}{\g} \eta_{r}\\[0.5em]
        \frac{\g}{\sqrt{2\pi}}\tilde{\eta}_{r}
    )
    +\ii 
    \mqty(
        0 & (-1)^{dp}*_{\Sigma}\dd_{\Sigma}\\
        (-1)^p*_{\Sigma}\dd_{\Sigma} & 0
    )
    \mqty(
        v_\inda\phi_\inda\\
        \tilde{v}_\inda\tilde{\phi}_\inda
    )~.
\end{equation}
Taking inner products with \(\phi_\inda\) and \(\tilde{\phi}_\inda\), respectively, gives the radial system of equations:
\begin{equation}\label{diffeq vs}
    \begin{split}
    \pd_r v_\inda+2\big\langle\phi_\inda,\pd_r \phi_\inda\big\rangle_{\Sigma_r} v_\inda&=-\ii (-1)^{p}\mu_{\inda} \tilde{v}_\inda\\[0.3em]
    \pd_r \tilde{v}_\inda+2\big\langle\tilde{\phi}_\inda,\pd_r \tilde{\phi}_\inda\big\rangle_{\Sigma_r} \tilde{v}_\inda &=\phantom{-}\ii (-1)^{p}\mu_{\inda} v_\inda~,
\end{split}
\end{equation}
where we used \(\ip{\phi_\inda}{\phi_\inda}_\Sigma =\ip{\tilde{\phi}_\inda}{\tilde{\phi}_\inda}_\Sigma =\frac{1}{2}\) and the eigenvalue equation
\begin{equation}\label{eigenvalue eq}
    \begin{split}
        *_{\Sigma}\sd\phi_{\inda +} &= \mu_{\inda} \tilde{\phi}_{\inda +} \\
        (-1)^{dp+p+1}*_{\Sigma}\sd\tilde{\phi}_{\inda +}&=\mu_{\inda}\phi_{\inda +}~.
    \end{split}
\end{equation}
The inner products in \cref{diffeq vs} are computed in \cref{app: inner prod computation} using explicit eigenforms of the generalized curl operator. The result is
\begin{equation}
    \begin{split}
        \big\langle \phi_{\inda +},\pd_r \phi_{\inda+}\big\rangle_{\Sigma}&=\frac{d-p-3-2\wdeg}{4r}\\
        \big\langle \tilde{\phi}_{\inda +},\pd_r \tilde{\phi}_{\inda +}\big\rangle_{\Sigma}&=\frac{1}{2}\frac{\pd_r \mu_{\inda}}{\mu_{\inda}}-\frac{d-p-3-2\wdeg}{4r}~.
    \end{split}
\end{equation}
Here, $\wdeg$ is an additional integral parameter, whose range is given in \cref{q range 1,q range 2}, that labels form degrees within the constructions of the eigenform $\Phi_{\inda\sigma}$. The radial dependence of the eigenvalue is known and given in \cref{mu and r}. By solving the radial system in \cref{diffeq vs}, we determine the radial evolution of \(\cB_\inda\).

We find this system has two families of solutions: one smooth at \(r\to 0\) and one divergent. The general solution takes the form:
\begin{equation}\label{solution for v}
    \begin{pmatrix}
        v_\inda(r)\\[0.1em]
        \tilde{v}_\inda(r)
    \end{pmatrix}=r^{-\frac{d-p-3-2\wdeg}{2}}\begin{pmatrix}
        1 &\\
        & \frac{i(-1)^{p}}{\mu_{\inda}(r)}
    \end{pmatrix} \begin{pmatrix}
        \cI_{\inda}(r) & \cK_{\inda}(r)\\
        \cI'_{\inda}(r) &\cK'_{\inda}(r)
    \end{pmatrix}\begin{pmatrix}
        c_I\\
        c_K
    \end{pmatrix}~,
\end{equation}
with \(c_I,c_K\in \mathbb{C}\). Here,
\begin{align}\label{modified modified bessels main}
	\cI_\inda(r) \coloneqq r^{\frac{d-p-2-2\wdeg}{2}}\, I_{\delta_\inda}\!\qty\big(\sqrt{\gamma_\inda} r) \qq{and} \cK_\inda(r)\coloneqq r^{\frac{d-p-2-2\wdeg}{2}}\, K_{\delta_\inda}\!\qty\big(\sqrt{\gamma_\inda} r)~,
\end{align}
where \(I_{\delta_\inda}(x)\) and \(K_{\delta_\inda}(x)\) are modified Bessel functions of order
\begin{equation}
\delta_\inda = \frac{1}{2}\sqrt{(d-p-2-2\wdeg)^2+4\,\nu_\inda}~,
\end{equation}
and \(\nu_\inda,\,\gamma_\inda\) are the eigenvalues of the transversal Laplacian on the \(S^{\pp}\) and \(S^p\) spheres, respectively. The function \(I_{\delta_\inda}(x)\) is regular as \(x\to 0\), while \(K_{\delta_\inda}(x)\) is singular. The regular solution therefore has \(c_K=0\). This fixes \(v_\inda(r)\) and \(\tilde{v}_\inda(r)\) up to one global normalization and phase.

Let us offer an alternative viewpoint to this calculation. Instead of evolving the dressing data \(\eta\), \(\tilde{\eta}\), one may evolve the currents \(J\), \(\widetilde{J}\), subject to the condition that the charges at \(r_0\) are the modes that built the spectrum in \cref{sec: the states}. Since the dressed charges are topological, the two radial evolutions must compensate each other and cancel. We check this explicitly in \cref{app: exact solutions} (detailing also further the derivation), where we happily find the inverse radial evolution to that in \cref{solution for v}.

What remains is to evaluate the regular solution on a Wilson state. Extending \cref{eq:Ba-def} to a bulk integral, as in \cref{sec: acting with zero modes}, gives
\begin{equation}
    \begin{split}
        \cB_\inda &=\int_{\Xi^d_{r_0}} \dd r\w 
        \left( \frac{\g}{\sqrt{2\pi}} \pd_r\qty(v_\inda \phi_\inda)\w *_{\Sigma} J_r+\frac{\sqrt{2\pi}}{\g}\pd_r\qty(\tilde{v}_\inda \tilde{\phi}_\inda)\w *_{\Sigma} \tilde{J}_r\right.\\
        &\hspace{6em}+\left. \frac{\g}{\sqrt{2\pi}}v_\inda \phi_\inda\w \pd_r*_{\Sigma} J_r+\frac{\sqrt{2\pi}}{\g}\tilde{v}_\inda \tilde{\phi}_\inda\w \pd_r*_{\Sigma} \tilde{J}_r\right)~.
    \end{split}
\end{equation}
The coefficients obey \(v_\inda(r_0)=v_\inda\) and \(\tilde{v}_\inda(r_0)=\tilde{v}_\inda\). This can be done consistently only if $v_\inda(r)$ and $\tilde{v}_\inda(r)$ are regular at $r=0$,%
\footnote{The limit \(r=0\) is special because this is where \(\cB\) is shrunken to live on \(S^p\), see \cref{fig: Baction}.}
a condition which we now impose, fixing \(c_K=0\) in \cref{solution for v}. Then, the bulk integral for $\cB_\inda$ can be repackaged as
 \begin{equation}
     \cB_\inda =\int_{\Xi^d_{r_0}}\qty((-1)^{p}\,  \eta \w \dd * J+(-1)^{d+p}\,\tilde{\eta}\w \dd * \tilde{J}\,)~.
 \end{equation}
with the identification \cref{identification v eta} now extended to  $r\in\closed{0}{r_0}$. Using the equation of motion with Wilson defect insertion \cref{eq:EOM-Wilson}, we find
\begin{equation}\label{eq:Ba-Wilson}
    \eval{\cB_{\inda}}_{W_e}=(-1)^{p(d+1)} e\frac{\g}{\sqrt{2\pi}}\int_{\Xi_{r_0}^{d}} \, v_\inda(r) \phi_{\inda +}\w \hat{\gamma}~.
\end{equation}
The eigenforms \(\Phi_{\inda +}\) are built from linear combinations of spherical harmonics. They integrate to zero over $\gamma=\S^p$ for every non-zero mode, and the regularity of \(v_\inda(r)\) prevents it from providing a compensating singular contribution from the origin. Therefore
\begin{equation}\label{W annihilation}
    \cB_{\inda}\ket{W_e}=0\qquad \forall \inda \neq 0 ~.
\end{equation}
Hence, \(\cB_\inda\) are annihilation operators with respect to which the Wilson defect is indeed primary. Note that the fact that the Wilson operator is inserted along $\S^p$ was crucial to obtain \cref{W annihilation}. If we choose instead a different $\gamma$, the Wilson operator insertion does not, in general, produce a primary state.

\subsection{Singular modes create descendants}

The regular solution produced the annihilation operator. Its raising partner is defined at the boundary $\Sigma_{r_0}$ by another linear combination
\begin{equation}\label{B prime}
    \cB_\inda'=\frac{v'_{\inda}+\tilde{v}'_{\inda}}{2}\J_{\inda +}+\frac{v'_{\inda}-\tilde{v}'_{\inda}}{2}\J_{\inda -}~.
\end{equation}
For any such pair, the algebra follows directly from the definitions. Indeed, if
\begin{equation}
    \begin{pmatrix}
        \cB_{\inda}\\
        \cB'_{\inda}
    \end{pmatrix}=M
    \begin{pmatrix}
        \J_{\inda +}\\
        \J_{\inda -}
    \end{pmatrix}~, \qq{with} M=\frac{1}{2}\begin{pmatrix}
        v_{\inda}+\tilde{v}_{\inda} & v_{\inda}-\tilde{v}_{\inda}\\
        v'_{\inda }+\tilde{v}'_{\inda } & v'_{\inda }-\tilde{v}'_{\inda}
    \end{pmatrix}~,
\end{equation}
the mode commutator \cref{mode commutator} gives
\begin{equation}\label{comm relation Bs}
    [\cB_{\inda},\cB'_{\inda}]=\det(M)\,[\J_{\inda +},\J_{\inda -}]=\frac{\ii (-1)^p}{4\pi}\mu_\inda \left( v_\inda\tilde{v}'_\inda  - \tilde{v}_\inda v'_\inda\right)~.
\end{equation}
We now choose \(\cB'_\inda\) so that it is the adjoint creation operator at the boundary.

The regular solution \(\cB_\inda\) still contains one normalization and one phase that we have yet to specify. We fix them by requiring that \(\cB_\inda\) and its boundary adjoint \(\cB^\dagger_\inda\) are related to the creation and annihilation operators of \cref{creation and annih} by a Bogoliubov transformation:
\begin{equation}\label{Bogoliubov}
    \begin{pmatrix}\cB_\inda \\ \cB^\dagger_\inda \end{pmatrix} = \begin{pmatrix}
        \ex{i\theta_\inda} & 0\\
        0 & \ex{-i\theta_\inda}
    \end{pmatrix}\begin{pmatrix} \cosh(\rho_\inda) & \ex{i\psi_\inda} \sinh(\rho_\inda) \\ \ex{-i\psi_\inda} \sinh(\rho_\inda)& \cosh(\rho_\inda)
    \end{pmatrix}\begin{pmatrix}
        \cA_\inda \\ \cA^\dagger_\inda 
    \end{pmatrix}~.
\end{equation}
This condition ensures both that \(\cB_\inda\) and \(\cB_\inda^\dagger\) are adjoints of each other, and that they obey the same algebra as the original set \(\cA_\inda\), \(\cA^\dagger_\inda\). Hence, this lets us organize the defect operators in the same way as the states.

To implement this condition, we write \(\cB_\inda\) in the \(\cA_\inda,\cA^\dagger_\inda\) basis:
\begin{equation}
    \begin{split}
        \cB_{\inda}&=\frac{v_{\inda}+\tilde{v}_{\inda}}{2}\J_{\inda +}+\frac{v_{\inda}-\tilde{v}_{\inda}}{2}\J_{\inda -}\\
        &=\frac{1}{2}\left(\ex{(-1)^p\ii \frac{\pi}{4}}v_{\inda}+\ex{-(-1)^p\ii \frac{\pi}{4}}\tilde{v}_{\inda}\right)\cA_{\inda}+\frac{1}{2}\left(\ex{-(-1)^p\ii \frac{\pi}{4}}v_{\inda}+\ex{(-1)^p\ii \frac{\pi}{4}}\tilde{v}_{\inda}\right)\cA_{\inda}^\dagger~.
    \end{split}
\end{equation}
Thus, requiring
\begin{equation}\label{cosh and sinh}
\begin{split}
    \frac{1}{2}\left(\ex{(-1)^p\ii\frac{\pi}{4}}v_{\inda}+\ex{-(-1)^p\ii\frac{\pi}{4}}\tilde{v}_{\inda}\right)&=\cosh(\rho_\inda)\ex{i\theta_\inda}\\
    \frac{1}{2}\left(\ex{(-1)^p\ii\frac{\pi}{4}}\tilde{v}_{\inda}+\ex{-(-1)^p\ii\frac{\pi}{4}}v_{\inda }\right)&=\sinh(\rho_\inda)\ex{i\theta_\inda+i\psi_\inda}~.
\end{split}
\end{equation}
The normalization condition then implies
\begin{equation}
    \abs{\cosh(\rho_\inda)}^2-\abs{\sinh(\rho_\inda)}^2=\frac{1}{4}\abs{\ex{(-1)^p\ii\frac{\pi}{2}}v_\inda+\tilde{v}_\inda}^2-\frac{1}{4}\abs{v_\inda+\ex{(-1)^p\ii\frac{\pi}{2}}\tilde{v}_\inda}^2=1~.
\end{equation}
This fixes the remaining constant, $c_I$ in \cref{solution for v} as
\begin{equation}
    \abs{c_I}^2=\frac{\mu_\inda r_0^{d-p-3-2\wdeg}}{\cI_\inda(r_0)\cI_\inda'(r_0)}~,
\end{equation}
where the \(\cI_\inda\)'s are the rescaled modified Bessel functions in \cref{modified modified bessels main} evaluated at $r_0$. In fact, $c_I$ is still defined up to a phase. We can use this freedom to remove the phase $\theta_\inda$ from \cref{cosh and sinh},%
\footnote{If we write $c_I = \ex{\ii \beta_\inda} \abs{c_I}$, the phase $\theta_\inda$ is removed by choosing $\beta_\inda = \theta_\inda +(-1)^{p+1}\pi/4$. \label{foot:theta-gone}}
while \(\rho_\inda\) reads
\begin{equation}\label{eq:chosrho}
    \cosh(\rho_\inda)=\frac{1}{2}\qty(\sqrt{\frac{\mu_\inda(r_0)\cI_{\inda}(r_0)}{\cI_{\inda}'(r_0)}}+\sqrt{\frac{\cI_{\inda}'(r_0)}{\mu_\inda(r_0)\cI_{\inda}(r_0)}})\,.
\end{equation}
The regular solution is then fully fixed:
\begin{equation}\label{vs fully specified}
    \begin{pmatrix}
        v_\inda\\
        \tilde{v}_\inda
    \end{pmatrix}= 
    \ex{-(-1)^p\ii\frac{\pi}{4}}\begin{pmatrix}
        1 & \\[0.3em]
        &  i(-1)^{p}
    \end{pmatrix}
    \begin{pmatrix}
        \ex{\rho_\inda} \\
        \ex{-\rho_\inda}
    \end{pmatrix}~,
\end{equation}
and so is its complex conjugate:
\begin{equation}
    \begin{pmatrix}
        v_\inda^*\\
        \tilde{v}_\inda^*
    \end{pmatrix}= \ex{(-1)^p\ii\frac{\pi}{4}}\begin{pmatrix}
        1 & \\
        &  -i(-1)^{p}
    \end{pmatrix}
    \begin{pmatrix}
        \ex{\rho_\inda} \\
        \ex{-\rho_\inda}
    \end{pmatrix}~.
\end{equation}
As a check,
\begin{equation}
    v_\inda\tilde{v}_\inda^*-\tilde{v}_\inda v_\inda^*=-2i(-1)^{p}~,
\end{equation}
which is exactly what is needed for the \(\cB\)'s to have the same algebra as the \(\cA\)'s, as follows by setting \(v'=v^*\) and \(\tilde{v}'=\tilde{v}^*\) in \cref{comm relation Bs}.

This lets us define the creation operator by radial evolving the boundary adjoint. Although \(\cB^{\dagger}_\inda\) is topological, its extension in the bulk is not the adjoint of \(\cB_\inda\), since \(\Xi^d_{r_0}\) is in Euclidean signature. That is, we extend to an operator \(\cB'_\inda(r)\), with
\begin{equation}
    \eval{\cB'_\inda(r)}_{r=r_0}=\cB^{\dagger}_\inda~.
\end{equation}
Using the same radial basis as in \cref{solution for v}, this operator has
\begin{equation}
    \begin{pmatrix}
        v'_\inda(r)\\
        \tilde{v}'_\inda(r)
    \end{pmatrix}=r^{-\frac{d-p-3-2\wdeg}{2}}\begin{pmatrix}
        1 &\\
        & \frac{i(-1)^{p}}{\mu_{\inda}(r)}
    \end{pmatrix} \begin{pmatrix}
        \cI_{\inda}(r) & \cK_{\inda}(r)\\
        \cI'_{\inda}(r) &\cK'_{\inda}(r)
    \end{pmatrix}\begin{pmatrix}
        c'_I\\
        c'_K
    \end{pmatrix}~,
\end{equation}
where
\begin{equation}
\begin{split}
    \begin{pmatrix}
        c'_I\\
        c'_K
    \end{pmatrix}&=r_0^{\frac{d-p-3-2\wdeg}{2}}\begin{pmatrix}
        \cI_{\inda}(r_0) & \cK_{\inda}(r_0)\\
        \cI'_{\inda}(r_0) &\cK'_{\inda}(r_0)
    \end{pmatrix}^{-1}\begin{pmatrix}
        1 &\\
        & -\mu_{\inda}(r_0)i(-1)^{p}
    \end{pmatrix}\begin{pmatrix}
        v_\inda^*\\
        \tilde{v}_\inda^*
    \end{pmatrix}\\
    &=-\ex{(-1)^p\ii\frac{\pi}{4}}\sqrt{\frac{\mu_\inda(r_0)r_0^{-(d-p-3-2\wdeg)}}{\cI_{\inda}(r_0)\cI'_{\inda}(r_0)}}\begin{pmatrix}
        \cI'_{\inda}(r_0)\cK_{\inda}(r_0)+\cK'_{\inda}(r_0)\cI_{\inda}(r_0)\\[0.2em]
        -2\cI_{\inda}(r_0)\cI'_{\inda}(r_0)
    \end{pmatrix}~.
\end{split}
\end{equation}
We observe that
\begin{equation}
    c'_K=2\frac{\mu_\inda(r_0)}{c_I}\neq 0~.
\end{equation}
Thus, \(\cB^\dagger_\inda\) contains the singular radial solution. When it is evolved to \(r=0\), this singularity produces a non-trivial \(p\)-dimensional operator, in line with the discussion after \cref{mode commutator}. In other words, \(\cB^{\dagger}_\inda\) does not annihilate \(\ket{W_e}\), but rather it creates a new state in the spectrum, corresponding to the defect obtained by shrinking \(\cB_\inda'(r)\) onto $W_e$.

\subsection{The commutator}

The path-integral construction should reproduce the algebraic commutator \cref{comm relation Bs}. Since \(\cB_\inda\ket{W_e}=0\) for all non-zero modes, the commutator acting on a Wilson state reduces to
\begin{equation}
    \cB_\inda\cB_\inda'\ket{W_e}~.
\end{equation}
It is therefore sufficient to evaluate \(\cB_\inda\) on-shell in the presence of both \(\cB'_\inda\) and \(\ket{W_e}\).

We first determine how a generic dressed charge changes the equations of motion. Consider inserting
\begin{equation}
    \ex{\cQ_\eta}=\exp(\int_{\Xi^d_{r_0}} \qty(\hat{\Sigma}\w \eta \w *J +\hat{\Sigma}\w \tilde{\eta}\w*\tilde{J}))~, 
\end{equation}
where $\hat{\Sigma}$ is the Poincaré dual of \(\Sigma\) in \(\Xi^d\). The boundary of a magnetic symmetry operator on an open manifold can be interpreted as a Wilson operator, possibly with non-integer charge. Similarly, the boundary of an electric symmetry operator can be interpreted as a 't Hooft operator, again possibly with non-integer charge \cite{Gaiotto:2014kfa}. Thus, inserting \(\ex{\mathcal{Q}_{\eta}}\) has the same effect as inserting the non-genuine operator
\begin{equation}\label{eq:non-genuine-ops}
    W_{e}\qty(\gamma_e)\, H_{m}\qty(\gamma_m)~,
\end{equation}
where $\gamma_e$ and $\gamma_m$ are the Poincaré duals of $\hat{\Sigma}\w \dd{\tilde{\eta}}$ and $\hat{\Sigma}\w \dd{\eta}$, respectively. Their (non-integer) charges are: 
\begin{equation}
    m=\frac{\ii }{2\pi}(-1)^{dp+d+1}~,\quad e=\frac{\ii }{2\pi}(-1)^{dp+1}~.
\end{equation}
Regardless, the conservation laws become
\begin{equation}
    \begin{split}
        d*J&=e\hat{\gamma}_e(-1)^{dp}\,=(-1)^{d+1}\frac{\ii }{2\pi}\hat{\gamma}_e \, \\
         d*\tilde{J}&=m\hat{\gamma}_m(-1)^{p}=(-1)^{p(d+1)+1}\frac{\ii }{2\pi}\hat{\gamma}_m~,
    \end{split}
\end{equation}
such that evaluating \(\cB_\inda\) on shell with this insertion yields
\begin{equation}
    \eval{\cB_\inda}_{\ex{\cQ_\eta}}=\int_{\Xi^d_{r_0}}  \hat{\Sigma}\w\qty( \frac{\g}{\sqrt{2\pi}}v_\inda (-1)^{d+1}\frac{\ii }{2\pi} \phi_{\inda} \w  \dd{\tilde{\eta}} +\frac{\sqrt{2\pi}}{\g}(-1)^{p(d+1)+1} \frac{\ii }{2\pi} \tilde{v}_\inda \tilde{\phi}_{\inda}\w \dd{\eta})~.
\end{equation}
We now specialize to the insertion of \(\ex{\cB'_\inda}\). Comparing \cref{B prime} to \(\cQ_\eta\), we identify
\begin{equation}
    \frac{\sqrt{2\pi}}{\g} \eta_{\Sigma}= v'_\inda\phi_\inda~,\qquad \frac {\g}{\sqrt{2\pi}} \tilde{\eta}_{\Sigma}= \tilde{v}'_\inda\tilde{\phi}_\inda~.
\end{equation}
Therefore,
\begin{equation}
    \begin{split}
        \eval{\cB_\inda}_{\ex{\cB'_\inda}} 
        &=\frac{\ii }{2\pi}(-1)^{dp+1} \int   \hat{\Sigma}\w\qty( (-1)^{dp} v_\inda\tilde{v}'_\inda \, \phi_{\inda}\w  \dd{\tilde{\phi}_{\inda}} +(-1)^{p+d}\, \tilde{v}_\inda v'_\inda\, \tilde{\phi}_{\inda}\w \dd{\phi_{\inda}})\\
&=\frac{\ii \mu_\inda}{4\pi} (-1)^{p}\qty( v_\inda\tilde{v}'_\inda  - \tilde{v}_\inda v'_\inda)~.
    \end{split}
\end{equation}
Equivalently,%
\footnote{From \cref{charged ops}, we see that the operator \(\ex{\cB'_\inda}\) is a non-local operator charged under the symmetry generated by \(\cB_\inda\). Stated differently, the operators charged under the 0-form symmetries constructed above are themselves built from combinations of the dressed charges. Moreover, the operator \(\ex{\alpha\cB'_\inda}\) is gauge invariant for any \(\alpha\in \C\), meaning that the charges of the symmetry generated by \(\cB_\inda\) are not quantized and thus \(\cB_\inda\) does not generate a \(\U(1)\) symmetry.
}
\begin{equation}\label{charged ops}
    \cB_{\inda}\, \ex{ \cB'_{\inda}(r)}\ket{W_e}=\frac{\ii  \mu_{\inda}}{4\pi}(-1)^{p}\qty( v_\inda \tilde{v}'_\inda  - \tilde{v}_\inda v'_\inda)\, \ex{ \cB'_{\inda}(r)} \ket{W_e}~,
\end{equation}
which implies
\begin{equation}
    \cB_{\inda} \cB'_{\inda}(r)\ket{W_e}=\frac{\ii \mu_{\inda}}{4\pi}(-1)^{p}\left( v_\inda\tilde{v}'_\inda  - \tilde{v}_\inda v'_\inda\right)\ket{W_e}~.
\end{equation}
This agrees with \cref{comm relation Bs}. The path-integral computation reproduces the operator algebra.

\subsection{Squeeze 'em}

The operators \(\cB_\inda,\cB^\dagger_\inda\) are a Bogoliubov transformation of the state operators \(\cA_\inda,\cA^\dagger_\inda\). Equivalently, the relation \cref{Bogoliubov} can be implemented by a unitary operator \(\cS_\inda(r_0)\) such that
\begin{equation}\label{squeezing transform}
    \begin{split}
        \cB_\inda &= \cS_\inda (r_0)\cA_\inda \cS^\dagger_\inda (r_0)\\
        \cB^\dagger_\inda &= \cS_\inda(r_0)\cA_\inda^\dagger \cS^\dagger_\inda (r_0)~.
    \end{split}
\end{equation}
The operator \(\cS_\inda (r_0)\) is known as a \emph{squeezing operator} \cite{Caves.31.3068, Bishop1988TwoModeSqueezed},%
\footnote{In principle, one could include an extra factor in \(\cS_\inda(r_0)\) of the form \(\exp(\frac{2\pi}{\mu_\inda}i\theta_\inda \cA_\inda^\dagger \cA_\inda)\). In our case, this factor can be removed without loss of generality, see \cref{foot:theta-gone}.} 
defined by
\begin{equation}
    \cS_\inda (r_0)=\exp[\frac{\pi}{\mu_\inda(r_0)} \qty(\zeta_\inda \qty\big(\cA_\inda^\dagger)^2 -\zeta_\inda^* \qty\big(\cA_\inda)^2)]~,
\end{equation}
where
\begin{equation}
    \zeta_{\inda}=\rho_{\inda}\ex{\ii \psi_{\inda}}~.
\end{equation}
Matching the Bogoliubov transformation with the radial solution gives
\begin{equation}
    \psi_{\inda}=0~,\quad \zeta_{\inda}=\frac{1}{2}\abs{\log(\frac{\mu_{\inda}(r_0)\cI_{\inda}(r_0)}{\cI'_{\inda}(r_0)})}~.
\end{equation}
Thus,
\begin{equation}
    \begin{split}
    \cS_\inda (r_0)&=\exp[\frac{\pi\, \zeta_\inda}{\mu_\inda(r_0)}\qty( \qty\big(\cA_\inda^\dagger)^2 -\qty(\cA_\inda)^2)]\\
    &=\exp[\frac{\ii (-1)^p\,\pi\, \zeta_\inda}{\mu_\inda(r_0)}\qty(\J_{\inda+}\J_{\inda-}+\J_{\inda-}\J_{\inda+})]~.
\end{split}
\end{equation}
Despite being defined on a codimension-1 manifold, this operator is equivalent to an operator inserted on $S^p$ at $r=0$. Indeed, as for $\cB_\inda$ and $\cB_\inda^\dagger$, this operator is built from $\J_{\inda\pm}$ operators for which  the radial evolution can be computed exactly (see \cref{app: exact solutions}). It is then possible to radially evolve this operator to shrink it onto $S^p$, where it will define a $p$-dimensional operator due to its singularities at $r=0$.
The extended operators of the theory are organized by creation and annihilation operators \(\cB_\inda\), \(\cB_\inda^\dagger\), into squeezed versions of the modes that organize the states.

As a sanity check, the above construction reduces to the usual 2d CFT state-operator correspondence. In 2d CFTs, necessarily with \(p=0\), the operators acting on local insertions in the Euclidean path integral are simply \(\cA_\inda,\cA^\dagger_\inda\). They can be radially evolved inward without any squeezing. Indeed, when \(p=0\), there is no \(S^p\) sphere, so \(\gamma_\inda=0\) (recall, $\gamma_\inda$ is the eigenvalue of the Laplacian on $\S^p$, cf. \cref{mu and r}). In this case
\begin{equation}\label{rho gamma=0}
    \rho_\inda = \frac{1}{2}\ln\left(\frac{d-p-2-2\wdeg}{2\sqrt{\nu_\inda}}+\sqrt{\left(\frac{d-p-2-2\wdeg}{2\sqrt{\nu_\inda}}\right)^2+1}\right)~,
\end{equation}
which is independent of \(r_0\). Imposing further \(d=2\), gives \(\wdeg=0\) and \(\pp=1\), hence \(\rho_\inda=0\). Therefore \(\cS_\inda=1\), and
\begin{equation}
    \eval{\cB_{\inda}}_{\substack{d=2 \\ p=0}} = \cA_\inda~,\qquad \eval{\cB^\dagger_{\inda}}_{\substack{d=2 \\ p=0}} = \cA^\dagger_\inda ~.
\end{equation}
In all other cases, including the conformal \(d=2p+2\), the squeezing persists.

The squeezing operator built above has the effect of squeezing a single mode, labeled by $\inda$. For future convenience, it will turn out useful to define a (decoupled) \emph{all-mode squeezing operator}, as 
\begin{equation}
    \cS(r_0) = \prod_{\inda\neq 0} \cS_\inda(r_0)~.
\end{equation}
As ladder operators with different quantum numbers commute, $\cS(r_0)$ has the effect of squeezing all oscillators simultaneously, namely:
\begin{equation}
    \cB_\inda = \cS(r_0)\, \cA_\inda\, \cS^\dagger(r_0), \qquad \t{for all}\ \inda~,
\end{equation}
and similarly for their adjoints.

\section{The state-defect correspondence}\label{sec: the correspondence}

Everything so far has been building up the pieces to define a state-defect correspondence for $p$-form Maxwell theories in general dimensions. Now, we can collect these pieces and fit them together to get the full picture, explicitly mapping out the relation between operators and states.

\subsection{Preparing the ground state}

Let us start off by showing the relation between the ground state $\ket{0}$ and the state prepared by the empty path integral. At finite volume, the ground state is unique. With respect to the local Hamiltonian \cref{hamiltonian}, this is the state, $\ket{0}_{\cA}$, satisfying
\begin{equation}
    \cA_\inda \ket{0}_{\cA}=0~ \quad \forall \inda~, \qq{and}\quad j_0\ket{0}_{\cA} = 0~.
\end{equation}
We have added a subscript to keep track of the set of oscillators that annihilate the state.

Interestingly, the above ground state does not exactly correspond to the empty path integral. Indeed, taking $e=0$ in the Wilson-operator state \cref{eq:Wilson-state} described above corresponds to the identity operator, and hence the empty path integral. As shown there, this state is annihilated by the $\cB$-set of oscillators. With the intuition about squeezing operators built above, this state corresponds to an \emph{all-mode squeezed vacuum} \cite{Hofman:2024oze,Vitouladitis:2025zoy},
\begin{equation}\label{eq:squeezed-vac}
    \ket{0}_{\cB} = \ket{\1} = \int \DD{a}\, \ex{-S[a]}\mathcal \,=\, \cS \ket{0}_{\cA}~.
\end{equation}
One can immediately see that this state has exactly the properties of $\ket{W_{e=0}}=\ket{\1}$. In particular:
\begin{equation}
    \cB_\inda \ket{0}_\cB = \cB_\inda\, \cS \ket{0}_{\cA} = \cS \cA_\inda \ket{0}_{\cA} = 0~, \qquad \t{for all}\ \inda~,
\end{equation}
as well as,
\begin{equation}
    j_0 \ket{0}_\cB = 0~,
\end{equation}
since $j_0$ commutes with all the oscillators.

Dually, the lowest energy state is prepared by inserting the squeezing operator itself into the path integral, and shrinking it onto $\S^p$. Inverting the above relation, one gets,
\begin{equation}\label{eq:0A-PI}
    \ket{0}_{\cA} =  \int \DD{a}\, \ex{-S[a]} \cS^\dagger~.
\end{equation}
Recall that the squeezing operator (and its adjoint) is made out of the modes $\cA_\inda$, or equivalently, out of the modes of the dressed charges. As such, it too can be used both as a codimension-1 operator acting on the Hilbert space, as in eq. \cref{eq:squeezed-vac}, and as a $p$-dimensional defect, upon collapsing it to its locus of singularities. The latter viewpoint is the one used in \cref{eq:0A-PI}. Explicitly, $\cS^\dagger$ takes the form 
\begin{equation}
    \cS^\dagger = \exp(-\sum_{\inda\neq 0} \frac{\ii (-1)^p\pi\zeta_\inda}{\mu_\inda(r_0)} \int_{\Sigma_r(x)}\int_{\Sigma_r(y)} \qty\big[f_{\inda+}(r,x)f_{\inda-}(r,y)+f_{\inda-}(r,x)f_{\inda+}(r,y)])
\end{equation}
with
\begin{equation}
    f_{\inda\pm}(r,x) = \frac{\g}{\sqrt{2\pi}} \phi_{\inda\pm}(r,x)\w *_{\Sigma_r} J_r(r,x) + \frac{\sqrt{2\pi}}{\g} \tilde{\phi}_{\inda\pm}(r,x) \w *_{\Sigma_r} \tilde{J}_r(r,x)~.
\end{equation}
In the limit $r\to 0$, it defines a genuine $p$-dimensional defect.

With this, we find that indeed,
\begin{equation}
    \cA_\inda \ket{0}_{\cA} = 0~, \qquad \forall\ \inda~,
\end{equation}
while, of course, we still have $j_0\ket{0}_{\cA} = 0$. Hence, the state prepared by \cref{eq:0A-PI} is the ground state of the Hamiltonian \cref{hamiltonian}.
We conclude that for the ground state, the correspondence is 
\begin{equation}
    \ket{0}_{\cA} \leftrightsquigarrow \cS^\dagger~.
\end{equation}
This result, while seemingly exotic, agrees with the existing literature. Indeed, in \cite{Belin2018} the authors attempt to define a state-operator correspondence for a 3d CFT with the spatial slice $T^2$, and find that the empty path integral does not correspond to the vacuum state for $\Sigma=S^1\times S^1$. Additionally, the result here reduces to and matches what was found in the preceding works \cite{Hofman:2018lfz,Vitouladitis:2025zoy}, which defined the state-operator correspondence for $d=4~,p=1$ and general $d,\,p=0$ respectively. 

\subsection{Building the Hilbert space}

Having understood the preparation of the ground state, building the rest of the Hilbert space is straightforward. To summarize, in \cref{sec: the states} we found that, when $d\neq 2p+2$, states are graded by their electric charges. Primary states $\ket{e}$ are those that are annihilated by all lowering operators,
\begin{equation}
    \cA_\inda \ket{e} = 0 \qquad \forall \inda\neq 0~,
\end{equation}
and acting on them with zero modes returns their electric charge,
\begin{equation}
    j_0 \ket{e} = \frac{\g}{\sqrt{2}\pi} \sqrt{\frac{\vol(S^p)}{\vol(S^{\pp}_{r_0})}}\, e \ket{e} ~.
\end{equation}
On the other hand, the states prepared by insertions of Wilson surfaces in the path integral are killed by the annihilation operator $\cB_\inda$,
\begin{equation}
    \cB_\inda \ket{W_e} = 0 \qquad \forall \inda \neq 0 ~,
\end{equation}
while the action of the zero mode $j_0$ also yields their electric charge 
\begin{equation}
    j_0 \ket{W_e} = \frac{\g}{\sqrt{2\pi}}\sqrt{\frac{\vol(S^p)}{\vol(S^{\pp}_{r_0})}}\, e\ket{W_e}~.
\end{equation}
This fact lends itself to the interpretation of Wilson surfaces as primary operators with respect to (a different representation of) the same Kac--Moody algebra.

The squeezing relation between the two sets of oscillators implies that
\begin{equation}
    \ket{W_e} = \cS \ket{e}~.
\end{equation}
In other words, primary $p$-dimensional operators prepare all-mode squeezed primary states, or dually, primary states correspond to Wilson surfaces dressed with squeezing operators:
\begin{equation}
    \ket{e} \leftrightsquigarrow \cS^\dagger\times W_e~.
\end{equation}

It is time to bring back the excluded case $d=2p+2$. Primary states are then characterized by both an electric and a magnetic flux: $\ket{e,m}$.%
\footnote{Adding a $\theta$-angle, as in \cref{eq:theta-term}, further shifts the electric zero-mode. The electric current is now $J+\frac{\theta}{2\pi} \tilde{J}$, while the magnetic current is unaffected. Correspondingly, the primary states have charges:
\[
    j_0 \ket{e,m} = \frac{\g}{\sqrt{2\pi}} \sqrt{\frac{\vol(\S^p)}{\vol(\S^{\pp}_{r_0})}}\qty(e+\frac{\theta}{2\pi}m)\ket{e,m}~, \qquad
    \tilde{\jmath}_0 \ket{e,m} = \frac{\sqrt{2\pi}}{\g}\sqrt{\frac{\vol(\S^p)}{\vol(\S^{\pp}_{r_0})}}\,m\ket{e,m}~.
\]}
Using the arguments presented above, together with those of \cite{Hofman:2018lfz}, it follows straightforwardly that the primary states $\ket{e,m}$ correspond to squeezed dyonic defects \cref{dyonic line}:
\begin{equation}
    \ket{e,m} \leftrightsquigarrow \mathcal{S}^\dagger\times V_{e,m}~.
\end{equation}

To complete the correspondence, all that remains is to relate the respective descendants. For the energy eigenstates, we can build descendants by acting with raising operators,
\begin{equation}
    \ket{N_{\inda}, e,m} = \prod_{\inda} \qty(\cA^{\dagger}_{\inda})^{N_{\inda}}\ket{e,m}~.
\end{equation}
We argued in \cref{sec: the ops} that smearing combinations of the currents---corresponding to singular solutions of the radial evolution equation---over the primary defects results in new states. A sensible combination of the currents that acts non-trivially is precisely the action of the creation operators $\cB^\dagger_\inda$,
\begin{equation}
    \qty(\cB^\dagger_\inda)^{N_\inda} \ket{V_{e,m}}~.
\end{equation}
Again, from the relation between the $A$ and $\cB$ oscillators, this allows us to identify,
\begin{equation}
    \qty(\cB^\dagger_\inda)^{N_\inda} \ket{V_{e,m}} = \cS\, \qty(\cA^\dagger_\inda)^{N_\inda} \ket{e,m}~,
\end{equation}
or equivalently
\begin{equation}
    \ket{N_\inda,e,m} \leftrightsquigarrow \cS^\dagger \times \prod_\inda  \qty(\cB^\dagger_\inda)^{N_\inda} \times V_{e,m}~.
\end{equation}
That is, we have identified extended operators smeared with conserved currents with squeezed descendant states, and dually, excited energy eigenstates with squeezed descendant defects. Our correspondence is summarized in \cref{table:summary}.

\begin{table}
\centering
\renewcommand{\arraystretch}{1.35}
\begin{tabular}{@{}|p{3cm}p{4.0cm}p{4cm}|@{}}
    \hline
    Type & State in \(\cH_{\S^p\times\S^m}\) & Defect insertion on \(\S^p\) \\
    \hline
    Ground state
        & \(\ket{0}\)
        & \(\cS^\dagger\) \\[0.7em]
    Squeezed vacuum
        & \(\ket{0}_\cB\)
        & \(1\) \\[0.7em]
    Primary
        & \(\ket{e,m}\)
        & \(\cS^\dagger \times V_{e,m}\) \\[0.5em]
    Descendant
        & \(\prod_{\inda}\qty(\cA^\dagger_\inda)^{N_\inda}\ket{e,m}\)
        & \(\cS^\dagger \times \prod_{\inda}\qty(\cB^\dagger_\inda)^{N_\inda}\times V_{e,m}\) \\[0.5em]
    \hline
\end{tabular}
\caption{A summary of the one-to-one correspondence between squeezed states (in the left column) and extended operator insertions (in the right column). The Kac--Moody algebra naturally organizes the states and defects respectively into ground states, primaries, and descendants. In dimensions \(d\neq 2p+2\), only the electric charge is present, and the magnetic labels should be omitted, with \(V_e=W_e\).}
\label{table:summary}
\end{table}

\section{Conclusions and future work}\label{sec:conclusions}

In this paper, we have studied $p$-form gauge theories with continuous symmetries and a mixed anomaly, and we have established that it is precisely this symmetry structure that allows for the state-defect correspondence in this (in general) non-conformal case. In fact, we saw that the existence of such a mixed anomaly implies that there exists an infinite family of conserved charges, $\cQ_\eta$, with an extended Abelian Kac--Moody-like algebra,
\begin{equation}
    [\cQ_\eta,\cQ_\zeta]=\frac{-\ii}{2\pi}\int_{\Sigma}\qty(\eta\w \dd\tilde{\zeta}- \zeta \w \dd\tilde{\eta})~.
\end{equation}
This set of charges is found to act non-trivially on both the states and the extended operators of the theory, allowing us to organize the spectrum into representations of this algebra and facilitating the identification between defects and states. One technical detail that allowed us not only to formulate a state-defect correspondence, but also to match individual states and operators, was that we were able to explicitly solve for the radial evolution of the aforementioned charges (or a combination thereof). 

In this paper, we focused mainly on the free case, while also commenting on the construction of dressed charges in a certain class of interacting theories that do not disrupt the overall symmetry structure of the theory. The success of this construction in extending the state-operator correspondence past the confines of conformal invariance (alongside \cite{Vitouladitis:2025zoy,Chen:2025ujx}) forces us to re-evaluate what underpins this correspondence. It seems probable that, given any QFT, a (typically infinite-dimensional) spectrum-generating algebra of charges would be sufficient.%
\footnote{Of course, it is by no means necessary.}
Such a statement is reminiscent of one of the defining features of classical two-dimensional integrability, namely the existence of infinitely many charges in involution. Here, in the free case, we saw that the spectrum-generating algebra implies the infinite tower of higher-spin charges. It would be interesting to see if the above observations have a deeper connection. Given this, it seems imperative to test the idea of a state-defect correspondence in other classes of theories with spectrum-generating algebras. In particular, there are two directions that we find particularly interesting for further study. 

First, it would be interesting to find the explicit map between states and (extended) operators in gapped theories---all of the examples provided here were gapless. A paradigmatic example would be to study three-dimensional Maxwell--Chern--Simons (MCS) theory, which has gained interest as a topologically massive gauge theory containing one massive propagating vector field with mass proportional to (the absolute value of) the Chern--Simons (CS) level, \(k\),
\begin{equation}
    S_\t{MCS}=\int_X \qty(\frac{1}{2\g^2} f\w* f + \frac{ik}{4\pi} a\w f)~.
\end{equation}
The addition of a Chern--Simons term serves to break the U(1) electric symmetry to a discrete $\bZ_k$ symmetry. This makes the construction of dressed charges more complicated due to now only having one continuous, conserved current, but it is still possible. The algebra in this case becomes:
\begin{equation}
    \comm{\cQ_{\eta}}{\cQ_{\zeta}}=\frac{-\ii}{2\pi }\int_{\Sigma} \left(\eta\w \sd\tilde{\zeta}  -\zeta \w \sd \tilde{\eta}\right)+\frac{\ii k}{2\pi}\int_{\Sigma} \eta\w \zeta ~.
\end{equation}
The first term is the same as in the pure Maxwell theory, while the second is new arises from the Chern--Simons coupling. Upon mode decomposition, this algebra implies that the zero modes of $Q_\eta$ will not commute among themselves. Relatedly, the Hilbert space of the theory has a more interesting spectrum of low-lying states. The corresponding map between states and operators remains a bit more subtle for Maxwell--Chern--Simons and other topologically massive theories. Apart from exemplifying the state-defect correspondence in gapped theories, this example is important for another reason. It lives in three dimensions and has $p=1$, meaning it is one of the cases that our construction was unable to reach ($d=2p+1$). The Chern--Simons coupling seems to be mandatory to obtain the correspondence. Presumably, the reason is that, viewed as an EFT of the magnetic conserved current, the CS term is more relevant than the usual Maxwell term, and so it must be included.%
\footnote{Of course, in the deep IR the CS term dominates, the theory is topological, and the state-defect correspondence is straightforward. The interesting observation in this case, is that it persists when the Maxwell term is also present.}
 This case, along with other instances of gapped systems, will be discussed in forthcoming work \cite{APV-MCS}.

Additionally, the construction of the dressed charges $\cQ_\eta$ here was carried out explicitly only for continuous \emph{invertible} symmetries. Non-invertible constructions of the dressed charges are certainly possible (for a construction in the conformal case see \cite{Hofman:2024oze}), starting from a variant of generalized Maxwell theory with gauged charge-conjugation. It would be interesting to develop the representation theory of such non-invertible Kac--Moody algebras and study its effects in detail on the spectrum of the theory---both on states and on operators. Relatedly, it would be interesting to extend the question of state-operator correspondence to generic symmetry breaking phases of
more general symmetries, including non-Abelian, and 2-group symmetries.

Apart from possible extensions of our construction, one may ask what it is good for. A first concrete application is entanglement. In CFTs, the state-operator correspondence aids entanglement computations, by replacing the density matrix with a path integral on a manifold with a cut (and a choice of boundary conditions to factorize the Hilbert space \cite{OhmoriTachikawa2015}) and, in the case of excited states, local operator insertions. The computation of Rényi entropies is then reduced to the computation of a correlation function of local operators. In gauge theories the situation is much more restrictive. Besides known issues with defining entanglement entropies unambiguously in the presence of edge modes \cite{Casini:2013rba,Donnelly:2011hn,Radicevic:2014kqa,Donnelly:2014gva}, there are comparatively few available computations \cite{Buividovich:2008gq,Agon:2013iva,Donnelly:2015hxa,Radicevic:2015sza,Soni:2015yga,Benini:2025hbj}, and even fewer away from the ground state \cite{Caputa:2014vaa,Ebner:2024mee}. Our construction provides the necessary tool for such calculations, and we expect it to have immediate application there. We note that the computation of entanglement away from the ground state is useful in computing refined entanglement quantities, such as symmetry-resolved entanglement \cite{Goldstein:2017bua,Kusuki:2023bsp,Capizzi:2022sreI,Capizzi:2022sreII}, and entanglement asymmetry \cite{Ares:2022qrd}.

Another direction suggested by our construction points to the fusion of defects and the operator product expansion (OPE). One of the standard arguments in CFT establishes the convergence of the OPE, as a consequence of the state-operator correspondence. See \cite{Polchinski:1998rq,Pappadopulo:2012jk}. As a result, one can reduce the calculation of an \(n\)-point function to a sum of \((n-1)\)-point functions:
\begin{align}
    \ev{\cO_1(x)\cO_2(y)\prod_{i}\cO_i(x_i)} = \sum_{k} C_{12}^{k} P(x-y,\pd_y) \ev{\cO_k(y)\prod_{i}\cO_i(x_i)}~.
\end{align}
Here \(\cO_k\) are conformal primary local operators, the function \(P(x-y,\pd_y)\) sprinkles descendants, and \(C_{12}^{k}\) are the ``OPE coefficients.'' Crucially, the state-operator correspondence guarantees that the above sum, i.e. the replacement 
\begin{equation}
    \cO_1(x)\cO_2(y) \sim \sum_{k} C_{12}^{k} P(x-y,\pd_y) \cO_k(y)~,
\end{equation}
is convergent so long as \(\abs{x-y}<\min_{i}\abs{x_i-y}\). By the same token, this convergence is presumably true for the non-conformal state-operator correspondence of \cite{Vitouladitis:2025zoy} (corresponding to the case \(p=0\) here). This is compatible with the broader viewpoint that the OPE is not intrinsically conformal, advocated in \cite{Hollands:2006ag,Hollands:2011gf,Hollands:2023txn}. More interesting is the case of extended operators. Our state-defect correspondence suggests that a similar, convergent fusion of defects may be possible, at least in the case of correlation functions of perfectly parallel defects. The argument follows straightforwardly from the local CFT argument \cite{Pappadopulo:2012jk}. For defects inserted at non-parallel loci, it is still plausible to engineer an analogous bound, though it would necessarily depend on the geometry of the defects' worldvolumes. The fusion of defects has been recently studied in CFTs in \cite{Soderberg:2021kne,Rodriguez-Gomez:2022gbz,SoderbergRousu:2023zyj,Diatlyk:2024zkk,Diatlyk:2024qpr,Kravchuk:2024qoh}. In light of our result, we speculate that a defect OPE with a non-zero radius of convergence may exist in the non-conformal setting too, at least in theories with a similar spectrum-generating algebra as ours.

\paragraph{Acknowledgements.} It is a pleasure to acknowledge stimulating discussions with Riccardo Argurio and Alejandro Vilar López. We especially thank Riccardo Argurio for comments on an earlier draft. E.P. and S.V. also thank the Galileo Galilei Institute for Theoretical Physics for hospitality during the workshop ``Defects and Extended Excitations in Quantum Field Theory, Quantum Matter and Statistical Models,''  where this work was completed.

\appendix
\crefalias{section}{appendix}
\crefalias{subsection}{appendix}
\addtocontents{toc}{\protect\setcounter{tocdepth}{2}}

\section{The transversal Laplacian}\label{app: transversal laplacian}

In this appendix, we examine the spectrum of the transversal Laplacian on \(\Sigma_{d-1} = \S^p\times\S^{\pp}_{r}\), with the product metric
\begin{align} \label{metric}
	\dd{s_\Sigma^2} = \dd{\Omega_p^2} + r^2\dd{\Omega_{\pp}^2}.
\end{align}
We take the radius of \(\S^p\) to be unity and that of \(\S^{\pp}\) to be \(r\), in accordance with the main text. (In reality, \(r\) is the dimensionless ratio between the radii of the two spheres; this is of course the only parameter physical quantities can depend on). We are concerned with the spectrum of \(\lapl_k\) on \(\Omega_\perp^k(\Sigma)\), the set of transversal (i.e., co-closed) $k$-forms on $\Sigma$. The reason for this restriction to transversal forms is motivated by the fact that the conserved currents in \cref{electric current} and \cref{magnetic current} are co-closed \cref{coclosed currents}. Moreover, the definition of the transversal Laplacian serves as motivation for the definition of the generalized curl operator, which we describe in detail in the following appendix. We are mostly interested in the values \(k=p\) and \(k=d-p-2\), but often we aim at keeping \(k\) in its full generality.

Before we jump in, we note a property of the Hodge star operator that will come in handy at several points later on. That is, its action splits when acting on product forms. Namely, if \(\f{\omega}{\ell}\in \Omega^\ell\qty(\S^p)\) and \(\f{\eta}{k}\in \Omega^{k}\qty(\S^{\pp})\), then
\begin{align}
	*_{\Sigma} \qty(\f{\omega}{\ell} \w \f{\eta}{k}) = (-1)^{k(p-\ell)} *_p \f{\omega}{\ell} \w *_{\pp}\, \f{\eta}{k}~.
\end{align}

\subsection{\tps{k}{k}-form spherical harmonics}
We begin by describing the spectrum of eigenvalues and eigenforms of \(\lapl_k\) on a single \(N\)-sphere, \(\S^N\). In other words, we are seeking a complete basis of eigenforms \(\Upsilon^{N}_{[k],\sfn}\in \Omega^k\qty(\S^N)\) such that
\begin{align}\label{eq:eigenproblem}
	\cdd\dd\, \Upsilon^{N}_{[k],\sfn} = \Lambda^{[k],N}_\sfn\, \Upsilon^{N}_{[k],\sfn}, \qq{with} \cdd \Upsilon^{N}_{[k],\sfn} = 0.
\end{align}
In the above, \(\sfn\) is a list of quantum numbers labeling \(\f{\Upsilon}{k}\), that is \(\sfn = (\ell_1,\ell_2,\cdots,\ell_N)\) where \(\ell_1\geq \ell_2 \geq \cdots \geq \ell_{N-1}\geq \abs{\ell_N}\). The quantum number \(\ell_1\) takes values in \(\Z_{\geq 1}\) if \(k>0\), and in \(\Z_{\geq 0}\) when \(k=0\).
The associated eigenvalues, \(\Lambda^{[k],N}_\sfn\), are
\begin{align}\label{eq:eigenvalue-sphere}
	\Lambda^{[k],N}_\sfn = (\ell_1+k)(\ell_1+N-k-1),
\end{align}
and they have the degeneracy \cite{Camporesi1994}
\begin{align}
	\operatorname{deg}\!\qty(\Lambda^{[k],N}_\sfn) = \frac{(2\ell_1+N-1)\fact{(\ell_1+N-1)}}{\fact{k}\fact{(N-k-1)}\fact{(\ell_1-1)}(\ell_1+k)(\ell_1+N-k-1)}~.
\end{align}

The eigenforms \(\Upsilon^{N}_{[k],\sfn}\) that solve the eigenvalue problem in \cref{eq:eigenproblem}, are known as \textit{\(k\)-form spherical harmonics} on \(\S^N\). They have been constructed in terms of harmonic polynomial forms in \cite{Ikeda1978}, and more explicitly in \cite{Camporesi1994}, in terms of lower-form spherical harmonics on lower-dimensional spheres, in a recursive manner. We review here some of the salient features of the construction presented in \cite{Camporesi1994}. 

First, we write the metric of \(\S^N\) as
\begin{align}
	\dd{s_{\S^N}^2} = \dd{\theta^2} + \sin^2\!{\theta}\, \dd{s_{\S^{N-1}}^2},
\end{align}
recursively, in terms of the line element of \(\S^{N-1}\). The eigenforms \(\Upsilon^{N}_{[k],\sfn}\) come in two families. The first possibility is
\begin{align}
	\Upsilon^{N}_{[k],\ell_1\ell_2\cdots} = (\sin\theta)^{k-2}\qty(f_{\ell_1\ell_2}(\theta) \dd{\theta}\w \Upsilon^{N-1}_{[k-1],\ell_2\cdots} + \sin^2\theta\; g_{\ell_1\ell_2}(\theta) \dd{\Upsilon^{N-1}_{[k-1],\ell_2\cdots}}),
\end{align}
where \(\Upsilon^{N-1}_{[k-1],\ell_2\cdots}\) are also eigenforms of the transversal Laplacian on $S^{N-1}$, and they satisfy:
\begin{align}
	\cdd\dd\, \Upsilon^{N-1}_{[k-1],\ell_2\cdots} = (\ell_2+k-1)(\ell_2+N-k-1)\, \Upsilon^{N-1}_{[k-1],\ell_2\cdots} \qq{and} \cdd \Upsilon^{N-1}_{[k-1],\ell_2\cdots} = 0~.
\end{align}
The functions \(f_{\ell_1\ell_2}(\theta)\) and \(g_{\ell_1\ell_2}(\theta)\) are given as:
\begin{align}
	f_{\ell_1\ell_2}(\theta) = (\sin\theta)^{\ell_2}\; {}_2F_1\qty(\ell_1+N+\ell_2-1,\ell_2-\ell_1;\ell_2+\frac{1}{2}N;\sin(\frac{\theta}{2})^2)~,
\end{align}
in terms of the \({}_2F_1\) hypergeometric function, and
\begin{align}
	g_{\ell_1\ell_2}(\theta) & = \frac{1}{(\ell_2+k-1)(\ell_2+N-k-1)}\qty{\dv{\theta} + (N-k-1)\cot\theta}f_{\ell_1\ell_2}(\theta)~.
\end{align}
The second family are the eigenforms constructed recursively as
\begin{align}
	\Upsilon^{N}_{[k],\ell_1\ell_2\cdots} = (\sin\theta)^k f_{\ell_1\ell_2}(\theta) \Upsilon^{N-1}_{[k],\ell_2\cdots}~,
\end{align}
where \(\Upsilon^{N-1}_{[k],\ell_2\cdots}\) satisfy
\begin{align}
	\cdd\dd\, \Upsilon^{N-1}_{[k],\ell_2\cdots} = (\ell_2+k)(\ell_2+N-k-2)\, \Upsilon^{N-1}_{[k],\ell_2\cdots}, \qq{and} \cdd \Upsilon^{N-1}_{[k],\ell_2\cdots} = 0~.
\end{align}

The \(k\)-form spherical harmonics can be made orthonormal with the normalization factor:
\begin{align}
	\cN^N_{[k],\ell_1\ell_2} = \qty(\frac{(\ell_2+k-1)(\ell_2+N-k-1)}{k(\ell_1+k)(\ell_1+N-k-1)})^{\frac{1}{2}}\frac{\qty(\frac{2\ell_1+N-1}{2}\frac{\fact{(\ell_1+\ell_2+N-2)}}{\fact{(\ell_1-\ell_2)}})^{\frac{1}{2}}}{2^{\ell_2+(N-2)/2} \Gamma\qty(\ell_2+\frac{1}{2}N)}~.
\end{align}
Explicitly, the harmonics $\hat{\Upsilon}^N_{[k],\sfn} = \cN^N_{[k],\sfn} \Upsilon^N_{[k],\sfn}$, are orthonormalized.
In what follows, we assume that \(\Upsilon^{N}_{[k],\sfn}\) already includes this prefactor (dropping the hat), such that
\begin{align}\label{laplace equation}
	\ip{\Upsilon^{N}_{[k],\sfn}}{\Upsilon^{N}_{[k],\sfm}} = \delta_{\sfn\sfm}~.
\end{align}
Since it will be important for the rest of the discussion, we point out that when \(\S^N\) has radius \(r\), the appropriately normalized spherical harmonics are further multiplied by a factor of \(r^{-N/2}\), and the eigenvalues are
\begin{align}
	\Lambda^{[k],N}_\ell = \frac{(\ell+k)(\ell+N-k-1)}{r^2}~,
\end{align}
again with \(\ell\in\Z_{\geq 1}\) for \(k>1\) and \(\ell\in\Z_{\geq 0}\) for \(k=0\).

\subsection{The spectrum on \tps{\S^p\times\S^{\pp}}{Sp x Sm}}\label{sec: spherical harmonics}

Now that we have established how to construct the $k$-form eigenforms of the transversal Laplacian on a single sphere, we turn to the product manifold of interest $\Sigma_r = S^p\times S^{\pp}_r$. That is, we are seeking all \(\Psi_{[k],\sfn}\in \Omega^k\qty(\Sigma_r)\),  such that
\begin{align}\label{eq:EPS}
	\cdd\dd\, \Psi_{[k],\sfn} = \lambda^{(k)}_\sfn\, \Psi_{[k],\sfn}, \qq{with} \cdd \Psi_{[k],\sfn} = 0~.
\end{align}
Again, we are primarily interested in the cases where \(k=p\) and \(k=d-p-2\).

\subsubsection{Zero modes}\label{laplacian zero modes}

Let us begin with the zero modes. These are given by the harmonic \(k\)-forms on \(\Sigma_r\). The existence of such harmonic forms on $\Sigma_r$ is dependent on the topology of the manifold, and there are two distinct cases, classified by their Betti numbers:
\begin{itemize}
	\item[] \textbf{\(d\neq 2p+1\):} In this case the Betti numbers of \(\Sigma_r\) are
	      \begin{align}
		      b_k(\Sigma) = \begin{cases} 1, & k = 0,\, p,\, \pp,\, d-1 \\ 0 & \t{otherwise}.
		                    \end{cases}
	      \end{align}
	\item[] \textbf{\(d=2p+1\):} Here \(d-p-1=p\) such that the Betti numbers of \(\Sigma_r\) are
	      \begin{align}
		      b_k(\Sigma) = \begin{cases} 1, & k = 0,\,d-1 \\ 2 & k = p \\ 0 & \t{otherwise}.\end{cases}
	      \end{align}
\end{itemize}

\subsubsection{Non-zero modes}\label{sec: laplacian nonzero}

Moving on to the non-zero eigenmodes, we start by finding eigen-\(p\)-forms. There is a first, obvious family of solutions to the eigenvalue equation in \cref{eq:EPS}. These are of the form
\begin{align}\label{eq:Psi-p-obvious}
	\phi^{(\uparrow,\wdeg)}_{\sfn}(r) = \cY_{[\wdeg],\sfn_1}\w \cW_{[p-\wdeg],\sfn_2}(r), \qquad \sfn = (\sfn_1,\sfn_2)~,
\end{align}
where we have denoted for notational clarity \(\cY_{[\wdeg],\sfn_1} = \Upsilon_{[\wdeg],\sfn_1}^{p}\), the spherical harmonics on the \(p\)-sphere, and similarly, \(\cW_{[p-\wdeg],\sfn_2} = \Upsilon_{[p-\wdeg],\sfn_2}^{\pp}\) as those on the \(\pp\)-sphere. We go out of our way to indicate the radial dependence of \(\cW\) since \(\S^{\pp}\) is taken to have radius \(r\). The eigenvalues of $\phi^{(\uparrow,\wdeg)}_{\sfn}(r)$ are
\begin{align}\label{eq:eigenvalue-product}
	\lambda_\sfn^{(p)}(r) = \gamma^{(\wdeg)}_{\sfn_1} + \frac{\nu^{(p-\wdeg)}_{\sfn_2}}{r^2}~.
\end{align}
In another attempt for notational clarity, we have distinguished the eigenvalues on the respective unit spheres as \(\gamma^{(\wdeg)}_{\sfn_1} = \Lambda^{(\wdeg),p}_{\sfn_1}\) and \(\nu^{(p-\wdeg)}_{\sfn_2} = \Lambda^{(p-\wdeg),\pp}_{\sfn_2}\). The above eigenform solutions exist within the range \(\max(0,2p+2-d)\leq \wdeg\leq p-1\).

For \((d-p-2)\)-forms we can define the eigenforms as:
\begin{align}\label{eigenform fam 1}
	\widetilde{\phi}^{(\downarrow,\wdeg)}_{\sfn}(r) = \cY_{[\wdeg],\sfn_1}\w \cW_{[d-p-2-\wdeg],\sfn_2}(r)~,
\end{align}
valid within the range \(0\leq \wdeg\leq \min(d-p-3,p)\) In this case, the eigenvalues are given by
\begin{align}\label{eq:eigenvalue-product 2}
	\tilde{\lambda}_\sfn^{(d-p-2)}(r) = \gamma^{(\wdeg)}_{\sfn_1} + \frac{\nu^{(d-p-2-\wdeg)}_{\sfn_2}}{r^2}~.
\end{align}
Note that \(\widetilde{\lambda}_\sfn\) does not coincide with \(\lambda_\sfn\), due to the fact that $\gamma_{\sfn_1}^{(\wdeg-1)}$ is always different than $\gamma_{\sfn_1}^{(\wdeg)}$ for all values of \(\deg\) that they can respectively be defined. 

Having constructed the eigenforms in \cref{eigenform fam 1}, we note that we can define a second family of eigen $p$-forms via, 
\begin{align}\label{eq:Psi-p-tricky}
	\phi^{(\downarrow,\wdeg)}_{\sfn}(r) \coloneqq \frac{1}{\sqrt{\widetilde{\lambda}_\sfn(r)}}*\dd{\tilde{\phi}^{(\downarrow,\wdeg)}_{\sfn}(r)}~.
\end{align}
This is indeed also an eigenform the transversal Laplacian, with eigenvalue \(\widetilde{\lambda}_{\sfn}(r)\). It is orthogonal to \cref{eq:Psi-p-obvious}, thus defining a second family of non-zero-mode \(p\)-form solutions to \cref{eq:EPS}. Similarly, we can define a second family of solutions for the \((d-p-2)\)-forms via a similar construction:
\begin{align}\label{eq:Psi-co-p-tricky}
	\tilde{\phi}^{(\uparrow,\wdeg)}_{\sfn}(r) \coloneqq \frac{1}{\sqrt{\lambda_\sfn(r)}}*\dd{\phi^{(\uparrow,\wdeg)}_{\sfn}(r)}.
\end{align}
One may check that altogether, these provide a full basis of solutions to \cref{eq:EPS}.

To lighten notation, we collect all the variable indices: \(i=\uparrow,\downarrow\) which labels the family of solutions, \(\wdeg\) which labels the different bigradings of the eigenform in the respective sphere factors, and \(\sfn\) that labels the quantum numbers of the eigenform, into a single multi-index, \(\inda\). Therefore, we have constructed the \(p\)- and \((d-p-2)\)- eigenforms on \(\Sigma_r\),
\begin{align}
	\phi_\inda \equiv \phi_{\sfn}^{(i,\wdeg)} \qq{and} \tilde{\phi}_\inda \equiv \tilde{\phi}_{\sfn}^{(i,\wdeg)}
\end{align}
that make up an orthogonal basis
\begin{align}
	\ip{\phi_\inda}{\phi_{\inda'}} = \frac{1}{2}\delta_{\inda\inda'} \qq{and} \ip{\tilde{\phi}_\inda}{\tilde{\phi}_{\inda'}} = \frac{1}{2}\delta_{\inda\inda'}~.
\end{align}
The funny normalization by $1/2$ is chosen for future convenience (cf. \cref{app: generalized curl}).

\section{The generalized curl operator}\label{app: generalized curl}

Here, we describe in more detail the generalized curl operator $\mathcal{D}_p$, introduced in the main text in \cref{generalized curl}. To start, we  present its general definition and the generalized curl eigenvalue equation. Then, we go on to explicitly construct the eigenforms of this operator---of which there are two families for the non-zero modes. Lastly, we calculate certain inner products of these eigenforms which will be useful for our computations of the radial evolution. 

Let \(\Sigma\) be a closed \(d-1\)-dimensional manifold, and consider the following space of co-closed $k$-forms:
\begin{align}
	\Omega_\perp^k(\Sigma)\coloneqq \set{\omega\in \Omega^k(\Sigma)\suchthat \dd*\omega =0}.
\end{align}
in line with the previous section, such forms are referred to as transversal. Acting on transversal forms, the Hodge Laplacian, \(\lapl_k = \cdd\dd+\dd\cdd\), reduces to \(\cdd\dd=(-1)^{k(d+1)+1}*\dd*\dd\), and is a self-adjoint operator.

For our purposes, it is useful to introduce a packaging of forms of different degree into one vector-like object. We notate this as,
\begin{align}
	\Omega_\perp^{k,\ell}(\Sigma)\coloneqq \Omega_\perp^{k}(\Sigma)\oplus \Omega_\perp^{\ell}(\Sigma)~,
\end{align}
where elements in \(\Omega_\perp^{k,\ell}\) are \emph{transversal \((k,\ell)\)-forms} (often we omit \textquote{transversal} for brevity).
It is valid to consider \(\Omega_\perp^{k,\ell}(\Sigma)\) as a Hilbert space, endowed with an inner product which is given by the inner products of \(\Omega_\perp^{k}\) and \(\Omega_\perp^\ell\) component-wise.

For such transversal $k,\ell$-forms, a useful operator is the \textquote{square-root} of the transversal Laplacian, defined as follows:
\begin{align}
	 & \B_k\coloneqq \mqty(0 & (-1)^{k(d-1)+1}*_\Sigma {\dd}_\Sigma\\ *_\Sigma {\dd}_\Sigma & 0):\quad \Omega_\perp^{k,d-k-2}(\Sigma)\to \Omega_\perp^{k,d-k-2}(\Sigma).
\end{align}
This operator can be thought of as a generalization of the familiar curl operator of three-dimensional calculus, \(\curl\sim *_\Sigma {\dd}_\Sigma\). In three dimensions, the operator \(*_\Sigma {\dd}_\Sigma\) is sometimes called the Beltrami operator \cite{Enciso:2010bel}. In general dimensions, we will refer to \(\B_k\) as the \emph{generalized curl operator}.

Indeed, by performing the generalized curl twice, we reproduce the Laplacians for the respective form components:
\begin{align}
	\B_k\B_k = \mqty(\dmat[0]{\lapl_k,\lapl_{d-k-2}})~.
\end{align}
Moreover, it is easy to show that \(\B_k\) is self-adjoint, with respect to the inner product on $\Sigma$. Self-adjointness is guaranteed by the choice of sign difference between the two non-zero components of \(\B_k\). This property implies that any element of \(\Omega_\perp^{k,d-k-2}\) can be validly expanded in eigen-\((k,d-k-2)\)-forms of \(\B_k\).

Another observation is that the spectrum of \(\B_k\) is symmetric around zero, due to the fact that it anticommutes with \(\parity\), the Pauli-\(z\) matrix,
\begin{align}
	\parity\B_k = - \B_k\parity~.
\end{align}
We can view the Pauli-$z$ matrix as a ``chirality'' operator, which grades eigenforms of the generalized curl operator---we shall henceforth call these eigenforms \textquote{curly eigenforms}---by their chirality. In particular, if \(\Phi_\inda\) is a curly eigenform, with non-zero eigenvalue \(\mu_{\inda}\), then it satisfies the eigenvalue equation:
\begin{align}\label{eq:curl-eigen}
	\B_k \Phi_{\inda} = \mu_{\inda} \Phi_{\inda},.
\end{align}
The combination \(\parity \Phi_{\inda}\) is a linearly independent eigenform of \(\B_k\), but instead has the eigenvalue \(-\mu_{\inda}\). With this in mind, we will denote the curly eigenforms as \(\Phi_{\inda \sigma}\), where \(\sigma= +,-\)  labels the chirality and therefore the sign of the eigenvalue \(\mu_{\inda\sigma}=\sigma \mu_\inda \). Both $\Phi_{\inda\pm}$ are eigenforms of the Laplacians \(\lapl_k\oplus \lapl_{D-k-1}\) with the same eigenvalue, \(\mu^2_\inda\).

\subsection{Curly eigenforms} \label{app: eigenform ansatz}

Now, let us move on to explicitly construct the spectrum of eigenforms and eigenvalues of \(\B_k\) on the manifold of interest, \(\Sigma=\S^p\times\S^{\pp}_r\).

\subsubsection{Zero modes}

Starting with the zero modes; that is, the transversal \((p,d-p-2)\)-forms which satisfy
\begin{equation}
    \B_k \Phi_0 = 0~.
\end{equation}
It is clear that the action of \(\B_k\) factorizes on the different form degree components,
\begin{equation}
    \Phi_0 = \mqty(\phi_0 \\[5pt] \tilde{\phi}_0)~.
\end{equation}
and is solved by taking \(\phi_0\) to be a harmonic \(p\)-form and \(\tilde{\phi}_0\) a harmonic \((d-p-2)\)-form. We refer the reader back to section \cref{laplacian zero modes}, where we describe the harmonic forms available on $\Sigma_r$. 

\subsubsection{Non-zero modes}

Moving on now to the non-zero modes, those satisfying \cref{eq:curl-eigen} with non-zero \(\mu_{\inda \sigma}\). Clearly, from the behavior under the action of $\vec{\sigma}_z$, the curly eigenforms take the form:%
\footnote{The reason we re-use the notation $\phi$ and $\tilde{\phi}$ here as previously used in \cref{sec: laplacian nonzero} is that when $q=p$ these are exactly the same objects, since eigenforms of the generalized curl operators are necessarily eigenforms of the Laplacian.}
\begin{equation}
	\Phi_{\inda \sigma} \coloneqq \mqty(\phi_{\inda} \\[5pt] \sigma\,\tilde{\phi}_{\inda})~,
\end{equation}
where \(\phi_\inda\) and \(\tilde{\phi}_\inda\) are eigen \(p\)- and \((d-p-2)\)-forms of the Laplacian, respectively, as described in \cref{sec: spherical harmonics}, and \(\sigma=\pm\) determines their chirality. 

In order to obtain more explicit expressions of \(\Phi_{\inda \sigma}\), we re-expand the multi-multi-index \(\inda\) into its constituent parts: \(\inda = (i,\wdeg,\sfn_1,\sfn_2)\). Here $i=\uparrow,\, \downarrow$ labels the family of eigenvalues---of which there are two---, $\wdeg$ is a form degree that is able to take a value in a given range depending on the family, and $\sfn_{1/2}$ are labels of the eigenvalues of the Laplacians for the respective spheres. 

\subsubsection*{First family: \tps{i=\uparrow}{i up}}

The first family of eigenforms we can construct are labeled by \(i=\uparrow\). As we are interested in co-closed forms, for $k \neq 0,\,p,\,2p-d-1$, there are no harmonic forms on either spheres, meaning that \(\cY_{[k],\sfn_1}\) and \(\cW_{[k],\sfn_2}\) must be co-exact:
\begin{align}
	\cY_{[k],\sfn_1} = *_p \sd \bbY_{[p-k-1],\sfn_1}~, \qquad \cW_{[k],\sfn_2} & = *_{\pp} \sd \bbW_{[d-p-2-k],\sfn_2}~,
\end{align}
where $\cY_{[k],\sfn_1}$ are eigenforms of the Laplacian on $S^p$ and $\cW_{[k],\sfn_2}$ eigenforms on $S^{\pp}$, and the double-stroke forms also denote eigenforms of the Laplacian.
From this, one can construct a curly eigenform as:
\begin{equation} \label{beltrami eigenfunction}
	\Phi^{(\uparrow,\wdeg)}_{\sfn,\sigma} = \cN_{\sfn}^{(\uparrow,\wdeg)}
	\begin{pmatrix}
		*_p \sd{\bbY_{[d-p-3-\wdeg],\sfn_1}} \w *_{\pp}\sd{\bbW_{[\wdeg],\sfn_2}} \\[5pt]
		\, \mu_{n\sigma}^{-1} *_{\Sigma} \sd \qty(*_p \sd{\bbY_{[d-p-3-\wdeg],\sfn_1}} \w *_{\pp}\sd{\bbW_{[\wdeg],\sfn_2}}\,)
	\end{pmatrix}
\end{equation}
where the associated eigenvalue is given by,
\begin{equation}\label{mu and r}
	\mu_{\sfn\sigma}= \sigma \sqrt{\gamma_{\sfn_1}^{(d-p-3-\wdeg)}+\frac{\nu_{\sfn_2}^{(\wdeg)}}{r^2}}~,
\end{equation}
and $\gamma_{\sfn_1}$ is the eigenvalue on $S^p$ and $\nu_{\sfn_2}$ that on $S^{\pp}$. The normalization factor,
\begin{align}
	\cN_{\sfn}^{(\uparrow,\wdeg)} = \qty(\frac{2\;\gamma_\sfn^{(d-p-3-\wdeg)}\; \nu_{\sfn_2}^{(\wdeg)}}{r^2})^{-1/2}~,
\end{align}
ensures that $\ip{\Phi_{\inda}}{\Phi_{\inda'}} =\delta_{\inda\inda'}$, and we remind the reader that spherical harmonics are already normalized to have unit norm on their respective spheres, including the factors of the radius. 
The second line of \cref{beltrami eigenfunction} can be expanded more explicitly as,
\begin{equation}\label{beltrami inner prod}
	\begin{split}
		*_{\Sigma} \sd \qty(*_p \sd{\bbY_{[d-p-3-\wdeg]\sfn_1}} \w *_{\pp}\sd{\bbW_{[\wdeg]\sfn_2}}) =                     \\
		(-1)^{d(p+\wdeg)-1}
		\bigg[ & (-1)^{d+p}\gamma_{\sfn_1}^{(d-p-3-\wdeg)} \bbY_{[d-p-3-\wdeg],\sfn_1}\w \sd \bbW_{[\wdeg],\sfn_2}             \\
		       & + (-1)^q\, \frac{\nu_{\sfn_2}^{(\wdeg)}}{r^2} \sd \bbY_{[d-p-3-\wdeg],\sfn_1}\w \bbW_{[\wdeg],\sfn_2}\bigg]~.
	\end{split}
\end{equation}
For consistency of the form degrees of $\bbY_{[d-p-3-\wdeg],\sfn_1}$ and $\bbW_{[\wdeg],\sfn_2}$, it is necessary that $h$ takes values in the range given by:
\begin{equation}\label{q range 1}
	\max(0,d-2p-3)\leq \wdeg \leq d-p-3~.
\end{equation}

\subsubsection*{Second family: \tps{i=\downarrow}{i down}}

$\uparrow$ is not the entire story, as we can construct a second family, labeled $i=\downarrow$, that is orthogonal to the first. This family is constructed as,
\begin{equation} \label{beltrami eigenfunction 2}
	\Phi^{(\downarrow,\wdeg)}_{\sfn,\sigma} = \mathcal{N}^{(\downarrow,\wdeg)}_{\sfn,\sigma}
	\begin{pmatrix}
		(-1)^{dp+p+1}\,\mu_{n\sigma}^{-1} *_{\Sigma} \sd \qty(*_p \sd{\bbY_{[p-1-\wdeg],\sfn_1}} \w *_{\pp} \sd{\bbW_{[\wdeg],\sfn_2}})\, \\[5pt]
		*_p \sd{\bbY_{[p-1-\wdeg],\sfn_1}} \w *_{\pp}\sd{\bbW_{[\wdeg],\sfn_2}}
	\end{pmatrix}~,
\end{equation}
with the normalization
\begin{align}
    	\cN_{\sfn}^{(\downarrow,\wdeg)} = \qty(\frac{2\;\gamma_\sfn^{(p-\wdeg-1)}\; \nu_{\sfn_2}^{(\wdeg)}}{r^2})^{-1/2}~. 
\end{align}
We can similarly expand out the first component more explicitly as,
\begin{equation}\label{beltrami inner prod 2}
	\begin{split}
		*_{\Sigma} \sd \qty(*_p \sd{\bbY_{[p-\wdeg-1],\sfn_1}} \w *_{\pp}\sd{\bbW_{[\wdeg],\sfn_2}})
		=   \\
           (-1)^{dp+1}\bigg[
		 & (-1)^{p}\gamma_{\sfn_1}^{(p-\wdeg-1)} \bbY_{[p-\wdeg-1],\sfn_1}\w \sd \bbW_{[\wdeg],\sfn_2}  \\
         &+ (-1)^{h}\frac{\nu_{\sfn_2}^{(\wdeg)}}{r^2} \sd \bbY_{[p-\wdeg-1],\sfn_1}\w \bbW_{[\wdeg],\sfn_2}
			\bigg]
            ~.
	\end{split}
\end{equation}
In this case, consistency of the form degrees of $\bbY_{[p-1-\wdeg],\sfn_1}$ and $\bbW_{[\wdeg],\sfn_2}$, requires
\begin{equation}\label{q range 2}
	0\leq \wdeg\leq \min(p-1,\pp)~.
\end{equation}

\subsection{A radial Berry connection}\label{app: inner prod computation}

With the explicit definitions of the eigenforms in hand, we can now compute inner products---or \textquote{radial Berry connections}---of the form,%
\footnote{Note that, more general inner products of the form \(\ip{\Phi_{\inda \sigma}}{\pd_r \Phi_{\inda' \sigma'}}\) only have non-trivial contributions if \(\inda'=\inda\) due to the orthonormality of the eigenforms. However, these inner products are not diagonal in the chirality plane, and therefore we consider those with differing $\sigma$ and $\sigma'$.}
\begin{align}
   \bbA^\inda_{\sigma\sigma'} =  \ip{\pd_r\Phi_{\inda \sigma}}{\Phi_{\inda \sigma'}}~.
\end{align}
While at present, interest in such inner products may seem rather arbitrary, these are an important ingredient in our computations of the radial evolution in \cref{diffeq vs}, as well as for the radial evolution of the currents n \cref{app: exact solutions}.

To go about computing these inner products, we employ the useful identity,
\begin{align} \label{inner prod relation}
	0 = \pd_r \delta_{\sigma \sigma'} = \pd_r \ip{\Phi_{\inda \sigma}}{\Phi_{\inda \sigma'}}_\Sigma  = \ip{\pd_r \Phi_{\inda \sigma}}{\Phi_{\inda \sigma'}}_\Sigma + \ip{\Phi_{\inda \sigma}}{*_{\Sigma}^{-1} \pd_r \qty(*_{\Sigma} \Phi_{\inda \sigma'})}_\Sigma~,
\end{align}
reducing our problem to computing $*^{-1}_{\Sigma} \pd_r (*_{\Sigma} \Phi_{\inda \sigma})$.

\subsubsection*{A Hodge-star identity}

To compute $*^{-1}_{\Sigma} \pd_r (*_{\Sigma} \Phi_{\inda \sigma})$, we first take a step back and consider a generic $k$-form $\f{f}{k} \in \Omega^k(\Sigma)$ with $\Sigma =S^p\times S^{\pp}_r$ with the metric,
\begin{equation}
	\dd{s^2} = \dd\Omega^2_{p} +r^2\dd\Omega_{\pp}^2~.
\end{equation}
A $k$-form can be decomposed into a sum over forms $\alpha_{[i]} \in \Omega^i(S^p)$ and $\beta_{[j]} \in \Omega^{j}(S^{\pp})$~,%
\footnote{Strictly speaking, $\alpha_{[i]} \in \Omega^i(S^p)\otimes \Omega^0(S^{\pp})$ and $\beta_{[j]} \in \Omega^0(S^p)\otimes \Omega^{j}(S^{\pp})$.}
\begin{equation}
	\f{f}{k}= \sum_{i+j=k} \alpha_{[i]}\w \beta_{[j]}~,
\end{equation}
such that taking $*_{\Sigma}$ of $\f{f}{k}$ splits into the Hodge stars over the respective spheres,
\begin{equation}
	*_{\Sigma} \f{f}{k} = \sum_{i+j=k}*_{\Sigma} (\alpha_{[i]}\w \beta_{[j]}) = \sum_{i+j=k} (-1)^{j(p-i)}r^{\pp-2j} \qty(*_p \alpha_{[i]}) \w \qty(*_{\pp} \beta_{[j]})~.
\end{equation}
Consequently,
\begin{equation}
	\pd_r \qty(*_{\Sigma} \f{f}{k}) = *_{\Sigma} \pd_r \f{f}{k}+\sum_{i+j=k} \frac{\pp-2j}{r}*_{\Sigma} (\alpha_{[i]}\w \beta_{[j]}) ~.
\end{equation}
This can be re-expressed as,
\begin{equation}\label{dr and decomp}
	\pd_r \qty(*_{\Sigma} \f{f}{k}) = *_{\Sigma} \pd_r \f{f}{k}+ \frac{\pp}{r}*_{\Sigma} \f{f}{k} -\frac{2}{r}*_{\Sigma} \mathcal{P}(\f{f}{k})~,
\end{equation}
where we have defined the operator $\mathcal{P}\;:\;\Omega^\ell (\Sigma) \to \Omega^\ell (\Sigma)$ which acts as,
\begin{equation}
	\mathcal{P}\left( \sum_{i+j=k} \alpha_{[i]} \w \beta_{[j]}  \right)= \sum_{i+j=k} j \alpha_{[i]} \w \beta_{[j]} ~.
\end{equation}
Alternatively, we can express this operator via $\Pi_j\;;\; \Omega^\ell(\Sigma) \to \Omega^{\ell-j}(S^p)\otimes \Omega^j(S^{\pp})$ which projects an $\ell$-form onto a degree-$j$ form on the $S^{\pp}$ sphere.
\begin{equation}
	\mathcal{P}=\sum_{j=0}^{\pp}j\Pi_j
\end{equation}
Here, $\Pi_j$ is a projector and is a self-adjoint operator with respect to the inner product on $\Sigma$.

\subsubsection*{Back to Berry}

Applying \cref{dr and decomp} to the problem at hand, we find that
\begin{equation}\label{eq:Phi-Dr-Phi}
\begin{split}
	\ip{\Phi_{\inda \sigma}}{*^{-1}_\Sigma\,\pd_r*_{\Sigma} \Phi_{\inda \sigma'}}_\Sigma &= \ip{\Phi_{\inda \sigma}}{\pd_r\Phi_{\inda \sigma'}}_\Sigma+\frac{\pp}{r}\ip{\Phi_{\inda \sigma}}{\Phi_{\inda \sigma'}}_\Sigma \\ 
    &\phantom{=~}-\frac{2}{r} \sum_{j=1}^{\pp} j \ip{\Pi_j \Phi_{\inda \sigma}}{\Pi_j \Phi_{\inda \sigma'}}_\Sigma~,
    \end{split}
\end{equation}
where we have utilized the self-adjointness of $\Pi_j$, and applied $\Pi_j$ component-wise on the components of the vector $\Phi_{\inda \sigma}\in \Omega^{p,d-p-2}(\Sigma)$, abusing notation to write $\text{diag}(\Pi_j,\Pi_j)$ as simply $\Pi_j$.

Applying this to either family of eigenforms, \cref{beltrami eigenfunction} or \cref{beltrami eigenfunction 2}, yields results with the same structure,
\begin{align}
    \sum_{j=1}^{\pp} j \ip{\Pi_j \Phi_{\inda \sigma}}{\Pi_j \Phi_{\inda \sigma}}_\Sigma = \frac{1}{2}\qty(\pp - \frac{\nu^{(\wdeg)}_{\sfn_2}/r^2}{\gamma^{(k)}_{\sfn_1}+\nu^{(\wdeg)}_{\sfn_2}/r^2})~,
\end{align}
where we introduced the shorthand
\begin{align}
    k=
    \begin{cases}
        d-p-3-\wdeg~,& i=\uparrow \\ 
        p-\wdeg-1~,& i=\downarrow~.    
    \end{cases}
\end{align}
Running through a similar computation for the opposite chirality, one finds
\begin{align}
    \sum_{j=1}^{\pp} j \ip{\Pi_j \Phi_{\inda \sigma}}{\Pi_j \Phi_{\inda\, -\sigma}}_\Sigma = \frac{1}{2}\qty(d-p-3-2\wdeg + \frac{\nu^{(\wdeg)}_{\sfn_2}/r^2}{\gamma^{(k)}_{\sfn_1}+\nu^{(\wdeg)}_{\sfn_2}/r^2})~.
\end{align}

Putting this altogether, and using \cref{inner prod relation}, we conclude that the radial Berry connection has components:
\begin{equation}\label{eq:berry}
\begin{aligned}
	\bbA^\inda_{\pm\pm} &= \ip{\Phi_{\inda \pm}}{\pd_r  \Phi_{\inda \pm}}_\Sigma  = \frac{1}{2}\frac{\pd_r \mu_\inda(r)}{\mu_\inda(r)}                     \\
	 \bbA^\inda_{\pm\mp} &=\ip{\Phi_{\inda \pm}}{\pd_r  \Phi_{\inda \mp}}_\Sigma  = -\frac{1}{2}\frac{\pd_r \mu_\inda(r)}{\mu_\inda(r)}+\frac{d-p-3-2\wdeg}{2r}~.
\end{aligned}
\end{equation}
The above inner products in turn imply that for the components of $\Phi_{\inda\sigma}$:
\begin{equation}\label{derivative Phi_i}
    \begin{split}
        \ip{\phi_{\inda +}}{\pd_r \phi_{\inda +}}_\Sigma&=\frac{d-p-3-2\wdeg}{4r}\\
        \ip{\tilde{\phi}_{\inda +}}{\pd_r \tilde{\phi}_{\inda +}}_\Sigma&=\frac{1}{2}\frac{\pd_r \mu_{\inda}}{\mu_{\inda}}-\frac{d-p-3-2\wdeg}{4r}~.
    \end{split}
\end{equation}
where we point out the components are normalized as $\ip{\phi_{\inda}}{\phi_{\inda'}}_\Sigma =\ip{\tilde{\phi}_{\inda}}{\tilde{\phi}_{\inda'}}_\Sigma =\frac{1}{2} \delta_{\inda \inda'}$, which ensures that $\ip{\Phi_\inda}{\Phi_{\inda'}}_\Sigma = \delta_{\inda\inda'}$.

\subsubsection*{Berry for zero modes}
Lastly, for completeness, we want to also mention similar inner products for the zero modes. Starting from the useful identity in \cref{inner prod relation} in the first line, and then using the Hodge star identity in the second, we find that for the $p$-form zero mode component $\phi_0$:
\begin{equation}
\begin{split}
    \pd_r \left\langle \phi_0,\phi_0\right\rangle_\Sigma &=\left\langle \phi_0,\pd_r\phi_0\right\rangle_\Sigma +\left\langle \phi_0,*^{-1}_{\Sigma}\pd_r *_{\Sigma}\phi_0\right\rangle_\Sigma \\
    &=2\left\langle \phi_0,\pd_r\phi_0\right\rangle_\Sigma +\frac{\pp}{r}\left\langle \phi_0,\phi_0\right\rangle_\Sigma -\frac{2}{r}\sum j\left\langle \Pi_j \phi_0,\Pi_j \phi_0\right\rangle_\Sigma 
\end{split}
\end{equation}
Since $\phi_0$ is a $p$-form proportional to the volume form of $S^p$, in this case $\Pi_j\phi_0=0$. Using that, $\left\langle \phi_0,\phi_0\right\rangle=1$, this implies
\begin{equation}\label{zero mode inner prod}
\begin{split}
    \left\langle \phi_0,\pd_r\phi_0\right\rangle_\Sigma =-\frac{\pp}{2r}\left\langle \phi_0,\phi_0\right\rangle_\Sigma ~,
\end{split}
\end{equation}
a result which we use in the computation of $\cB_0$ in \cref{B0 intermediate}.

\section{Radial evolution of the currents} \label{app: exact solutions}

In this appendix, we provide the details on the differential equation governing the radial evolution of the radial component of the currents $J_r,\,\tilde{J}_r$, and its explicit solution in terms of Bessel functions.

This should be viewed as an alternate, complementary approach to the radial evolution of the $v,\,\tilde{v}$ appearing in \cref{solution for v}, which is used for the construction of the operators $\cB_\inda$. There, the $\cB_\inda$ operators are defined at $r_0$ as,
\begin{equation}\label{eq:Ba-app-def}
    \cB_\inda =\frac{v_{\inda}+\tilde{v}_{\inda}}{2}\J_{\inda +}+\frac{v_{\inda}-\tilde{v}_{\inda}}{2}\J_{\inda -}~.
\end{equation}
To radially evolve the charges $\cB$ inward to act on states, one way is to give $v_\inda$ and $\tilde{v}_\inda$ a radial profile, viewed as functions on $\S^p\times B^{q+1}$, with boundary conditions reproducing \cref{eq:Ba-app-def}. This is what was done in  \cref{sec: the ops}. Alternatively, recalling that $\J_{\inda\sigma}$ is composed out $J_r$ and $\tilde{J}_r$ by taking specific inner products \cref{decomp to eigenmodes}, we can solve for the radial profile of the currents instead. These two approaches are complementary. Here we detail the second approach. We note that for the operator $\cB_\inda$ to be topological, the radial evolution matrix of the currents must be the inverse of that of the coefficients $\qty(v_\inda,\tilde{v}_\inda)$.

\subsection{The radial evolution equation}

We begin by decomposing the conserved currents into their radial and spatial parts:
\begin{equation}
	\f{J}{p+1}=\dd{r}\w  J_r+J_{\Sigma}\, \quad , \quad \f{\tilde{J}}{\pp}= \dd{r}\w \tilde{J}_r+\tilde{J}_{\Sigma}~,
\end{equation}
where $\tilde{J}_{\Sigma}$ and $J_{\Sigma}$ do not have non-zero components along $\dd{r}$. 
The two currents are related to one-another by
\begin{equation}
    *J=\frac{\ii 2\pi}{\g^2}\tilde{J}~.  
\end{equation}
With the above decomposition, this implies,
\begin{equation}\label{current relations}
    \begin{split}
        *_\Sigma J_r &= \frac{2\pi i }{g^2}\tilde{J}_\Sigma \\
        \dd{r}*_\Sigma J_\Sigma &=  (-1)^{p+1} \frac{2\pi i }{g^2}\dd{r} \tilde{J}_r ~.
    \end{split}
\end{equation}
Using these relations, the conservation law of the currents can now be re-expressed as, 
\begin{align}
	\begin{cases}
		\dd*J=0 \\[0.5em]
		\dd*\tilde{J}=0
	\end{cases}\Longrightarrow \begin{cases}
	    \sd *_\Sigma J_r -\dfrac{2\pi \ii}{\g^2}\dd{r}\w \sd \tilde{J}_r +\dd{r}\w \pd_r *_\Sigma J_r =0 \\[0.5em]
        \sd *_\Sigma \tilde{J}_r +\dd{r}\w \pd_r *_\Sigma \tilde{J}_r +(-1)^{(d-1)(p-1)}\dfrac{\ii \g^2}{2\pi}\dd{r}\w \sd J_r =0~,
	\end{cases}
\end{align}
where we have decomposed the exterior derivative as $\dd=\dd{r}\w \pd_r +\sd$. Separating out the radial and $\Sigma$ components, the above relation gives the two Gauss' laws:
\begin{equation}
\begin{split}
    \sd*_\Sigma J_r =0~,\\
		\sd *_\Sigma\tilde{J}_r = 0~,
\end{split}
\end{equation}
and two equations governing the radial evolution of the current components:
\begin{equation}\label{eq:curren-r-eq-1}
    \begin{split}
        \pd_r *_\Sigma J_r-\frac{2\pi \ii}{\g^2} \sd \tilde{J}_r =0\\
         \pd_r *_\Sigma \tilde{J}_r +(-1)^{(d-1)(p-1)}\frac{\ii g^2}{2\pi} \sd J_r =0~.\\
    \end{split}
\end{equation}
    
Enforcing the two Gauss' laws, i.e. co-closedness of the radial components of the currents, allows us to decompose $J_r$ and $\tilde{J}_r$ in terms of eigenmodes of the generalized curl operator (cf. \cref{app: generalized curl}):
\begin{equation}\label{current decomp}
\begin{pmatrix}
    \frac{\g}{\sqrt{2\pi}}J_r\\[0.5em]
\frac{\sqrt{2\pi}}{\g}\tilde{J}_r
\end{pmatrix}=\sum_{\inda \sigma}\J_{\inda,\sigma}\Phi_{\inda \sigma}~,
\end{equation}
where again the factors of $\g$ are determined from dimensional analysis. Plugging \cref{current decomp} into the radial equation \cref{eq:curren-r-eq-1} implies a radial evolution equation for the current modes:  
\begin{equation}\label{radial evolution}
     *_{\Sigma}^{-1}\pd_r\qty\big( \J_{\inda,\sigma}*_{\Sigma}\Phi_{\inda,\sigma})+(-1)^p\,\ii \J_{\inda,\sigma} \vec{\sigma}_z \B_p\, \Phi_{\inda,\sigma}=0~.
\end{equation}
Taking the inner product of this with $\Phi_{\inda+}$ and $\Phi_{\inda-}$, respectively, yields a system of differential equations:
\begin{equation}
    \begin{split}
        \pd_r \J_{\inda +}- \left\langle \Phi_{\inda +},\pd_r\Phi_{\inda,+}\right\rangle_\Sigma  \J_{\inda+}- \left\langle \Phi_{\inda +},\pd_r\Phi_{\inda,-}\right\rangle_\Sigma  \J_{\inda-} -(-1)^p\ii \mu_{\inda} \J_{\inda,-} &= 0\\
        \pd_r \J_{\inda -}- \left\langle \Phi_{\inda -},\pd_r\Phi_{\inda,+}\right\rangle_\Sigma  \J_{\inda+}- \left\langle \Phi_{\inda -},\pd_r\Phi_{\inda,-}\right\rangle_\Sigma  \J_{\inda-} +(-1)^p\ii  \mu_{\inda}\J_{\inda,+} &= 0~,
    \end{split}
\end{equation}
where we have also used \cref{inner prod relation}. Now we can assemble $\J_{\inda+}$ and $\J_{\inda-}$ into a vector, $\vec{\mathrm{J}}_\inda$ and with the aid of the radial Berry connection, $\bbA^\inda$, with components $\bbA_{\sigma\sigma'}^\inda$ as introduced in \cref{eq:berry}, these differential equations can be written compactly as,
\begin{equation}\label{radial evo eq}
    \qty(\pd_r - \bbA^\inda) \vec{\mathrm{J}}_{\inda}(r) + (-1)^{p} \mu_\inda(r) \, \vec{\sigma}_y \vec{\mathrm{J}}_{\inda}(r) = 0~.
\end{equation}

\subsection{The zero modes}

Here, we first solve the radial evolution for the zero modes $j_0$. This serves as a warm-up for the general solution for non-zero modes which we tackle in the next section. Note that we stick to the case $d\neq 2p+2$ such that for this choice of $\Sigma = S^p\times S^{\pp}$ we only have the $j_0$ zero mode component.

Starting from the differential equation in \cref{radial evolution}, and using the fact that for zero modes $\B_p \phi_0=0$, the radial equation in simplifies to,
\begin{equation}
    \qty(\pd_r j_0 )\phi_0 + j_0 *_\Sigma^{-1} \pd_r *_\Sigma \phi_0=0~.
\end{equation}
The inner product with $\phi_0$, together with \cref{zero mode inner prod} implies
\begin{equation}
    \pd_r j_0 +\frac{\pp}{2r}j_0=0~,
\end{equation}
and has the general solution $j_0(r) = c\, r^{-q/2}$, such that in all,
\begin{equation}
    j_0(r) = \left(\frac{r_0}{r}\right)^{\frac{\pp}{2}}j_0(r_0) ~.
\end{equation}
This solution for general $d$ matches with the results in \cite{Hofman:2024oze, Vitouladitis:2025zoy} for the specific cases treated in those papers, $p=1,\,d=4$ and $p=0$ with general $d$, respectively.

\subsection{The general solution}\label{gen solution radial}

In order to solve the differential equation for the radial evolution for non-zero modes, we will go through various redefinitions and changes of basis, to massage \cref{radial evo eq} into a convenient form. To begin, we note that the Berry connection \cref{eq:berry} can be diagonalized using a constant Hadamard rotation:
\begin{align}
	\sfH^{-1} \bbA^\inda\, \sfH = \bbA^\inda_\t{diag} = \mqty(\dmat{\dfrac{d-p-3-2\wdeg}{2r},\dfrac{\pd_r \mu_\inda}{\mu_\inda}-\dfrac{d-p-3-2\wdeg}{2r}})~,\quad \,\quad  \sfH=\frac{1}{\sqrt{2}}\mqty(1 & 1 \\ 1 & -1)~.
\end{align}
Observe, further that the diagonalized Berry connection is, in fact, pure gauge. In particular,
\begin{equation}
    \bbA^\inda_\t{diag} =  \left(\pd_r \sfD_\inda(r)\right)\sfD_\inda^{-1}(r)~, \qq{where} \sfD_\inda(r) = \begin{pmatrix}
	    r^{(d-p-3-2\wdeg)/2} & 0 \\ 0 & \mu_\inda(r)\; r^{-(d-p-3-2\wdeg)/2}
	\end{pmatrix}~,
\end{equation}
such that it can be removed by a gauge transformation. In our variables this amounts to redefining
\begin{equation}
    \vec{\mathrm{G}}_\inda = \sfD_\inda(r)^{-1}\qty(\sfH\, \vec{\mathrm{J}}_\inda)~.
\end{equation}
In terms of this new variable, the radial equation reads:
\begin{align}
    \pd_r \vec{\mathrm{G}}_\inda + (-1)^p \,\mu_\inda(r)\, \qty[\sfD_\inda(r)^{-1} \, \sfH \vec{\sigma}_y \sfH \sfD_\inda (r)] \vec{\mathrm{G}}_{\inda}(r) &= 0~,
\intertext{or equivalently}
    \label{diffeq G}
    \pd_r\vec{\mathrm{G}}_\inda + (-1)^{p} \mqty(\admat{\ii\, r^{-(d-p-3-2\wdeg)} \mu_\inda^2(r),-\ii\, r^{d-p-3-2\wdeg}})\vec{\mathrm{G}}_\inda &= 0~.
\end{align}
Re-expanding out $\vec{\mathrm{G}}_\inda$ into its components $\mathtt{G}_{\inda\pm}$ gives

\begin{equation}\label{eq:pde-G-components}
    \begin{split}
        \pd_r \mathtt{G}_{\inda+}+\ii \,(-1)^{p}\mu_\inda^2(r) r^{-(d-p-3-2\wdeg)}\mathtt{G}_{\inda-} &= 0~, \\
        \pd_r \mathtt{G}_{\inda -} +  \ii\, (-1)^{p+1}\, r^{(d-p-3-2\wdeg)}\mathtt{G}_{\inda +} &= 0~.
    \end{split}
\end{equation}
Substituting the second line into the first, yields a second order ODE entirely in terms of $\mathtt{G}_{\inda-}$,
\begin{equation}
    \pd_r^2 \mathtt{G}_{\inda -} - \frac{d-p-3-2\wdeg}{r}\; \pd_r \mathtt{G}_{\inda -} - \mu_\inda^2(r) \mathtt{G}_{\inda -} = 0~.
\end{equation}
From here, one can see that with a further rescaling of \(\mathtt{G}_\inda\) by a factor of \(r^{(d-p-2-2\wdeg)/2}\) to absorb the awkward coefficient in the first derivative term, the resulting ODE looks like the modified Bessel equation. There are two independent solutions to this equation:
\begin{align}\label{modified modified bessels}
	\cI_\inda(r) \coloneqq r^{\frac{d-p-2-2\wdeg}{2}}\, I_{\delta_\inda}\!\qty\big(\sqrt{\gamma_\inda} r) \qq{and} \cK_\inda(r)\coloneqq r^{\frac{d-p-2-2\wdeg}{2}}\, K_{\delta_\inda}\!\qty\big(\sqrt{\gamma_\inda} r)~,
\end{align}
where \(I_\delta(x)\) and \(K_\delta(x)\) are the modified Bessel functions of order \(\delta_\inda = \frac{1}{2}\sqrt{(d-p-2-2\wdeg)^2+4\,\nu_\inda}\).%
\footnote{Using the explicit value of $\nu_\inda$ in \cref{eq:eigenvalue-sphere}, we can also write $\delta_\inda=\ell_1+(d-p-2)/2$.}
\(I_\delta(x)\) is regular at \(x\to 0\), while \(K_\delta(x)\) is singular; this is an important point as it implies that the operator defined as a combination of the currents (these are the $\cB_\inda$ and $\cB^\dagger_\inda$ operators) when shrunken to $r=0$ are only trivial if there are no divergent contributions at $r=0$. These solutions completely determine \(\mathtt{G}_{\inda -}(r)\) as a general linear combination of the two independent solutions,
\begin{equation}
    \mathtt{G}_{\inda-}(r)=c_1\, \cI_\inda(r) + c_2\, \cK_\inda(r)
\end{equation}
where the coefficients will be determined by imposing the relevant boundary conditions for our problem. This solution, along with the first line of \cref{eq:pde-G-components}, allows us to then determine $\mathtt{G}_{\inda+}$ in terms of rescaled, modified Bessel functions. 

Having obtained the solution to the radial evolution of $\mathtt{G}_{\inda\pm}$, we have, in principle, achieved our goal. The only remaining task is to rotate back to the original currents \(\vec{\mathrm{J}}_\inda\) and impose the boundary conditions, at $r_0$:
\begin{equation}\label{eq:BC-J}
	\vec{\mathrm{J}}_\inda(r_0) = \vec{Q}_\inda~.
\end{equation}
Doing this, we obtain
\begin{align}\label{eq:J-sol}
	\vec{\mathrm{J}}_\inda (r) = U_\inda(r,r_0) \vec{Q}_\inda~,
\end{align}
where \(U_\inda(r,r_0)\) is the radial evolution matrix of the form :
\begin{align}
	U_\inda(r,r_0) = \tilde{U}_\inda(r)\, \tilde{U}_\inda^{-1}(r_0)~,
\end{align} 
with
\begin{align}
	\tilde{U}_\inda(r) = \sfH\; \sfD_\inda(r)\; \sfM_\inda(r)~.
\end{align}
\(\sfM_\inda(r)\) is a fundamental matrix for the system \cref{diffeq G}, given by:
\begin{align}
	\sfM_\inda(r) = \mqty(\ii\, (-1)^{p}\, r^{-(d-p-3-2\wdeg)} \cI'_\inda(r) & \ii\, (-1)^{p}\, r^{-(d-p-3-2\wdeg)} \cK'_\inda(r) \\[7pt] \cI_\inda(r) & \cK_\inda(r))~,
\end{align}
with the prime denoting the derivative with respect to \(r\). Carrying out the remaining matrix multiplication, the radial evolution matrix \(\tilde{U}_\inda\) takes the final form:
\begin{align}\label{U radial evo}
	\tilde{U}_\inda(r) = \frac{r^{-(d-p-3-2\wdeg)/2}}{\sqrt{2}}
	\mqty(\phantom{-}\mu_\inda(r) \cI_\inda(r) + \ii (-1)^p \cI'_\inda(r) & \phantom{-}\mu_\inda(r) \cK_\inda(r) + \ii (-1)^p \cK'_\inda(r) \\[8pt] 
	-\mu_\inda(r) \cI_\inda(r) + \ii (-1)^p \cI'_\inda(r) & -\mu_\inda(r) \cK_\inda(r) + \ii (-1)^p \cK'_\inda(r))~.
\end{align} 
One can verify that this evolution matrix \(\vec{\mathrm{J}}_\inda\) as in \cref{eq:J-sol} indeed solves the radial evolution equation and the boundary conditions desired  for $\vec{\mathrm{J}}_\inda$.

From the differential equation defining $U(r,r_0)$, it is straightforward to get the explicit expression for its determinant:\begin{equation}\label{determinant U}
	\det(U_\inda(r,r_0)) = \exp(\int_{r_0}^r \tr \bbA^\inda) = \frac{\mu_\inda(r)}{\mu_\inda(r_0)}~.
\end{equation}
Armed with this determinant, we are able to then determine the commutation relation of the $\J_{\inda\sigma}$ valued at any $r$.
\begin{equation}
	\begin{split}
		\comm{\J_{\inda\sigma}(r)}{\J_{\inda' \sigma'}(r)} & =\det(U_\inda(r,r_0))\comm{\J_{\inda\sigma}(r_0)}{\J_{\inda' \sigma'}(r_0)}      \\
		 & =\frac{i}{2\pi}\sigma\mu_{\inda}(r)\delta_{\inda,\inda' }\delta_{\sigma,-\sigma'}
	\end{split}
\end{equation}

\section{Spin-\texorpdfstring{\(4\)}{4} current from the dressed currents}\label{app:spin-four}

In this appendix we spell out all the details to obtain the spin-\(4\) current of \cite{Anselmi:1999bb} from our dressed currents. We work in flat space, in the conformal case
\(d=2p+2\). We write \(\vec{\kappa}=[\kappa_1\cdots \kappa_p]\), and repeated \(\vec{\kappa}\)'s include the normalization factor \(1/p!\) in their contraction.

The conserved spin-4 current for a conformal \(p\)-form gauge field is \cite{Anselmi:1998bh,Anselmi:1999bb}:
\begin{align}\label{eq:anselmi-spin-four}
	T^{(4)}_{\mu\nu\rho\sigma}
	=
	\qty[
	f^+_{\rho\vec{\kappa}}
	\overleftrightarrow{\pd_\mu}\,
	\overleftrightarrow{\pd_\nu}
	f^-_{\sigma}{}^{\vec{\kappa}}
	-
	\frac{1}{d+3}\,
	\Pi_{\mu\nu}
	\qty(f^+_{\rho\vec{\kappa}}f^-_{\sigma}{}^{\vec{\kappa}})
    ]_\t{symm}~.
\end{align}
In the above, \(f^\pm\) denotes the (anti-)self-dual field strength%
\footnote{For odd \(p\) the relevant projector should be complexified \cite{Hofman:2024oze}.}
\(f^\pm \coloneqq \frac{1}{2}\qty(f\pm * f)\), the bidirectional derivative, \(\overleftrightarrow{\pd_\mu}\), acts as
\begin{align}
    	A\overleftrightarrow{\pd_\mu} B
	=
	A\,\pd_\mu B
	-\pd_\mu A\; B~,
\end{align}
\(\Pi_{\mu\nu}\) is a projector defined as
\begin{align}
    \Pi_{\mu \nu} = \pd_\mu \pd_\nu - \delta_{\mu \nu}\, \pd^2~,   
\end{align}
and \(\qty[\,\cdot\,]_\t{symm}\) denotes symmetrization in all free indices, i.e. a sum over the necessary permutations divided by the number of permutations. 

Translated to our currents, \(J\) and \(\tilde{J}\) from \cref{electric current,magnetic current} the spin-4 current reads:%
\footnote{This is not to say that \(f_{\rho\vec{\kappa}}^+ f_{\sigma}^{-\vec{\kappa}}\) is proportional to the stress tensor, but that their symmetrized parts match: \(\qty[\Pi_{\mu\nu} (f_{\rho\vec{\kappa}}^+ f_{\sigma}^{-\vec{\kappa}})]_\t{symm} \propto \qty\big[\Pi_{\mu\nu}T_{\rho\sigma}^{(2)}]_\t{symm}\).}
\begin{align}\label{eq:anselmi-spin-four-J}
	T^{(4)}_{\mu\nu\rho\sigma}
	=
	\qty[
	-K_{\mu\nu\rho\sigma}
	-\frac{\g^2}{2}
	\frac{1}{d+3}\,\Pi_{\mu\nu}T^{(2)}_{\rho\sigma}
    ]_\t{symm}~,
\end{align}
where
\begin{align}
   	K_{\mu\nu\rho\sigma}
	&\coloneqq
	\frac{\g^4}{4}\,
	J_{\rho\vec{\kappa}}
	\overleftrightarrow{\pd_\mu}\,
	\overleftrightarrow{\pd_\nu}
	J_{\sigma}{}^{\vec{\kappa}}
	+
	\pi^2\,
	\tilde{J}_{\rho\vec{\kappa}}
	\overleftrightarrow{\pd_\mu}\,
	\overleftrightarrow{\pd_\nu}
	\tilde{J}_{\sigma}{}^{\vec{\kappa}}~,
\end{align}
and \(T^{(2)}_{\rho \sigma}\) is the stress tensor \cref{eq:stress-tensor}. We have already shown how to obtain the stress tensor from the dressed currents in the main text. Thus, we only need to show how to obtain \(K_{\mu \nu \rho \sigma}\).

Let \(\xi^{\alpha\beta\gamma}\) be a constant symmetric tensor. First take
\begin{align}\label{eq:spin-four-eta-zero}
	\eta^{(0)}_{\vec{\kappa}}
	&=
	-\frac{\g^4}{2}\,
	\xi^{\alpha\beta\gamma}
	\pd_\alpha\pd_\beta J_{\gamma\vec{\kappa}}~,
	\\
	\tilde{\eta}^{(0)}_{\vec{\kappa}}
	&=
	-2\pi^2\,
	\xi^{\alpha\beta\gamma}
	\pd_\alpha\pd_\beta \tilde{J}_{\gamma\vec{\kappa}}~.
\end{align}
These parameters satisfy \cref{condition eta}. The resulting dressed current gives the endpoint splitting,
\begin{align}\label{eq:spin-four-endpoint}
	\cJ^{(0)}_\mu = \xi^{\alpha\beta\gamma} Z^{(0)}_{\mu\alpha\beta\gamma}~, \qquad
	Z^{(0)}_{\mu\alpha\beta\gamma} = -2
	\qty(
	\frac{\g^4}{4}\,
	J_{\mu\vec{\kappa}}\pd_\alpha\pd_\beta J_{\gamma}{}^{\vec{\kappa}}
	+
	\pi^2\,
	\tilde{J}_{\mu\vec{\kappa}}\pd_\alpha\pd_\beta
	\tilde{J}_{\gamma}{}^{\vec{\kappa}}
    )~.
\end{align}
The middle splitting uses the second family. The ansatz
\cref{eq:family-2-ansatz} contains one constant vector \(v\). To contract
with a general \(\xi^{\alpha\beta\gamma}\), take the linear combination with a basis vector \(v=e_\alpha\), with components \(e_\alpha^\mu=\delta^\mu_\alpha\), and sum over
\(\alpha\). Thus, \(\pounds_{e_\alpha}=\pd_\alpha\). We then choose
\begin{align}\label{eq:spin-four-eta-one}
	\eta^{(1,\alpha)}_{\vec{\kappa}}
	&=
	\frac{\g^4}{2}\,
	\xi^{\alpha\beta\gamma}
	\pd_\beta J_{\gamma\vec{\kappa}}~,
	\\
	\tilde{\eta}^{(1,\alpha)}_{\vec{\kappa}}
	&=
	2\pi^2\,
	\xi^{\alpha\beta\gamma}
	\pd_\beta \tilde{J}_{\gamma\vec{\kappa}}~.
\end{align}
For each fixed \(\alpha\), these parameters also satisfy \cref{condition eta}. The corresponding dressed current is
\begin{align}\label{eq:spin-four-middle}
	\cJ^{(1)}_\mu
	= 
	\xi^{\alpha\beta\gamma} Z^{(1)}_{\mu\alpha\beta\gamma}~,
    \qquad
    Z^{(1)}_{\mu\alpha\beta\gamma}
    =\qty(
	\frac{\g^4}{2}\,
	\pd_\beta J_{\gamma\vec{\kappa}}\,
	\pd_\alpha J_{\mu}{}^{\vec{\kappa}}
	+
	2\pi^2\,
	\pd_\beta\tilde{J}_{\gamma\vec{\kappa}}\,
	\pd_\alpha\tilde{J}_{\mu}{}^{\vec{\kappa}}
    )~.
\end{align}

We now have all the ingredients. To spell out the match, let \(B\) denote either \(J\) or
\(\tilde{J}\), with coefficient
\begin{align}
	c_B=
	\begin{cases}
		\g^4/4~, & B=J~,\\
		\pi^2~, & B=\tilde{J}~.
	\end{cases}
\end{align}
The \(k=0\) family contributes
\begin{align}
	Z^{(0)}_{\mu\alpha\beta\gamma} = \sum_{B} \qty(-2c_B\,
	B_{\mu\vec{\kappa}}\pd_\alpha\pd_\beta
	B_{\gamma}^{~\vec{\kappa}})~,
\end{align}
while the \(k=1\) family gives
\begin{align}
	Z^{(1)}_{\mu\alpha\beta\gamma} = \sum_{B} \qty( 2c_B\,
	\pd_\beta B_{\gamma\vec{\kappa}}\,
	\pd_\alpha B_{\mu}^{~\vec{\kappa}})~.
\end{align}
On the other hand,
\begin{align}
    K_{\mu \nu \rho \sigma} = \sum_{B} \qty(c_B\,B_{\rho\vec{\kappa}}
	\overleftrightarrow{\pd_\mu}\,
	\overleftrightarrow{\pd_\nu}
	B_{\sigma}{}^{\vec{\kappa}})~,
\end{align}
with
\begin{align}
	c_B\,B_{\rho\vec{\kappa}}
	\overleftrightarrow{\pd_\mu}\,
	\overleftrightarrow{\pd_\nu}
	B_{\sigma}^{~\vec{\kappa}}
	&=
	c_B\,B_{\rho\vec{\kappa}}\pd_\mu\pd_\nu B_{\sigma}^{~\vec{\kappa}}
	-c_B\,\pd_\mu B_{\rho\vec{\kappa}}\pd_\nu B_{\sigma}^{~\vec{\kappa}} \\
    &\phantom{=~}
	-c_B\,\pd_\nu B_{\rho\vec{\kappa}}\pd_\mu B_{\sigma}^{~\vec{\kappa}}
	+c_B\,\pd_\mu\pd_\nu B_{\rho\vec{\kappa}}B_{\sigma}^{~\vec{\kappa}}~.
\end{align}
The first and last terms are equal after symmetrization, and so are the
two middle terms. Hence, we get precisely:
\begin{align}\label{eq:spin-four-reproduced}
	\qty[Z^{(0)}_{\mu \nu \rho \sigma} + Z^{(1)}_{\mu \nu \rho \sigma}]_\t{symm} 
	= \qty[
	-K_{\mu\nu\rho\sigma}^{\vphantom{{(0)}}}
    ]_\t{symm}~.
\end{align}
This concludes the proof that the spin-4 current is reproduced from the two families of dressed currents.


\printbibliography
\end{document}